\documentclass[a4paper,aps,prd,preprint,nofootinbib,superscriptaddress,showpacs,showkeys]{revtex4-1}
\usepackage{epsf,graphicx,amssymb,amsfonts,amsmath,bm,bbm,hyperref,slashed,xcolor}
\usepackage[nodisplayskipstretch]{setspace}

\newcommand{\etab}{\bm{\eta}}
\newcommand{\xib}{\bm{\xi}}
\newcommand{\kb}{\bm{k}}
\newcommand{\Kb}{\bm{K}}

\newcommand{\Db}{\bm{D}}
\newcommand{\Eb}{\bm{E}}
\newcommand{\Bb}{\bm{B}}
\newcommand{\rb}{\bm{r}}
\newcommand{\Rb}{\bm{R}}
\newcommand{\nablab}{\bm{\nabla}}
\newcommand{\sigmab}{\bm{\sigma}}
\newcommand{\Dlr}{\overleftrightarrow{\bm{D}}}
\newcommand{\Dlrn}{\overleftrightarrow{D}}

\allowdisplaybreaks
\begin{document}

\preprint{TUM-EFT 74/15}

\title{Poincar\'e invariance in NRQCD and pNRQCD revisited}
\author{Matthias Berwein}
\email{matthias.berwein@riken.jp}
\affiliation{Nishina Center, RIKEN, Wako, Saitama 351-0198, Japan}
\affiliation{Physik-Department, Technische Universit\"{a}t M\"{u}nchen, James-Franck-Str.~1, 85748 Garching, Germany}
\author{Nora Brambilla}
\email{nora.brambilla@ph.tum.de}
\affiliation{Physik-Department, Technische Universit\"{a}t M\"{u}nchen, James-Franck-Str.~1, 85748 Garching, Germany}
\affiliation{Institute for Advanced Study, Technische Universit\"{a}t M\"{u}nchen, Lichtenbergstra{\ss}e 2a, 85748 Garching, Germany}
\author{Sungmin Hwang}
\email{egilels@hotmail.com}
\affiliation{Physik-Department, Technische Universit\"{a}t M\"{u}nchen, James-Franck-Str.~1, 85748 Garching, Germany}
\affiliation{Max-Planck-Institut f\"{u}r Physik, F\"{o}hringer Ring~6, 80805 M\"{u}nchen, Germany}
\author{Antonio Vairo}
\email{antonio.vairo@ph.tum.de}
\affiliation{Physik-Department, Technische Universit\"{a}t M\"{u}nchen, James-Franck-Str.~1, 85748 Garching, Germany}

\begin{abstract}
We investigate how fields transform under the Poincar\'e group in nonrelativistic effective field theories of QCD. In constructing these transformations, we rely only on symmetries and field redefinitions to limit the number of allowed terms. By requiring invariance of the action under these transformations, nontrivial relations between Wilson coefficients for both nonrelativistic QCD and potential nonrelativistic QCD are derived. We show explicitly how the Poincar\'e algebra is satisfied, and how this gives complementary information on the Wilson coefficients. We also briefly discuss the implications of our results, as well as the possibility of applying this method to other types of effective field theories.
\end{abstract}


\maketitle

\section{Introduction and outline}

Effective field theories (EFTs) are a standard tool for particle and nuclear physics and have been for at least forty years~\cite{Weinberg:1978kz}. Low energy EFTs have been constructed for different sectors of the Standard Model to describe specific low energy systems: for example, EFTs of Quantum Electrodynamics (QED) to describe atomic physics, or EFTs of Quantum Chromodynamics (QCD) to describe hadronic and nuclear physics. High energy EFTs, on the other hand, provide a systematic framework for investigating physics beyond the Standard Model. In this case, however, there is not yet an experimentally confirmed underlying renormalizable theory from which to derive the EFT. Hence, high energy EFTs are built relying only on symmetry arguments. In all cases, the construction of an EFT, not being bound by renormalizability, requires an increasing number of operators when going to higher orders in the expansion, which may limit its predictive power.

In this paper, we focus on nonrelativistic EFTs. Nonrelativistic EFTs describe systems where the mass $M$ of the heavy particle(s) is much larger than any other relevant energy scale of the system, including the scale $\Lambda_{\text{QCD}}$ of confinement in QCD. The fact that the mass $M$ is much larger than the momentum and energy of the heavy particle implies that its velocity is much smaller than the velocity of light, qualifying the particle as nonrelativistic. There exists a wide variety of such EFTs. Among them we will concentrate on the Heavy Quark Effective Theory (HQET), nonrelativistic QCD (NRQCD), and potential NRQCD (pNRQCD), which are nonrelativistic EFTs of QCD. The HQET~\cite{Isgur:1989vq,Eichten:1989zv,Georgi:1990um,Manohar:2000dt} is a low energy EFT for heavy-light mesons, NRQCD~\cite{Caswell:1985ui,Bodwin:1994jh} provides a nonrelativistic effective description for the dynamics of heavy quarks and antiquarks, and pNRQCD~\cite{Pineda:1997bj,Brambilla:1999xf,Brambilla:2004jw} is an effective theory for heavy quark-antiquark bound states (heavy quarkonia).

The Lagrangian of an EFT is organized as an expansion in the inverse of the high energy scale that has been integrated out; in the case of nonrelativistic EFTs, this is the quark mass $M$. It contains all terms allowed by the symmetries of the EFT and can be schematically written as
\begin{equation}
\mathcal{L}_{\text{EFT}} = \sum_n c_n\frac{\mathcal{O}_n}{M^{d_n-4}}\,,
\end{equation}
where the operators $\mathcal{O}_n$, made up of the fields that describe the effective degrees of freedom, are of mass dimension $d_n$, and the $c_n$ are the \emph{matching} or \emph{Wilson coefficients} of the EFT. These coefficients contain all the information from the high energy scale. They are determined by \emph{matching} to the underlying theory.

In the case of pNRQCD, one proceeds further by integrating out all energy scales larger than the binding energy of the bound state. The originating Wilson coefficients are called \emph{potentials}, because they are the potentials appearing in the Schr\"odinger equation. How precisely pNRQCD is constructed depends on the relation among the three different energy scales: $p\sim Mv, E\sim Mv^2$, and $\Lambda_{\text{QCD}}$, where $v \ll 1$ is the heavy (anti)quark relative velocity, $p$ is the heavy (anti)quark relative momentum, and $E$ is the binding energy. For systems satisfying the hierarchy $\Lambda_{\text{QCD}}\lesssim Mv^2$ (these are top-antitop states near threshold and possibly the lowest lying bottomonium and charmonium states), the integration of the relative momentum can be done in perturbation theory. The case $\Lambda_{\text{QCD}}\lesssim Mv^2$ is called \emph{weak-coupling} case. If the binding energy is smaller than the typical hadronic scale, $\Lambda_{\text{QCD}}\gg Mv^2$, which is called the \emph{strong-coupling} case, then $\Lambda_\mathrm{QCD}$ is integrated out as well. Although the matching is in this case nonperturbative, the resulting effective Lagrangian is somewhat simpler than in the weakly-coupled case. The reason is the absence of colored degrees of freedom (quark-antiquark color octet states and gluons).

The Wilson coefficients of these EFTs have to be determined by matching to the underlying theory, i.e., QCD. Beyond leading order in the coupling or in the expansion parameter, this can easily become technically involved, and even more so for a theory with more than one expansion parameter like pNRQCD. For this reason, one would like to exploit as much prior knowledge on the Wilson coefficients as possible before commencing the matching calculation.

Due to the nonrelativistic expansions, Poincar{\'e} invariance is no longer manifest in a nonrelativistic EFT. A physical system is symmetric under Poincar{\'e} transformations when its action is invariant under spacetime translations, rotations, and boost transformations. Dirac showed that the potentials of a quantum mechanical Hamiltonian satisfy nontrivial relations if one imposes the Poincar{\'e} algebra~\cite{Dirac:1949cp}. His analysis was extended to interacting relativistic composite systems~\cite{Foldy:1960nb,Krajcik:1974nv,Pauri:1975mr,Pauri:1976kz,Sebastian:1979rr}, where relations between the relativistic corrections were derived using the Poincar{\'e} algebra. As these quantum mechanical systems can be generalized into EFTs, it is natural to expect that also some nontrivial relations between the Wilson coefficients of nonrelativistic EFTs can be deduced in a systematic way from Poincar\'e invariance. It is well justified to assume this invariance, as the EFT is by construction equivalent at each order of the expansion to the original quantum field theory, which is invariant under Poincar\'e transformations.

In~\cite{Brambilla:2003nt} (see also~\cite{Brambilla:2001xk,Vairo:2003gx}), Poincar{\'e} invariance has been imposed on NRQCD and pNRQCD by constructing all generators of the symmetry group in these EFTs. The generators corresponding to spacetime translations and rotations have been obtained in the usual closed form from the associated conserved Noether currents. The generators of boosts, on the other hand, have been derived from a general ansatz that includes all operators allowed by the other symmetries (such as parity, $P$, charge conjugation, $C$, and time reversal, $T$) up to a certain order in the expansion; in other words, the general principles for the construction of the EFT Lagrangian have been applied also to the boost generators. Demanding that all generators satisfy the commutation relations of the Poincar{\'e} algebra provides some exact constraints on the Wilson coefficients of the EFTs.

Reparametrization invariance is a symmetry found in low energy EFTs of QCD, like the HQET or Soft Collinear Effective Theory (SCET)~\cite{Bauer:2000ew, Bauer:2000yr, Bauer:2001ct, Bauer:2001yt}. In these EFTs, the momentum of the high energy particles is separated into a large and a small component. This separation is arbitrary by a small shift of the momenta that preserves the hierarchy of the energy scales. Requiring the Lagrangian to be invariant with respect to this shift leads to a number of nontrivial relations between the Wilson coefficients~\cite{Luke:1992cs, Manohar:2002fd}. In the HQET case, these relations have been shown to be equivalent to the ones obtained from Poincar\'e invariance~\cite{Brambilla:2003nt}. This is not surprising, since both approaches are closely related: a shift in the parametrization of the high energy momentum may be interpreted as a change of the reference frame. Whereas the implementation of reparametrization invariance might have some advantages, its applications are limited. Poincar\'e invariance, on the other hand, is a general principle that all quantum field theories have to obey.

Recently, another approach has been suggested for deriving constraints in EFTs through Poincar\'e invariance that employs Wigner's \emph{induced representation}~\cite{Wigner:1939cj} (see also~\cite{Weinberg:1995mt} for a textbook presentation). It has been proposed in~\cite{Heinonen:2012km} that a free nonrelativistic field $\phi$, which has a well defined transformation behavior under rotations $R$ as $\phi(x)\to D[R]\phi(R^{-1}x)$, should transform under a generic Lorentz transformation $\Lambda$ as
\begin{equation}\label{induced}
 \phi(x)\rightarrow D[W(\Lambda, i\partial)]\phi(\Lambda^{-1}x)\,.
\end{equation}
The transformation $W$ is a particular rotation depending on the Lorentz transformation $\Lambda$ and also the momentum of the field $\phi$. The resulting expression is then expanded in powers of derivatives (momenta) according to the nonrelativistic power counting.

While this seems to work well for noninteracting fields, some issues arise in an interacting gauge theory. First, the boost transformation~\eqref{induced} does not have the right behavior under gauge transformations. One would like to have the boosted field to transform in the same way as the original field would at the new coordinates, but this is not possible because of the derivatives in the induced representation. Promoting the derivatives to gauge covariant derivatives fixes the problem, but it introduces an ambiguity in how the covariant derivatives are ordered. It is also necessary to add additional gauge field dependent terms to the boost in order to cancel some terms that would prevent the EFT Lagrangian from being invariant. Ultimately, the constraints obtained in this way agree with previous results in NRQCD and nonrelativistic QED (NRQED)~\cite{Manohar:1997qy}; new results in NRQED were derived in~\cite{Hill:2012rh}.

In this paper, we address some of the questions left open by this method. 
First, the appearance of additional terms in the boost transformation, whose coefficients turn out to be linear combinations of the Wilson coefficients from the Lagrangian, is very reminiscent of the construction of EFT Lagrangians, where one includes all terms allowed by the symmetries of the theory. However, terms with derivatives originating from the induced representation were assumed to be free of radiative corrections. We address the question if the specific choice of coefficients made in~\cite{Heinonen:2012km} for the terms appearing in the boost transformation can be justified, fully or partially, through  constraint enforced by Poincar\'e invariance.

A second question is the following. In~\cite{Heinonen:2012km}, the addition of new gauge field dependent terms to the boost transformation, which do not arise from just promoting derivatives in the induced representation to covariant derivatives, seems to be born out of necessity: without them the nonrelativistic Lagrangian could not be made invariant under the boost transformation. It is conceivable, however, that other terms could be added to the boost, which are allowed by the symmetries, but are not strictly necessary for an invariant Lagrangian. They may have the effect to relax some of the constraints on the Wilson coefficients, since each extra term introduces a new parameter.

Thus, we take from~\cite{Heinonen:2012km} that the boost transformation of the nonrelativistic field is realized in a nonlinear way, and that requiring the invariance of the Lagrangian under this boost leads to constraints on the Wilson coefficients, but apart from that we will not refer to the induced representation. Instead, we follow the EFT logic of~\cite{Brambilla:2003nt} by including all possible terms in the boost transformation that are allowed by the other symmetries of the theory (such as $P$, $C$, $T$) and by assigning a Wilson coefficient to each of them. Even though we start with a general expression, we will exploit the possibility to redefine the effective fields in order to remove terms from the general ansatz.

Lastly, since the boost generator for the field transformation has to satisfy the Poincar\'e algebra, we show how the commutation relations have to be implemented in the case of a nonlinear boost generator. Requiring all commutators of the Poincar\'e algebra to be satisfied leads to additional constraints on the boost parameters as well as on the Wilson coefficients of the Lagrangian. It is important to note here that, like in~\cite{Brambilla:2003nt}, this approach is defined for bare fields and couplings. However, our results hold also in the renormalized EFTs if Poincar\'e invariance is not broken by quantum effects (anomalies)~\cite{Brambilla:2004jw}.

In summary, we provide a tool for the construction of EFTs in which not all of the fundamental symmetries are manifest. Since these latent symmetries nevertheless emerge in the form of constraints on the Wilson coefficients, thereby limiting the number of independent parameters, the computational or experimental effort required to determine the EFT at a given order is reduced considerably if these constraints are taken into account. Apart from answering the fundamental question of how such latent symmetries are realized in the EFTs in terms of nonlinear field transformations, this is expected to be of relevance whenever new EFTs need to be developed, such as in beyond the Standard Model physics, or already established EFTs need to be extended to higher orders. We explicitly demonstrate the method for the two examples of NRQCD and pNRQCD, where the latent symmetry corresponds to boost transformations. Since, however, the method relies only on the single assumption that a nonlinear realization of the latent symmetries exists, 
it should be possible to extend it to other theories and symmetries as well.

The paper is organized as follows. In Sec.~\ref{consNRQCD}, we study NRQCD, first discussing the boost generators of the Poincar\'e group in the EFT approach in Sec.~\ref{NRQCDgen} and how they satisfy the Poincar\'e algebra. Generators for spacetime translations as well as rotations are found in Appendix~\ref{PoincareAlgebra}, as the derivation is well-known and not directly related to the main discussion. We then derive the constraints for the Wilson coefficients of the Lagrangian up to order $M^{-3}$ in the two-fermion sector in Sec.~\ref{NRQCDinv}, up to order $M^{-4}$ in the four-fermion sector in Sec.~\ref{NRQCD4f}, and up to order $M^{-5}$ in the four-fermion sector in Appendix~\ref{appendix}. We also write the Noether charges obtained from the Poincar\'e transformations in Sec.~\ref{Noether}, and show that they correspond to the quantum field generators constructed in~\cite{Brambilla:2003nt}. We then study the pNRQCD case in Sec.~\ref{conspNRQCD}. First, we derive how quarkonium fields transform under boosts in Sec.~\ref{pNRQCDcoord}, then we use field redefinitions to remove terms from the most general boost generators in Sec.~\ref{pNRQCDredef}, and, finally, we obtain constraints on the Wilson coefficients of the pNRQCD Lagrangian in Sec.~\ref{pNRQCDinv}. We conclude the paper in Sec.~\ref{discussion} with a summary and a short outlook on possible applications to other effective field theories.

\section{Constraints in NRQCD}\label{consNRQCD}

Nonrelativistic QCD (NRQCD) is the EFT obtained from QCD after integrating out modes associated with the scale of the heavy quark mass $M$~\cite{Caswell:1985ui,Bodwin:1994jh}. The effective degrees of freedom are nonrelativistic Pauli spinor fields $\psi$ and $\chi$, where $\psi$ annihilates a heavy quark and $\chi$ creates a heavy antiquark, as well as gluon fields $A_\mu$ and light quark fields $q_l$ (which will be assumed massless) with four-momenta constrained to take values much smaller than $M$. Its Lagrangian up to $\mathcal{O}\left(M^{-2}\right)$ is given by
\bigskip
\begin{align}\label{LNRQCD}
 \mathcal{L}_\mathrm{NRQCD}={}&\psi^\dagger\left\{iD_0+\frac{c_2}{2M}\Db^2+\frac{c_F}{2M}g\Bb\cdot\sigmab+\frac{c_D}{8M^2}\bigl[\Db\cdot,g\Eb\bigr]+\frac{ic_S}{8M^2}\bigl[\Db\times,g\Eb]\cdot\sigmab\right\}\psi\notag\\
 &+\chi^\dagger\left\{iD_0-\frac{c_2}{2M}\Db^2-\frac{c_F}{2M}g\Bb\cdot\sigmab+\frac{c_D}{8M^2}\bigl[\Db\cdot,g\Eb\bigr]+\frac{ic_S}{8M^2}\bigl[\Db\times,g\Eb]\cdot\sigmab\right\}\chi\notag\\
 &+\mathcal{L}^{(2)}\bigr|_{4f}+\text{Tr}\left[\Eb^2-\Bb^2\right]-\frac{d_2}{M^2}\left[\Eb^a\cdot\left(D^2\Eb\right)^a-\Bb^a\cdot\left(D^2\Bb\right)^a\right]\notag\\
 &-\frac{d_3}{M^2}gf^{abc}\left[3\left(\Eb^a\times\Eb^b\right)\cdot\Bb^c- \left(\Bb^a\times\Bb^b\right)\cdot\Bb^c\right]+\mathcal{L}^{(2)}_{\text{light}}+\mathcal{O}(1/M^3)\,.
\end{align}

\bigskip\noindent
$\mathcal{L}^{(2)}\bigr|_{4f}$ contains terms made of four quark fields up to order $M^{-2}$. These may consist of heavy (anti)quark fields and/or light quark fields. We explicitly write the four fermion terms made of two heavy quark and two heavy antiquark fields in the following Eq.~\eqref{L24f}. In the rest of the paper, we will neglect other four fermion terms (see~\cite{Bauer:1997gs,Moreno:2017sgd}), because we are not interested in the Poincar\'e invariance constraints from those sectors. $\mathcal{L}^{(2)}_{\text{light}}$ is the light quark sector of the Lagrangian, which contains all terms with two light quark fields (and gluons) up to order $M^{-2}$, whose leading term is $\sum_l\bar{q}_li\slashed{D}q_l$. Note that terms made exclusively of light fields have to be constructed in a manifestly Poincar\'e invariant way, so they do not give rise to constraints. We denote the Wilson coefficients in the heavy quark sector by $c$ and the ones in the gluon sector by $d$~\cite{Cho:1994yu, Novikov:1983gd, Balk:1993ev}.

We use the convention $D_0=\partial_0+igA_0$ and $\bm{D}=\bm{\nabla}-ig\bm{A}$ for the sign of the coupling constant $g$ in the covariant derivatives, from which one obtains the chromoelectric and chromomagnetic fields as $\Eb=\dfrac{1}{ig}\bigl[D_0,\bm{D}\bigr]$ and $\Bb=\dfrac{i}{2g}\bigl\{\bm{D}\times,\bm{D}\bigr\}$, while $\bm{\sigma}$ denotes the vector made of the three Pauli matrices; in addition, we define $D^2 = D_0^2 - \boldsymbol{D}^2$. The commutator or anticommutator with a cross product is defined as\footnote{Because of the antisymmetry of the cross product, the roles of commutator and anticommutator are in fact reversed: $\{\bm{X}\times,\bm{Y}\}_i=\epsilon_{ijk}[X_j,Y_k]$ and $[\bm{X}\times,\bm{Y}]_i=\epsilon_{ijk}\{X_j,Y_k\}$.}
\begin{equation}
 \bigl[\bm{X}\times,\bm{Y}\bigr]=\bm{X}\times\bm{Y}-\bm{Y}\times\bm{X}\hspace{20pt}\mathrm{and}\hspace{20pt}\bigl\{\bm{X}\times,\bm{Y}\bigr\}=\bm{X}\times\bm{Y}+\bm{Y}\times\bm{X}\,,
\end{equation}
and equivalently for the dot product. We have made use of the equations of motion\footnote{This is equivalent to performing certain field redefinitions, as shown in~\cite{Arzt:1993gz}. The field redefinitions we will discuss in detail in this work, however, are of a different kind not related to the equations of motion.} to remove all higher time derivatives; we have also removed the constant term $-M\psi^\dagger\psi+M\chi^\dagger\chi$ through the field redefinitions $\psi\to e^{-iMt}\psi$ and $\chi\to e^{iMt}\chi$.

\subsection{Poincar\'e algebra for boost transformations}\label{NRQCDgen}

Before we go into details, we should clarify our notion of transformation. In general, performing a field transformation means to replace any field $\phi$, as well as its derivatives, in the Lagrangian or in other field-dependent quantities by a new field $\phi'$, called the transformed field, and its derivatives. In the quantized theory, this corresponds to a change of variable $\phi\to\phi'$ in the path integral. The transformation constitutes a symmetry if the action remains invariant under this change of variable (in the quantized theory also the path integral measure needs to be considered).

In the case of coordinate transformations, we write
\begin{equation}
 \phi_i(x)\stackrel{\Lambda}{\longrightarrow}\phi'_i(x)\equiv\Lambda_{ij}^{(R)}\phi_j\left(\Lambda^{-1}x\right)\,,
 \label{gentrafo}
\end{equation}
where $\Lambda$ denotes a generic spacetime transformation and the representation $R$ depends on the spin of the field $\phi$. Note that the arrow in Eq.~\eqref{gentrafo} represents the change of the \textit{function} $\phi$ to $\phi'$ and the relation between the two fields is written on the right-hand side of Eq.~\eqref{gentrafo}: the value of the transformed field at position $x$ is given by the value of the original field at the same point, which in the old coordinates corresponds to $\Lambda^{-1}x$, while its orientation is also adapted to the new axes by $\Lambda^{(R)}$. Also note that the notion of active or passive transformations\footnote{An active transformation changes the position or orientation of a physical object with respect to a fixed coordinate system, while a passive transformation keeps the object fixed and changes the coordinates.} does not affect the form of Eq.~\eqref{gentrafo}, it just changes the sign of the generators of $\Lambda$. 
For the record, we will assume passive transformations.

The spacetime transformations contained in the Poincar\'e group are translations, rotations, and boosts, which will be the main subject of this paper. Translations and rotations act on all fields of NRQCD in the usual way (see Appendix~\ref{PoincareAlgebraNRQCD}), so we do not need to discuss them here further. However, boost transformations are a priori not defined in an obvious manner for the heavy (anti)quark fields. We will show how such transformations can be constructed for the heavy (anti)quark fields, but they will no longer be linear in the fields. Gluons and light quarks are still relativistic fields and transform in the usual way under boosts. We will not discuss the light quark fields here, as they do not appear in any operator of interest in this paper, but we will consider the transformations of the gluon fields, distinguishing between their space and time components, as the NRQCD Lagrangian is written in an explicitly nonrelativistic fashion.

The coordinates $(t,\bm{r})$ in a reference frame moving with the infinitesimal velocity $\bm{\eta}$ correspond to $(t+\bm{\eta}\cdot\bm{r},\bm{r}+\bm{\eta}t)$ in a resting frame, where we will always neglect terms of $\mathcal{O}\left(\eta^2\right)$ or higher. The gluons are described by vector fields, whose transformations are identical to those of the coordinates, so Eq.~\eqref{gentrafo} implies that
\begin{align}
 A'_0(t,\bm{r})&=A_0(t+\bm{\eta}\cdot\bm{r},\bm{r}+\bm{\eta}t)-\bm{\eta}\cdot\bm{A}(t,\bm{r})\,,\label{A0boost}\\
 \bm{A}'(t,\bm{r})&=\bm{A}(t+\bm{\eta}\cdot\bm{r},\bm{r}+\bm{\eta}t)-\bm{\eta}A_0(t,\bm{r})\,.\label{Aboost}
\end{align}
It is convenient to perform a Taylor expansion to first order in $\bm{\eta}$ on all fields with transformed coordinates, in order to consistently work only with at most linear terms of the infinitesimal parameter. Since the gluon fields never appear individually in the Lagrangian but always inside covariant derivatives, we also write explicitly the boost transformations for those:
\begin{align}
 D'_0&=\partial_0+igA'_0=D_0+[\etab\cdot t\nablab+\etab\cdot\rb\partial_0,D_0]+\etab\cdot\Db\,,\label{D0boost}\\
 \Db'&=\nablab-ig\bm{A}'=\Db+[\etab\cdot t\nablab+\etab\cdot\rb\partial_0,\Db]+\etab D_0\,.\label{Dboost}
\end{align}
Notice that the sign of the last terms has changed compared to Eqs.~\eqref{A0boost} and \eqref{Aboost}, which is a consequence of the fact that the gauge fields in $D_0$ and $\bm{D}$ have opposite signs. The commutator in the middle terms serves two purposes: first, the commutator with the gauge field ensures that the derivatives (from the Taylor expansion) act exclusively on the gauge field and not on any other field that may be present in the Lagrangian.  Second, the commutator with the derivative cancels the derivative in the last term, ensuring that overall the derivatives on both sides of Eqs.~\eqref{D0boost} and \eqref{Dboost} match.\footnote{As stated above, we replace derivatives of fields in the Lagrangian by derivatives of the transformed fields: $\partial_x\phi(x)\stackrel{\Lambda}{\longrightarrow}\partial_x\phi'(x)$. The typical transformation of derivatives as vectors arises when the transformed fields are replaced by the right-hand side of Eq.~\eqref{gentrafo}.} Finally, the transformations for the chromoelectric and chromomagnetic fields follow directly from their expressions in terms of the covariant derivatives:
\smallskip
\begin{align}
 \Eb'&=\Eb+[\etab\cdot t\nablab+\etab\cdot\rb\partial_0,\Eb]+\etab\times\Bb\,,\\
 \Bb'&=\Bb+[\etab\cdot t\nablab+\etab\cdot\rb\partial_0,\Bb]-\etab\times\Eb\,.
\end{align} 

\smallskip
In the following, we will use $\bm{K}$ to denote the generators of boosts as an operator that may act on any kind of field, and $\bm{k}_\phi$ to denote the explicit expression of $\bm{K}$ when acting on the field $\phi$:
\smallskip
\begin{align}
 \phi(x)\stackrel{K}{\longrightarrow}\phi'(x)&\equiv(1-i\bm{\eta}\cdot\bm{K})\phi(x)\notag\\
 &=(1-i\bm{\eta}\cdot\bm{k}_\phi)\phi(x)\,.
\end{align}
Since each field has the same coordinate transformations, the term $it\bm{\nabla}+i\bm{r}\partial_0$ appears in any $\bm{k}_\phi$, so we can write
\begin{equation}
 \bm{k}_\phi=it\bm{\nabla}+i\bm{r}\partial_0+\bm{\hat{k}}_\phi\,.
\end{equation}
Now $\bm{\hat{k}}_\phi$ denotes the part of the boost transformation acting only on the components of $\phi$ and not the coordinates. The previously introduced notation of writing the coordinate transformations as commutators is particularly convenient when considering transformations of products of fields, 
as by the product rule of commutators we can write
\begin{align}
 \phi'_1\phi'_2={}&\phi_1\phi_2+[\etab\cdot t\nablab+\etab\cdot\rb\partial_0,\phi_1]\phi_2+\phi_1[\etab\cdot t\nablab+\etab\cdot\rb\partial_0,\phi_2]\notag\\
 &+(-i\etab\cdot\bm{\hat{k}}_1\phi_1)\phi_2+\phi_1(-i\etab\cdot\bm{\hat{k}}_2\phi_2)\notag\\
 ={}&\phi_1\phi_2+[\etab\cdot t\nablab+\etab\cdot\rb\partial_0,\phi_1\phi_2]+(-i\etab\cdot\bm{\hat{k}}_1\phi_1)\phi_2+\phi_1(-i\etab\cdot\bm{\hat{k}}_2\phi_2)\,.
\end{align}
In this way, the coordinate transformations can be decoupled from the component transformations, also when performing several consecutive transformations.

For relativistic fields, $\bm{\hat{k}}_\phi$ is some constant matrix, but for the heavy (anti)quark fields, which are nonrelativistic, it takes the form of a function depending on all fields or their derivatives. Apart from the coordinate transformations, all derivatives have to be covariant, so we can write
\begin{align}
 \psi(x)&\stackrel{K}{\longrightarrow}\psi'(x)=\Bigl(1-i\etab\cdot\kb_\psi(D_0,\Db,\Eb,\Bb,\psi,\chi,x)\Bigr)\psi(x)\,,\\
 \chi(x)&\stackrel{K}{\longrightarrow}\chi'(x)=\Bigl(1-i\etab\cdot\kb_\chi(D_0,\Db,\Eb,\Bb,\psi,\chi,x)\Bigr)\chi(x)\,.
\end{align}
In principle, $\bm{\hat{k}}_\phi$ depends on the coordinates only implicitly through the fields; however, the field redefinitions we have performed in order to remove the heavy mass terms from the Lagrangian also affect the boost transformations. So instead of the usual coordinate transformations generated by $it\bm{\nabla}+i\bm{r}\partial_0$, we have
\begin{equation}
 e^{\pm iMt}(it\bm{\nabla}+i\bm{r}\partial_0)e^{\mp iMt}=it\bm{\nabla}+i\bm{r}\partial_0\pm M\bm{r}\,,
\end{equation}
and we will include the terms $M\bm{r}$ and $-M\bm{r}$ in the definitions of $\bm{\hat{k}}_\psi$ and $\bm{\hat{k}}_\chi$, respectively.

Besides the mass terms $\pm M\bm{r}$, there is no obvious expression for the remaining terms of $\bm{\hat{k}}_\psi$ and $\bm{\hat{k}}_\chi$. We have no choice but to make a general ansatz for them and determine the exact parameters through further calculations. Fortunately, the number of possible operators in the boost generators at a given order in $1/M$ is limited, in particular, by the discrete transformations $P$, $C$, and $T$ (parity, charge conjugation, and time reversal). Under them, the coordinates and fields transform as
\begin{align}
 (t,\bm{r})&\stackrel{P}{\longrightarrow}(t,-\bm{r})\,, & (t,\bm{r})&\stackrel{C}{\longrightarrow}(t,\bm{r})\,, & (t,\bm{r})&\stackrel{T}{\longrightarrow}(-t,\bm{r})\,,\label{PCTstart}\\
 \psi&\stackrel{P}{\longrightarrow}\psi\,, & \psi&\stackrel{C}{\longrightarrow}-i\sigma_2\chi^*\,, & \psi&\stackrel{T}{\longrightarrow}i\sigma_2\psi\,,\\
 \chi&\stackrel{P}{\longrightarrow}-\chi\,, & \chi&\stackrel{C}{\longrightarrow}i\sigma_2\psi^*\,, & \chi&\stackrel{T}{\longrightarrow}i\sigma_2\chi\,,\\
 D_0&\stackrel{P}{\longrightarrow}D_0\,, & D_0&\stackrel{C}{\longrightarrow}D_0^*\,, & D_0&\stackrel{T}{\longrightarrow}-D_0\,,\\
 \Db&\stackrel{P}{\longrightarrow}-\Db\,, & \Db&\stackrel{C}{\longrightarrow}\Db^*\,, & \Db&\stackrel{T}{\longrightarrow}\Db\,,\\
 \Eb&\stackrel{P}{\longrightarrow}-\Eb\,, & \Eb&\stackrel{C}{\longrightarrow}-\Eb^*\,, & \Eb&\stackrel{T}{\longrightarrow}\Eb\,,\\
 \Bb&\stackrel{P}{\longrightarrow}\Bb\,, & \Bb&\stackrel{C}{\longrightarrow}-\Bb^*\,, & \Bb&\stackrel{T}{\longrightarrow}-\Bb\,.\label{PCTend}
\end{align}

We expect the boosted fields to transform in exactly the same way under these discrete symmetries, i.e.,

\begin{align}
 P\phi'&=P(1-i\etab\cdot\Kb)\phi=(1-i(-\etab)\cdot(P\Kb))P\phi\,,\\
 C\phi'&=C(1-i\etab\cdot\Kb)\phi=(1-i\etab\cdot(C\Kb))C\phi\,,\\
 T\phi'&=T(1-i\etab\cdot\Kb)\phi=(1+i(-\etab)\cdot(T\Kb))T\phi\,,
\end{align}
where we have also reversed the direction of the infinitesimal velocity $\etab$ for $P$ and $T$.\footnote{Also remember that $T$ takes the complex conjugate of numerical coefficients.} We take from this that the boost generators for the heavy (anti)quark fields need to satisfy
\begin{align}
 P\kb_\psi&=-\kb_\psi\,, & C\kb_\psi&=-\sigma_2\kb_\chi^*\sigma_2\,, & T\kb_\psi&=\sigma_2\kb_\psi\sigma_2\,, \\
 P\kb_\chi&=-\kb_\chi\,, & C\kb_\chi&=-\sigma_2\kb_\psi^*\sigma_2\,, & T\kb_\chi&=\sigma_2\kb_\chi\sigma_2\,,
\end{align}
where the expressions on the left-hand sides mean that the transformed fields and coordinates according to Eqs.~\eqref{PCTstart}-\eqref{PCTend} are to be inserted into the expressions for $\bm{k}_\psi$ and $\bm{k}_\chi$.

General expressions for $\kb_\psi$ and $\bm{k}_\chi$ satisfying these conditions up to $\mathcal{O}(M^{-3})$ are\footnote{Note that, in particular, $\bm{\hat{k}}_{\psi/\chi}=\pm i\sigmab$ are not allowed, even though they would satisfy all commutators of the Poincar\'e algebra, just because they do not reproduce the right $P$ or $T$ transformation behavior. They would be appropriate for Weyl spinors, but here we deal with Pauli spinors.}
\begin{align}\label{kpsi}
 \kb_\psi={}&it\nablab+i\rb\partial_0+M\rb-\frac{k_D}{2M}\Db-\frac{ik_{DS}}{4M}\Db\times\sigmab+\frac{k_E}{8M^2}g\Eb+\frac{ik_{D0}}{8M^2}\left\{D_0,\Db\right\}\notag\\
 &+\frac{ik_{ES}}{8M^2}g\Eb\times\sigmab-\frac{k_{DS0}}{8M^2}\left\{D_0,\Db\times\sigmab\right\}-\frac{k_{D3}}{8M^3}\left\{\Db,(\Db^2)\right\}-\frac{ik_{D3S}}{32M^3}\left\{(\Db\times\sigmab),(\Db^2)\right\}\notag\\
 &+\frac{ik_{B1}}{16M^3}[\Db\times,g\Bb]+\frac{ik_{B2}}{16M^3}\{\Db\times,g\Bb\}+\frac{k_{BS1}}{16M^3}[\Db,(g\Bb\cdot\sigmab)]+\frac{k_{BS2}}{16M^3}\{\Db,(g\Bb\cdot\sigmab)\}\notag\\
 &+\frac{k_{BS3}}{16M^3}[(\Db\cdot\sigmab),g\Bb]+\frac{k_{BS4}}{16M^3}\{(\Db\cdot\sigmab),g\Bb\}+\frac{k_{BS5}}{16M^3}\{\Db\cdot,g\Bb\}\,\sigmab\notag\\
 &+\frac{k_{D00}}{16M^3}\left\{D_0,\left\{D_0,\Db\right\}\right\}+\frac{ik_{E01}}{16M^3}\left[D_0,g\Eb\right]+\frac{ik_{E02}}{16M^3}\left\{D_0,g\Eb\right\}\notag\\
 &+\frac{ik_{DS00}}{16M^3}\left\{D_0,\left\{D_0,\Db\times\sigmab\right\}\right\}+\frac{k_{ES01}}{16M^3}\left[D_0,g\Eb\times\sigmab\right]+\frac{k_{ES02}}{16M^3}\left\{D_0,g\Eb\times\sigmab\right\}\,,
\end{align}
\begin{align}\label{kchi}
 \kb_\chi={}&it\nablab+i\rb\partial_0-M\rb+\frac{k_D}{2M}\Db+\frac{ik_{DS}}{4M}\Db\times\sigmab+\frac{k_E}{8M^2}g\Eb+\frac{ik_{D0}}{8M^2}\left\{D_0,\Db\right\}\notag\\
 &+\frac{ik_{ES}}{8M^2}g\Eb\times\sigmab-\frac{k_{DS0}}{8M^2}\left\{D_0,\Db\times\sigmab\right\}+\frac{k_{D3}}{8M^3}\left\{\Db,(\Db^2)\right\}+\frac{ik_{D3S}}{32M^3}\left\{(\Db\times\sigmab),(\Db^2)\right\}\notag\\
 &-\frac{ik_{B1}}{16M^3}[\Db\times,g\Bb]-\frac{ik_{B2}}{16M^3}\{\Db\times,g\Bb\}-\frac{k_{BS1}}{16M^3}[\Db,(g\Bb\cdot\sigmab)]-\frac{k_{BS2}}{16M^3}\{\Db,(g\Bb\cdot\sigmab)\}\notag\\
 &-\frac{k_{BS3}}{16M^3}[(\Db\cdot\sigmab),g\Bb]-\frac{k_{BS4}}{16M^3}\{(\Db\cdot\sigmab),g\Bb\}-\frac{k_{BS5}}{16M^3}\{\Db\cdot,g\Bb\}\,\sigmab\notag\\
 &-\frac{k_{D00}}{16M^3}\left\{D_0,\left\{D_0,\Db\right\}\right\}-\frac{ik_{E01}}{16M^3}\left[D_0,g\Eb\right]-\frac{ik_{E02}}{16M^3}\left\{D_0,g\Eb\right\}\notag\\*
 &-\frac{ik_{DS00}}{16M^3}\left\{D_0,\left\{D_0,\Db\times\sigmab\right\}\right\}-\frac{k_{ES01}}{16M^3}\left[D_0,g\Eb\times\sigmab\right]-\frac{k_{ES02}}{16M^3}\left\{D_0,g\Eb\times\sigmab\right\}\,.
\end{align}
Note that we are no longer able to remove terms with temporal derivatives through the equations of motion, because the corresponding field redefinitions have already been used to eliminate such terms in the Lagrangian; performing further redefinitions to remove temporal derivatives from the boost would reintroduce them in the Lagrangian. Terms that do not transform as a vector under rotations have already been removed, as they would violate one of the relations of the Poincar\'e algebra. However, non-Hermitian terms are in general allowed in the boost generator for a field, although they will eventually cancel in the associated Noether charge, which is always Hermitian (see Sec.~\ref{Noether}). In the following, it will be sufficient to discuss only the heavy quark sector, since the antiquark sector follows directly from charge conjugation.

The nonlinear boost transformations constructed in this way have to satisfy the Poincar\'e algebra
\begin{align}\label{PoincareAlgebraRelations}
 \bigl[P_0,P_i\bigr]&=0\,, & \bigl[P_0,J_i\bigr]&=0\,, & \bigl[P_0,K_i\bigr]&=-iP_i\,, \notag\\
 \bigl[P_i,P_j\bigr]&=0\,, & \bigl[P_i,J_j\bigr]&=i\epsilon_{ijk}P_k\,, & \bigl[P_i,K_j\bigr]&=-i\delta_{ij}P_0\,, \notag\\
 \bigl[J_i,J_j\bigr]&=i\epsilon_{ijk}J_k\,, & \bigl[K_i,J_j\bigr]&=i\epsilon_{ijk}K_k\,, & \bigl[K_i,K_j\bigr]&=-i\epsilon_{ijk}J_k\,,
\end{align}
where $P_0$ is the generator of time translations, $\bm{P}$ is the generator of space translations, and $\bm{J}$ is the generator for rotations.\footnote{We reserve covariant notation for Greek indices, writing purely spatial vector indices $i,j,k,\dots$ always as lower indices.}

The commutation relations of the Poincar\'e algebra not involving a boost generator are trivially satisfied for the generators specified in Appendix~\ref{PoincareAlgebraNRQCD}, but the remaining relations involving a boost provide nontrivial information. It is straightforward to check that the commutators of a boost generator with the generators of spacetime translations or rotations are satisfied since $\kb_\psi$ and $\kb_\chi$ depend explicitly on the coordinates only through the terms generated by the coordinate transformation and they have been written in terms of vectors under rotations. The commutator of two boosts, however, gives new constraints on the parameters of the boost generator.

The commutator between any two transformations is defined as the difference in performing them in reverse orders. In the case of linear transformations, this can be written simply as the commutator of two matrices, but for the nonlinear transformations corresponding to the boost of a heavy (anti)quark field, one has to be careful to express the second transformation in terms of fields that have already undergone the first transformation. Consider the commutator of two infinitesimal boost transformations:
\begin{equation}\label{commutation relation}
 [1-i\xib\cdot\bm{K},1-i\etab\cdot\bm{K}]=i(\xib\times\etab)\cdot\bm{J}\,.
\end{equation}
Applying the left-hand side to the heavy quark field requires computing the two successive boosts 
\begin{align}
 &\psi(x)\stackrel{K_\eta}{\longrightarrow}\psi'_{\eta}(x)=\Bigl(1-i\etab\cdot\kb_\psi(D_0,\Db,\Eb,\Bb,\psi,\chi,x)\Bigr)\psi(x)\,,\\
 &\psi'_\eta(x)\stackrel{K_\xi}{\longrightarrow}\psi''_{\xi\eta}(x)=\Bigl(1-i\xib\cdot\kb_\psi(D'_{0\eta},\Db'_\eta,\Eb'_\eta,\Bb'_\eta,\psi'_\eta,\chi'_\eta,x)\Bigr)\psi'_\eta(x)\,.
\end{align}
Expanding the commutator to linear order in $\xib$ and $\etab$ gives
\begin{align}\label{successiveboosts}
 &[1-i\xib\cdot\bm{K},1-i\etab\cdot\bm{K}]\psi(x)=\psi''_{\xi\eta}(x)-\psi''_{\eta\xi}(x)\notag\\
 &=\Bigl(1-i\xib\cdot\kb_\psi(D'_{0\eta},\Db'_\eta,\Eb'_\eta,\Bb'_\eta,\psi'_\eta,\chi'_\eta,x)\Bigr)\psi'_\eta(x)\notag\\
 &\phantom{=}-\Bigl(1-i\etab\cdot\kb_\psi(D'_{0\xi},\Db'_\xi,\Eb'_\xi,\Bb'_\xi,\psi'_\xi,\chi'_\xi,x)\Bigr)\psi'_\xi(x)\notag\\
 &=(\xib\times\etab)\cdot(\rb\times\nablab)\psi(x)\notag\\
 &\phantom{=}-\left[\xib\cdot\bm{\hat{k}}_\psi(D_0,\Db,\Eb,\Bb,\psi,\chi,x),\etab\cdot\bm{\hat{k}}_\psi(D_0,\Db,\Eb,\Bb,\psi,\chi,x)\right]\psi(x)\notag\\
 &\phantom{=}-i\left(\xib\cdot\bm{\hat{k}}_\psi\bigr|_\eta(D_0,\Db,\Eb,\Bb,\psi,\chi,x)-\etab\cdot\bm{\hat{k}}_\psi\bigr|_\xi(D_0,\Db,\Eb,\Bb,\psi,\chi,x)\right)\psi(x)\,,
\end{align}

\smallskip\noindent
where in the last line
\begin{align}
 \bm{\hat{k}}_\psi\bigr|_\eta(D_0,\Db,\Eb,\Bb,\psi,\chi,x)=\eta_i\Bigl[\nabla_{\tilde{\eta}\,i}\,\bm{\hat{k}}_\psi\bigl(&D_0+\bm{\tilde{\eta}}\cdot\Db,\bm{D}+\bm{\tilde{\eta}} D_0,\Eb+\bm{\tilde{\eta}}\times\Bb,\Bb-\bm{\tilde{\eta}}\times\Eb,\notag\\
 &\,(1-i\bm{\tilde{\eta}}\cdot\bm{\hat{k}}_\psi)\psi,(1-i\bm{\tilde{\eta}}\cdot\bm{\hat{k}}_\chi)\chi,x\bigr)\Bigr]_{\tilde{\eta}=0}\,,
\end{align}
and analogously for $\bm{\hat{k}}_\psi\bigr|_\xi$, such that only linear orders of $\xib$ and $\etab$ have been kept in~\eqref{successiveboosts}. This last line contains new terms (compared to the naive application of the commutator in the previous line) arising from the nonlinear nature of the boost transformation.

Inserting the explicit expression of Eq.~\eqref{kpsi} into Eq.~\eqref{successiveboosts}, we obtain the following expression at $\mathcal{O}(M^{-2})$:

\begin{align}\label{boostcommuteNRQCD}
 \psi''_{\xi\eta}(x)-\psi''_{\eta\xi}(x)={}&i(\xib\times\etab)\cdot\left\{\rb\times(-i\nablab)+\frac{k_{DS}}{2}\sigmab+\frac{i}{2M}(k_{DS}-k_{DS0})D_0\,\sigmab\right.\notag\\
 &+\frac{1}{4M^2}(k_{D3S}+k_{DS0})\left(\Db^2\right)\sigmab+\frac{1}{16M^2}(k_{DS}^2-k_{D3S}-2k_{DS0})\left\{\Db,\Db\cdot\sigmab\right\}\notag\\
 &+\frac{1}{2M^2}(k_{DS0}-k_{DS00})D_0^2\,\sigmab-\frac{1}{16M^2}(4k_E-4k_D^2+k_{DS}^2+4k_{B1})\,g\Bb\notag\\
 &-\left.\frac{i}{8M^2}(k_{ES}-k_Dk_{DS}-k_{BS4}+k_{BS5})\left(g\Bb\times\sigmab\right)\right\}\psi(x)\,.
\end{align}
This expression has to satisfy the commutation relation of Eq.~\eqref{commutation relation}, i.e.,
\begin{equation}
 \psi''_{\xi\eta}(x)-\psi''_{\eta\xi}(x)\stackrel{!}{=}i(\xib\times\etab)\cdot{}\left\{\rb\times(-i\nablab)+\frac{1}{2}\bm{\sigma}\right\}\psi(x)\,.
\end{equation}
This gives the following relations:
\begin{equation}\label{boostalgebraconsNRQCD1}
 k_{DS}=1\,,\hspace{10pt}k_{DS0}=1\,,\hspace{10pt}k_{D3S}=-1\,,\hspace{10pt}k_{DS00}=1\,,
\end{equation}
\begin{equation}\label{boostalgebraconsNRQCD2}
 k_{B1}=k_D^2-k_E-\frac{1}{4}\,,\hspace{10pt}k_{BS5}=k_{BS4}+k_D-k_{ES}\,.
\end{equation}
With this result, already four of the boost parameters are completely fixed. Also note that the system at $\mathcal{O}\left(M^{-2}\right)$ is overcomplete: there are 4 equations depending only on the 3 paramters $k_{DS}$, $k_{DS0}$, and $k_{D3S}$, so the fact that all can be satisfied simultaneously is a nontrivial result.

\subsection{Invariance of the Lagrangian}\label{NRQCDinv}

Now that we have constructed a nonlinear boost transformation for the heavy (anti)quark field that satisfies the Poincar\'e algebra, we can proceed to check which conditions need to be satisfied in order for the Lagrangian to be invariant under this transformation. We start with the bilinear terms in the heavy quark sector. The Lagrangian at $\mathcal{O}(M^{-2})$ was already given in Eq.~\eqref{LNRQCD}, but in order to study the transformed Lagrangian at this order, we also need to include
\begin{align}
 \mathcal{L}^{(3)}\Bigr|_{\bm{D}}=\psi^\dagger&\biggl\{\frac{c_4}{8M^3}(\bm{D}^2)^2+\frac{c_{W1}}{8M^3}\bigl\{\Db^2,g\Bb\cdot\sigmab\bigr\}-\frac{c_{W2}}{4M^3}D_i(g\Bb\cdot\sigmab)D_i\notag\\
 &\,\left.+\frac{c_{p'p}}{16M^3}\Bigl\{(\Db\cdot\sigmab),\bigl\{\Db\cdot,g\Bb\bigr\}\Bigr\}+\frac{ic_M}{8M^3}\Bigl\{\Db\cdot,\bigl\{\Db\times,g\Bb\bigr\}\Bigr\}\right\}\psi\,,
\label{psiM3}
\end{align}
which consists of all $\mathcal{O}(M^{-3})$ terms that contain a derivative:\footnote{The term $-iM\bm{r}$ from the boost transformation adds a power of $M$ to the $\mathcal{O}(M^{-3})$ Lagrangian, but the commutator with this term vanishes unless there is a derivative.}

Strictly speaking, it is not the Lagrangian that needs to be invariant under a transformation but the action. So when we speak about an invariant Lagrangian, we mean that the difference between transformed and original Lagrangian is at most an overall derivative, which we denote as $\partial_\mu\Delta^\mu\mathcal{L}$. Overall derivatives are often implicitly omitted, as all they contribute to the action is a vanishing surface term. We will include them here for the sake of completeness, and because they play a role in the calculation of the conserved Noether currents and charges.

The heavy quark Lagrangian defined in Eqs.~\eqref{LNRQCD} and~\eqref{psiM3} transforms in the following way at $\mathcal{O}(M^{-2})$:
\begin{align}\label{DeltaL}
 \partial_\mu&\Delta^\mu\mathcal{L}=\mathcal{L}\bigl(D'_0,\Db',\Eb',\Bb',\psi',\chi',x)-\mathcal{L}\bigl(D_0,\Db,\Eb,\Bb,\psi,\chi,x)\notag\\
 ={}&\etab\cdot(\rb\partial_0+t\nablab)\mathcal{L}+\frac{ik_{D0}}{4M^2}\partial_0\,\etab\cdot\nablab\,\psi^\dagger D_0\psi-\frac{ik_{D0}}{4M^2}\partial_0\,\etab\cdot\psi^\dagger\Db D_0\psi-\frac{ik_{D0}}{4M^2}\etab\cdot\nablab\,\psi^\dagger D_0^2\psi\notag\\
 &+\frac{k_{DS0}}{4M^2}\partial_0\,\etab\times\nablab\cdot\psi^\dagger D_0\,\sigmab\psi-\frac{k_{DS0}}{4M^2}\partial_0\,\etab\cdot\psi^\dagger(\Db\times\sigmab)D_0\psi-\frac{k_{DS0}}{4M^2}\etab\times\nablab\cdot\psi^\dagger D_0^2\,\sigmab\psi\notag\\
 &+\etab\cdot\nablab\,\psi^\dagger\left\{\frac{k_D}{2M}D_0-\frac{ic_2k_D}{4M^2}\Db^2-\frac{ic_Fk_D}{4M^2}g\Bb\cdot\sigmab\right\}\psi\notag\\
 &+\etab\times\nablab\cdot\psi^\dagger\left\{\frac{ik_{DS}}{4M}D_0\,\sigmab+\frac{c_2k_{DS}}{8M^2}(\Db^2)\sigmab+\frac{c_Fk_{DS}}{8M^2}g\Bb+\frac{ic_Fk_{DS}}{8M^2}g\Bb\times\sigmab\right\}\psi\notag\\
 &+\etab\cdot\psi^\dagger\left\{i(1-c_2)\Db+\frac{1}{2M}(c_2-k_D)\bigl\{D_0,\Db\bigr\}+\frac{1}{4M}(k_{DS}-2c_F+c_S)\,g\Eb\times\sigmab\right.\notag\\
 &+\frac{ik_{D0}}{8M^2}\bigl\{D_0,\{D_0,\Db\}\bigr\}+\frac{1}{8M^2}(c_D+k_E)\bigl[D_0,g\Eb\bigr]\hspace{-2pt}+\frac{i}{8M^2}(c_S+k_{ES}-k_{DS0})\bigl\{D_0,g\Eb\bigr\}\hspace{-2pt}\times\sigmab\notag\\
 &+\frac{i}{4M^2}(c_2k_D-c_4)\bigl\{\Db,\Db^2\bigr\}+\frac{1}{8M^2}(2c_M-c_D+c_Fk_{DS})\bigl\{\Db\times,g\Bb\bigr\}\notag\\
 &+\frac{i}{8M^2}(c_S-c_Fk_{DS}-c_{p'p})\bigl\{\Db\cdot,g\Bb\bigr\}\sigmab+\frac{i}{8M^2}(c_Fk_{DS}-c_2k_{DS}-c_{p'p})\bigl\{(\Db\cdot\sigmab),g\Bb\bigr\}\notag\\
 &\left.+\frac{i}{8M^2}(c_2k_{DS}+2c_Fk_D-c_S-2c_{W1}+2c_{W2})\bigl\{\Db,(g\Bb\cdot\sigmab)\bigr\}\right\}\psi\,.
\end{align}
The first four lines of this result consist of overall derivatives, which (apart from the coordinate transformations) arise from the boost transformation of $\psi^\dagger$ through integration by parts, e.g., $(\Db\psi)^\dagger=\nablab\psi^\dagger-\psi^\dagger\Db$.

All terms which are not overall derivatives have to vanish, otherwise the Lagrangian (or rather the action) would not be invariant, from which the following constraints on the coefficients are obtained:
\begin{equation}\label{consNRQCDafterboost1}
 c_2=1\,,\hspace{12pt}k_D=1\,,\hspace{12pt}k_{DS}=1\,,\hspace{12pt}k_{D0}=0\,,\hspace{12pt}k_E=-c_D\,,\hspace{12pt}k_{ES}=k_{DS0}-c_S\,,
\end{equation}
\begin{equation}\label{consNRQCDafterboost2}
 c_4=1\,,\hspace{10pt}c_S=2c_F-1\,,\hspace{10pt}2c_M=c_D-c_F\,,\hspace{10pt}c_{p'p}=c_F-1\,,\hspace{10pt}c_{W2}=c_{W1}-1\,.
\end{equation}
These coincide with the constraints derived in HQET via reparametrization invariance~\cite{Manohar:1997qy,Heinonen:2012km}.\footnote{As noted in~\cite{Hill:2011wy,Hill:2012rh}, the relation for $c_M$ in~\cite{Manohar:1997qy} and~\cite{Heinonen:2012km} differs by a sign. Our relation, $2c_M=c_D-c_F$, agrees with~\cite{Heinonen:2012km}. It is also compatible with the one loop expression of the coefficients in QED~\cite{Hill:2011wy,Hill:2012rh,Moreno:2017sgd}.} We also see that the constraint $k_{DS}=1$ is consistent with the result obtained from the Poincar{\'e} algebra and the commutation of two boost generators (but now it is obtained at a higher order in $1/M$). By combining both results, Eqs.~\eqref{boostalgebraconsNRQCD1}, \eqref{boostalgebraconsNRQCD2}, \eqref{consNRQCDafterboost1}, and \eqref{consNRQCDafterboost2}, we can simplify the remaining constraints to
\begin{equation}\label{consWilsonboost}
 k_{ES}=2(1-c_F)\,,\hspace{10pt}k_{B1}=c_D+\frac{3}{4}\,,\hspace{10pt}k_{BS5}=k_{BS4}+2c_F-1\,.
\end{equation}

The boost parameter $k_{D3}$ has not been fixed yet. Its value can easily be derived from the contribution of the corresponding boost term to $\partial_\mu\Delta^\mu\mathcal{L}$ at $\mathcal{O}\left(M^{-3}\right)$,
\begin{equation}
\etab\cdot \psi^\dagger\left[\frac{1}{8M^3}\left(c_4+\frac{c_2k_{D0}}{4}-k_{D3}\right)\Bigl\{\bigl\{D_0,\Db\bigr\},\Db^2\Bigr\}+\frac{i}{8M^3}(k_{D3}+\dots)\Bigl[\Db,\bigl\{\Db\cdot,g\Eb\bigr\}\Bigr]\right]\psi\,.
\label{M3cons1}
 \end{equation}
The second term will contribute to other constraints, but in the first we have included all possible terms consisting only of covariant derivatives, where exactly one is temporal, so its coefficient has to vanish.

Similar observations can be made for terms with two or more temporal derivatives. For such terms, no knowledge of the full $\mathcal{O}(M^{-3})$ Lagrangian or its derivative terms at $\mathcal{O}(M^{-4})$ is required, because we have defined it such that temporal derivatives do not appear, and an infinitesimal boost can at most introduce one temporal derivative. Hence, the contributions to $\partial_\mu\Delta^\mu\mathcal{L}$ at $\mathcal{O}\left(M^{-3}\right)$ with two or three temporal derivatives are
\begin{align}
 \etab\cdot\psi^\dagger&\left[\frac{k_{D00}}{16M^3}\{D_0,\{D_0,\{D_0,\Db\}\}\}+\frac{i}{16M^3}(k_{E01}+k_{E02})\{D_0,[D_0,g\Eb]\}\right.\notag\\
 &+\left.\frac{k_{ES01}}{16M^3}[D_0,[D_0,g\Eb\times\sigmab]]+\frac{1}{16M^3}(k_{ES02}-k_{DS00})\{D_0,\{D_0,g\Eb\times\sigmab\}\}\right]\psi\,.
 \label{M3cons2}
\end{align}

With $k_{DS00}=1$ from Eq.~\eqref{boostalgebraconsNRQCD1}, and $k_{D0}=0$ and $c_4=1$ from Eqs.~\eqref{consNRQCDafterboost1} and~\eqref{consNRQCDafterboost2}, the vanishing of the terms in Eqs.~\eqref{M3cons1} and~\eqref{M3cons2} require
\begin{equation}\label{consNRQCDafterboost1bis}
 k_{D3}=1\,,\hspace{10pt}k_{D00}=0\,\hspace{10pt}k_{E02}=-k_{E01}\,,\hspace{10pt}k_{ES01}=0\,,\hspace{10pt}k_{ES02}=1\,.
\end{equation}
In particular, we see that the inclusion of temporal derivative terms in the boost generator is not ruled out, even though the Lagrangian does not contain such terms at any order beyond the leading one.

If we compare the expression for the boost that follows from \eqref{boostalgebraconsNRQCD1}, \eqref{consNRQCDafterboost1} and \eqref{consNRQCDafterboost1bis} with the one in~\cite{Heinonen:2012km},\footnote{The context of~\cite{Heinonen:2012km} is NRQED, not NRQCD, but both calculations are analogous at low orders in $1/M$.} we see some differences. The reason for that is that in the derivation of the induced representation in~\cite{Heinonen:2012km} the equations of motion were used in order to remove temporal derivatives. If we do the same, i.e., insert
\begin{equation}\label{NRQCDeom}
 D_0\psi=\left[\frac{ic_2}{2M}\Db^2+\frac{ic_F}{2M}g\Bb\cdot\sigmab+\mathcal{O}(M^{-2})\right]\psi
\end{equation}
into the heavy quark boost operator, we can replace
\begin{align}
 \frac{ik_{D0}}{8M^2}\{D_0,\Db\}\psi={}&-\frac{k_{D0}}{8M^2}g\Eb\psi+\frac{ik_{D0}}{4M^2}\Db(D_0\psi)\notag\\
 ={}&-\frac{k_{D0}}{8M^2}g\Eb\psi-\frac{k_{D0}c_2}{16M^3}\{\Db,(\Db^2)\}\psi+\dots\,,\\
 -\frac{k_{DS0}}{8M^2}\{D_0,\Db\times\sigmab\}\psi={}&-\frac{ik_{DS0}}{8M^2}(g\Eb\times\sigmab)\psi-\frac{k_{DS0}}{4M^2}(\Db\times\sigmab)(D_0\psi)\notag\\
 ={}&-\frac{ik_{DS0}}{8M^2}(g\Eb\times\sigmab)\psi-\frac{ik_{DS0}c_2}{16M^3}\{(\Db\times\sigmab),(\Db^2)\}\psi+\dots\,,
\end{align}
where the dots contain $\mathcal{O}(M^{-3})$ terms with a magnetic field and higher order terms. We need not study them further, because they affect none of the terms given in~\cite{Heinonen:2012km}. For the terms with two or more temporal derivatives in the boost generator, also the equations of motion for gauge fields will become necessary.

These replacements change the other boost parameters in the following way:
\begin{equation}
 k_E\to k_E-k_{D0}\,,\hspace{10pt}k_{ES}\to k_{ES}-k_{DS0}\,,\hspace{10pt}k_{D3}\to k_{D3}+\frac{k_{D0}c_2}{2}\,,\hspace{10pt}k_{D3S}\to k_{D3S}+2k_{DS0}c_2\,.
\end{equation}
Changing \eqref{boostalgebraconsNRQCD1}, \eqref{consNRQCDafterboost1}, and \eqref{consNRQCDafterboost1bis} accordingly, the obtained boost parameters agree with the ones given in~\cite{Heinonen:2012km}.\footnote{We have chosen to write the terms with three derivatives in the form of an anticommutator in order to work with explicitly (anti-)Hermitian terms, while~\cite{Heinonen:2012km} does not. The difference between both ways of writing are terms with magnetic fields not listed in~\cite{Heinonen:2012km}.} While this may not be a general proof that coefficients obtained from the induced representation (here $k_D$, $k_{DS}$, $k_{D3}$, and $k_{D3S}$) can be assumed to be 1 also in the interacting theory at all orders, it seems to show that at this order there is no contradiction between the two approaches. The constraints on the Wilson coefficients and also the boost Noether charge (see below) are the same, as the equations of motion also enter in the derivation of the conserved Noether current. Note, however, that a boost generator without temporal derivatives, as used in~\cite{Heinonen:2012km}, does not satisfy the Poincar\'e algebra for the commutator of two boosts, unless the equations of motion are used again. This would suggest that the Poincar\'e algebra is only satisfied for on-shell heavy (anti)quark fields, a restriction that does not apply to our approach. Finally, we remark that most of the constraints on the boost coefficients at $\mathcal{O}(M^{-3})$ presented in Eqs.~\eqref{consWilsonboost} and~\eqref{consNRQCDafterboost1bis} are new.

\subsection{Noether charges}\label{Noether}

Now that the boost transformations of the heavy quark and antiquark fields have been determined, we can write the corresponding Noether charge $\bm{\mathcal{K}}$:
\begin{align}\label{boostcharge}
 \etab\cdot\bm{\mathcal{K}}=&\int d^3r\left[\frac{\partial\mathcal{L}}{\partial(\partial_0\phi_i)}(-i\etab\cdot\kb_\phi\phi_i)-\Delta^0\mathcal{L}\right]\notag\\
 =&\int d^3r\left[\etab\cdot\left(\frac{\partial\mathcal{L}}{\partial(\partial_0\phi_i)}(t\nablab+\rb\partial_0)\phi_i+\psi^\dagger\bm{\hat{k}}_\psi\psi+\chi^\dagger\bm{\hat{k}}_\chi\chi-\bm{\Pi}^aA_0^a-\rb\mathcal{L}\right)-\widehat{\Delta}^0\mathcal{L}\right]\notag\\
 =&-t\etab\cdot\bm{\mathcal{P}}+\int d^3r\left[\etab\cdot\rb h+\psi^\dagger\etab\cdot\bm{\hat{k}}_\psi\psi+\chi^\dagger\etab\cdot\bm{\hat{k}}_\chi\chi-\widehat{\Delta}^0\mathcal{L}\right]\,,\notag\\
 \bm{\mathcal{K}}=&-t\bm{\mathcal{P}}+\frac{1}{2}\int d^3r\left\{\rb,h+M\psi^\dagger\psi-M\chi^\dagger\chi\right\}-\int d^3r\,\psi^\dagger\left[\frac{i}{4M}\Db\times\sigmab+\frac{c_D}{8M^2}g\Eb\right]\psi\notag\\
 &+\int d^3r\,\chi^\dagger\left[\frac{i}{4M}\Db\times\sigmab-\frac{c_D}{8M^2}g\Eb\right]\chi+\mathcal{O}(M^{-3})\,,
\end{align}
where $\phi_i$ stands for all three types of field, $\psi$, $\chi$, and $\bm{A}$, while $\bm{\Pi}$ is the canonical momentum field conjugated to $\bm{A}$:
\begin{equation}
 \Pi^a_i=\frac{\partial\mathcal{L}}{\partial(\partial_0A_i^a)}=-E_i^a+\mathcal{O}(M^{-2})\,.
\end{equation}
We have also defined $\widehat{\Delta}^0\mathcal{L}$ in analogy to $\bm{\hat{k}}_{\psi/\chi}$ as the part of $\Delta^0\mathcal{L}$ that does not originate from the coordinate transformations:
\begin{equation}
 \widehat{\Delta}^0\mathcal{L}=-\frac{1}{4M^2}\psi^\dagger(\Db\times\sigmab)D_0\psi-\frac{1}{4M^2}\chi^\dagger(\Db\times\sigmab)D_0\chi+\mathcal{O}(M^{-3})\,.
\end{equation}
We can ignore here terms with an overall spatial derivative in Eq.~\eqref{DeltaL}, because they vanish in the integral of the Noether charge.

In addition, $\bm{\mathcal{P}}$ is the Noether charge associated with spatial translations
\begin{align}
 \bm{\mathcal{P}}&=\int d^3r\left[\frac{\partial\mathcal{L}}{\partial(\partial_0\phi_i)}(-\nablab)\phi_i\right]\notag\\
 &=\int d^3r\left(\psi^\dagger(-i\bm{D})\psi+\chi^\dagger(-i\bm{D})\chi-\mathrm{Tr}\bigl[\bm{\Pi}\times,\Bb\bigr]\right)\,,
\end{align}
where the equations of motion,
$\bigl[\bm{D}\cdot,\bm{\Pi}\bigr]=-\left(\psi^\dagger gT^a\psi+\chi^\dagger gT^a\chi\right)T^a$, have been used in order to make the expression explicitly gauge invariant.\footnote{
In fact, the equations of motion are Gauss's law.}
The Hamiltonian density $h$ is given through the Hamiltonian
\begin{align}
 \mathcal{H}&=\int d^3r\left[\frac{\partial\mathcal{L}}{\partial(\partial_0\phi_i)}\partial_0\phi_i
-\mathcal{L}\right]=\int d^3r\left(\psi^\dagger h_\psi\psi+\chi^\dagger h_\chi\chi+\mathrm{Tr}\bigl[\bm{\Pi}^2+\Bb^2\bigr]\right)\notag\\*
 &\equiv\int d^3r\, h\,,
\end{align}
where $h_\psi$ and $h_\chi$ are defined through the Lagrangian density,
\begin{align}
\mathcal{L}=\psi^\dagger(iD_0-h_\psi)\psi+\chi^\dagger(iD_0-h_\chi)\chi+\mathrm{Tr}\bigl[\Eb^2-\Bb^2\bigr]\,,
\end{align}
and we have made use of Gauss's law again.

The initial expression $\displaystyle\partial\mathcal{L}/\partial(\partial_0\phi_i)\,\partial_0\phi_i-\mathcal{L}$ and the Hamiltonian density, as defined in the final expression, differ by a derivative term, ${-\bm{\nabla}\cdot(\bm{\Pi}^aA_0^a)}$, which vanishes in $\mathcal{H}$, but gives a contribution to $\bm{\mathcal{K}}$ that exactly cancels the $\bm{\Pi}^aA_0^a$ term in Eq.~\eqref{boostcharge}. In the last expression for $\bm{\mathcal{K}}$, we have replaced $\rb h$ by $\{\rb,h\}/2 - [h,\rb]/2$ in order to obtain an explicitly Hermitian expression. The anti-Hermitian terms from $\psi^\dagger\bm{\hat{k}}_\psi\psi$ and $\chi^\dagger\bm{\hat{k}}_\chi\chi$ as well as the terms with a temporal derivative cancel against $[h,\rb]/2$ and $\widehat{\Delta}^0\mathcal{L}$. At $\mathcal{O}(M^{-2})$, there is exactly one kind of term:
\begin{equation}
 \frac{i}{8M^2}(2(1-c_F)+c_S-1)\left[\psi^\dagger g\Eb\times\sigmab\psi+\chi^\dagger g\Eb\times\sigmab\chi\right]=0\,.
\end{equation}
The coefficients add up to zero according to the constraints~\eqref{consNRQCDafterboost2}, where the first comes from the $k_{ES}$ term in the boost generator, the second from $[h,\rb]/2$, and in the third $\widehat{\Delta}^0\mathcal{L}$ has been combined with the $k_{DS0}$ boost term, turning the anticommutator into a commutator that gives the electric field.

The Noether charge $\bm{\mathcal{K}}$ corresponds exactly to the boost operator of the quantized theory obtained in~\cite{Brambilla:2003nt} and extends it up to $\mathcal{O}(M^{-2})$. Note that the field redefinitions that remove the $\mathcal{O}(M)$ terms from the Lagrangian have not been performed in~\cite{Brambilla:2003nt}, hence the definition of $h$ in~\cite{Brambilla:2003nt} differs from ours by $M\psi^\dagger\psi-M\chi^\dagger\chi$. This term appears explicitly in our expression for $\bm{\mathcal{K}}$. Accordingly, the generators for time translations are given by $i\partial_0\pm M$ after the redefinition of $\psi$ and $\chi$, so that the proper Noether charge of time translations is given by the above Hamiltonian $\mathcal{H}$ plus $M\displaystyle\int d^3r\,(\psi^\dagger\psi-\chi^\dagger\chi)$, which coincides with the expression in~\cite{Brambilla:2003nt}. Another way of obtaining $\bm{\mathcal{K}}$ at tree level [in Eq.~\eqref{boostcharge} this corresponds to setting $c_D=1$] is to perform Foldy--Wouthuysen transformations on the QCD Noether charge~\cite{Vairo:2003gx}.

\subsection{The four-fermion Lagrangian}\label{NRQCD4f}

We now turn to the four-fermion part of the NRQCD Lagrangian, or more specifically the part consisting of two heavy quark and two heavy antiquark fields. The lowest order terms of the Lagrangian are given by
\begin{align}\label{L24f}
 \mathcal{L}^{(2)}\Bigr|_{4f}=\frac{1}{M^2}&\Bigl\{f_1(^1S_0)\,\psi^\dagger\chi\chi^\dagger\psi+f_1(^3S_1)\,\psi^\dagger\sigmab\chi\cdot\chi^\dagger\sigmab\psi\Bigr.\notag\\
 &+\Bigl.f_8(^1S_0)\,\psi^\dagger T^a\chi\chi^\dagger T^a\psi+f_8(^3S_1)\,\psi^\dagger\sigmab T^a\chi\cdot\chi^\dagger\sigmab T^a\psi\Bigr\}\,.
\end{align}
The Wilson coefficients $f$ are related by Poincar\'e invariance to the coefficients of the next order four-fermion Lagrangian, which is $\mathcal{O}(M^{-4})$~\cite{Brambilla:2008zg}. It is straightforward to see that the $\mathcal{O}(M)$ terms of $\kb_\psi$ and $\kb_\chi$ cancel each other in the boost transformation of the leading order part of this Lagrangian, so the first constraints can be obtained at $\mathcal{O}(M^{-3})$.

At $\mathcal{O}(M^{-3})$, the $\mathcal{O}(M^{-4})$ Lagrangian contributes only with the $\mathcal{O}(M)$ terms of $\kb_\psi$ and $\kb_\chi$, which are given by $\pm M\rb$. Since the boost of operators with two left-right derivatives, like $\psi^\dagger\Dlr\chi\cdot\chi^\dagger\Dlr\psi$ (see Eqs.~\eqref{leftrightderivativedef} and~\eqref{leftrightderivativedef1} for the definition of left-right derivatives), or with a chromomagnetic field $\Bb$, cancels at $\mathcal{O}(M)$, only operators with at least one ``center-of-mass''~(cm) derivative (i.e., a derivative acting on two heavy fields like $\bm{\nabla}\chi^\dagger\psi$) give nonvanishing contributions. Including only such terms, the four-fermion part of the Lagrangian at $\mathcal{O}(M^{-4})$ is given by
\begin{align}
 \mathcal{L}^{(4)}\Bigr|_{4f,\,{\rm cm}}={}&-\frac{if_{1\,{\rm cm}}}{2M^4}\bigl(\psi^\dagger(\Dlr\times\sigmab)\chi\cdot\nablab\chi^\dagger\psi+(\nablab\psi^\dagger\chi)\cdot\chi^\dagger(\Dlr\times\sigmab)\psi\bigr)\notag\\
 &-\frac{if_{8\,{\rm cm}}}{2M^4}\bigl(\psi^\dagger(\Dlr\times\sigmab)T^a\chi\cdot\Db^{ab}\chi^\dagger T^b\psi+(\Db^{ab}\psi^\dagger T^b\chi)\cdot\chi^\dagger(\Dlr\times\sigmab)T^a\psi\bigr)\notag\\
 &+\frac{if'_{1\,{\rm cm}}}{2M^4}\bigl(\psi^\dagger\Dlr\chi\cdot(\nablab\times\chi^\dagger\sigmab\psi)+(\nablab\times\psi^\dagger\sigmab\chi)\cdot\chi^\dagger\Dlr\psi\bigr)\notag\\
 &+\frac{if'_{8\,{\rm cm}}}{2M^4}\bigl(\psi^\dagger\Dlr T^a\chi\cdot(\Db^{ab}\times\chi^\dagger\sigmab T^b\psi)+(\Db^{ab}\times\psi^\dagger\sigmab T^b\chi)\cdot\chi^\dagger\Dlr T^a\psi\bigr)\notag\\
 &+\frac{g_{1a\,{\rm cm}}}{M^4}\left(\nabla_i\psi^\dagger\sigma_j\chi\right)\left(\nabla_i\chi^\dagger\sigma_j\psi\right)+\frac{g_{8a\,{\rm cm}}}{M^4}\left(D_i^{ab}\psi^\dagger\sigma_jT^b\chi\right)\left(D_i^{ac}\chi^\dagger\sigma_jT^c\psi\right)\notag\\
 &+\frac{g_{1b\,{\rm cm}}}{M^4}\left(\nablab\cdot\psi^\dagger\sigmab\chi\right)\left(\nablab\cdot\chi^\dagger\sigmab\psi\right)+\frac{g_{8b\,{\rm cm}}}{M^4}\left(\Db^{ab}\cdot\psi^\dagger\sigmab T^b\chi\right)\left(\Db^{ac}\cdot\chi^\dagger\sigmab T^c\psi\right)\notag\\
 &+\frac{g_{1c\,{\rm cm}}}{M^4}\left(\nablab\psi^\dagger\chi\right)\cdot\left(\nablab\chi^\dagger\psi\right)+\frac{g_{8c\,{\rm cm}}}{M^4}\left(\Db^{ab}\psi^\dagger T^b\chi\right)\cdot\left(\Db^{ac}\chi^\dagger T^c\psi\right)\,,
\end{align}
where covariant derivatives with color indices are understood in the adjoint representation. The relevant left-right derivatives are defined as follows (see~\cite{Brambilla:2008zg}):
\begin{align}
 \psi^\dagger\Dlrn_iT\chi&=-\left(D_i\psi\right)^\dagger T\chi+\psi^\dagger TD_i\chi\,,
 \label{leftrightderivativedef}\\
 \psi^\dagger\Dlrn_i\Dlrn_jT\chi&=\left(D_iD_j\psi\right)^\dagger T\chi-\left(D_i\psi\right)^\dagger TD_j\chi-\left(D_j\psi\right)^\dagger TD_i\chi+\psi^\dagger TD_iD_j\chi\,,
 \label{leftrightderivativedef1}
\end{align}
where $T$ stands for either the unit or a color matrix. Thus, we obtain the following expression at $\mathcal{O}(M^{-3})$ after the boost transformation:
\begin{align}\label{4foverallLT}
 \partial_\mu\widehat{\Delta}^\mu\mathcal{L}\Bigr|_{4f} =&-\frac{1}{2M^3}\left(f_1(^1S_0)+4g_{1c\,{\rm cm}}\right)\left[(\etab\cdot i\nablab\psi^\dagger\chi)\chi^\dagger\psi+H.c.\right]\notag\\
 &-\frac{1}{2M^3}\left(f_8(^1S_0)+4g_{8c\,{\rm cm}}\right)\left[(\etab\cdot i\Db^{ab}\psi^\dagger T^b\chi)\chi^\dagger T^a\psi+H.c.\right]\notag\\
 &+\frac{1}{4M^3}\left(f_1(^1S_0)-f_{1\,{\rm cm}}\right)\left[\psi^\dagger\etab\cdot(\Dlr\times\sigmab)\chi\chi^\dagger\psi+H.c.\right]\notag\\
 &+\frac{1}{4M^3}\left(f_8(^1S_0)-f_{8\,{\rm cm}}\right)\left[\psi^\dagger\etab\cdot(\Dlr\times\sigmab)T^a\chi\chi^\dagger T^a\psi+H.c.\right]\notag\\
 &-\frac{1}{2M^3}\left(f_1(^3S_1)+4g_{1a\,{\rm cm}}\right)\left[(\etab\cdot i\nablab\psi^\dagger\sigma_i\chi)\chi^\dagger\sigma_i\psi+H.c.\right]\notag\\
 &-\frac{1}{2M^3}\left(f_8(^3S_1)+4g_{8a\,{\rm cm}}\right)\left[(\etab\cdot i\Db^{ab}\psi^\dagger\sigma_iT^b\chi)\chi^\dagger\sigma_iT^a\psi+H.c.\right]\notag\\
 &+\frac{1}{4M^3}\left(f_1(^3S_1)-f'_{1\,{\rm cm}}\right)\left[\psi^\dagger(\etab\times\Dlr)\chi\cdot\chi^\dagger\sigmab\psi+H.c.\right]\notag\\
 &+\frac{1}{4M^3}\left(f_8(^3S_1)-f'_{8\,{\rm cm}}\right)\left[\psi^\dagger(\etab\times\Dlr)T^a\chi\cdot\chi^\dagger\sigmab T^a\psi+H.c.\right]\notag\\
 &+\frac{2}{M^3}g_{1b\,{\rm cm}}\left[\psi^\dagger(\etab\cdot\sigmab)\chi(i\nablab\cdot\chi^\dagger\sigmab\psi)+H.c.\right]\notag\\
 &+\frac{2}{M^3}g_{8b\,{\rm cm}}\left[\psi^\dagger(\etab\cdot\sigmab)T^a\chi(i\Db^{ab}\cdot\chi^\dagger\sigmab T^b\psi)+H.c.\right],
\end{align}
where we have neglected the terms from the coordinate transformations. As none of the terms in \eqref{4foverallLT} has the form of an overall derivative, all coefficients have to vanish, which implies:
\begin{align}
 g_{1a\,{\rm cm}}&=-\frac{1}{4}f_1(^3S_1),&g_{1c\,{\rm cm}}&=-\frac{1}{4}f_1(^1S_0),&g_{8a\,{\rm cm}}&=-\frac{1}{4}f_8(^3S_1),&g_{8c\,{\rm cm}}&=-\frac{1}{4}f_8(^1S_0),\label{4fermionconstraints1}\\
 f_{1\,{\rm cm}}&=\frac{1}{4}f_1(^1S_0),&f'_{1\,{\rm cm}}&=\frac{1}{4}f_1(^3S_1),&f_{8\,{\rm cm}}&=\frac{1}{4}f_8(^1S_0),&f'_{8\,{\rm cm}}&=\frac{1}{4}f_8(^3S_1),\label{4fermionconstraints2}
\end{align}
\begin{equation}
 g_{1b\,{\rm cm}}=g_{8b\,{\rm cm}}=0\,.\label{4fermionconstraints3}
\end{equation}
These relations were first derived in~\cite{Brambilla:2008zg} and later confirmed in~\cite{Hill:2012rh} for NRQED, which at this order is equivalent to the singlet sector of NRQCD.

At $\mathcal{O}(M^{-4})$, the boost generators contain terms involving the heavy (anti)quark fields themselves. This is a novel feature if one follows the line of argument in~\cite{Hill:2012rh}, where the appearance of gauge field operators has been explained with the ambiguities related to the ordering of derivatives when promoted to covariant derivatives. Of course, an argument could be made based on the fact that gauge fields and heavy (anti)quark fields are related through the equations of motion; in the EFT approach used in this paper, however, the appearance of heavy (anti)quark fields in the boost generators is natural and requires no further justification. We discuss the effect of these terms in Appendix~\ref{appendix}.

\section{Constraints in pNRQCD}\label{conspNRQCD}

Potential nonrelativistic QCD (pNRQCD) is a low energy EFT obtained from NRQCD after integrating out the scale $Mv$ of the relative momentum between a heavy quark and an antiquark~\cite{Pineda:1997bj, Brambilla:1999xf,Brambilla:2004jw}.\footnote{In this paper, we only consider a heavy quark and an antiquark with the same flavor.} In weakly-coupled pNRQCD, we also assume $\Lambda_{\text{QCD}}\lesssim Mv^2$, which implies that the matching can be carried out perturbatively. Since the relative momentum scale is of the same order as the inverse of the quark-antiquark distance $r$, integrating out this scale corresponds to multipole expanding. The effective degrees of freedom are heavy quarkonium fields instead of separate heavy quark and antiquark fields, as the other degrees of freedom (ultrasoft gluons and light quarks) can no longer resolve the individual heavy particles after integrating out the scale $1/r$.

In SU(3) the heavy quark and antiquark may form either a color singlet or octet state, hence they appear in pNRQCD either as a singlet field $\mathrm{S}$ or an octet field $\mathrm{O}$. These fields are the only ones that depend on the relative distance $\bm{r}$ as well as the center-of-mass coordinate $\bm{R}$, while all other fields depend on $\bm{R}$ only. The Lagrangian of weakly-coupled pNRQCD can be written schematically as 
\begin{equation}\label{SchematicpNRQCD}
 \mathcal{L}^{\text{weak}}_{\text{pNRQCD}} = \int d^3r\,\mathrm{Tr}\left[\mathrm{S}^\dagger(i\partial_0-h_{S})\mathrm{S} + \mathrm{O}^\dagger iD_0{O}-\left(\mathrm{O}^\dagger h_{O}\mathrm{O}+c.c.\right) 
 -\left(\mathrm{S}^\dagger h_{SO}\mathrm{O}+H.c.\right)\right] + \dots ,
\end{equation}
where the heavy quark-antiquark fields are matrices in color space:
\begin{equation}
 \mathrm{S}=\frac{1}{\sqrt{3}}S\,\mathbbm{1}\,,\hspace{30pt}\mathrm{O}=\sqrt{2}O^aT^a\,.
\end{equation}
The trace is understood both in spin and in color spaces, and the coefficients for the matrices have been chosen in such a way that the traces over two fields are properly normalized. The ellipsis stands for the gluon and light quark sectors, which can be read from the NRQCD Lagrangian~\eqref{LNRQCD} and following discussion. The covariant derivatives are understood as commutators with all terms to their right. Furthermore, $H.c.$ in Eq.~\eqref{SchematicpNRQCD} stands for the Hermitian conjugate,  and $c.c.$ for the charge conjugate of the preceding term within the parentheses. The explicit expressions of $h_{S}$, $h_{O}$, and $h_{SO}$ are not immediately required for the following discussions, so we postpone them until they become relevant: $h_{S}$ and $h_{O}$ are found in Eqs.~\eqref{ss lagrangian}, \eqref{oo lagrangian}, respectively, and $h_{\rm SO}$ is given by Eqs.~\eqref{soh lagrangian} and \eqref{soa lagrangian}. As usual, they contain all terms allowed by the symmetries.

On the other hand, when the hierarchy of scales is given by $\Lambda_{\text{QCD}}\gg Mv^2$, the theory enters the strong-coupling regime. In this case, the pNRQCD Lagrangian is obtained after integrating out the hadronic scale $\Lambda_{\text{QCD}}$, which means that all colored degrees of freedom are absent~\cite{Brambilla:2000gk,Pineda:2000sz,Brambilla:2003mu,Brambilla:2004jw}:
\begin{align}
\mathcal{L}_{\text{pNRQCD}}^{\text{strong}}=\int d^3r\,\text{Tr}\left[\mathrm{S}^{\dagger}\left(i\partial_0-h_{S}\right)\mathrm{S}\right] + \dots\,,
\end{align}
where the ellipsis denotes now terms that, in the simplest setting, depend on the light-quark fields only in the form of light mesons, and $h_{S}$ has the same form as in weakly-coupled pNRQCD but with all gluonic operators removed. The reason is that, once the effective degrees of freedom have been established, the allowed terms in the effective Lagrangian depend only on the symmetries, which are the same for weakly- and strongly-coupled pNRQCD. Therefore, although the Wilson coefficients of strongly-coupled pNRQCD are different from those of weakly-coupled pNRQCD and need to be determined in a nonperturbative matching, nevertheless any weak-coupling result can be immediately extended to the strong-coupling case by setting to zero all coefficients from $h_{O}$, $h_{SO}$, and from gluonic operators in $h_{S}$. In case the hadronic scale also factorizes from the soft scale (i.e., for the hierarchy $1/r\sim Mv\gg\Lambda_{\text{QCD}}\gg Mv^2$), the Wilson coefficients can be expanded in $r\Lambda_{\text{QCD}}$, so that the constraints have to be satisfied order by order in this expansion. Thus, it suffices to study only the weakly-coupled case and any superscript on the Lagrangian will be omitted from now on.

The matching between NRQCD and  weakly-coupled pNRQCD is performed through interpolating fields~\cite{Brambilla:1999xf}
\begin{align}
 &\chi^\dagger(\bm{R}-\bm{r}/2)\,W(\bm{R}-\bm{r}/2,\bm{R}+\bm{r}/2)\,\psi(\bm{R}+\bm{r}/2)\notag\\
 &\hspace{4cm}\to Z_{S}^{(0)}(r)S(\bm{r},\bm{R})+Z_{O}^{(2)}(r)r\bm{r}\cdot g\Eb^a(\bm{R})O^a(\bm{r},\bm{R})+\mathcal{O}(r^3)\,,\\
 &\chi^\dagger(\bm{R}-\bm{r}/2)\,W(\bm{R}-\bm{r}/2,\bm{R})\,T^a\,W(\bm{R},\bm{R}+\bm{r}/2)\,\psi(\bm{R}+\bm{r}/2)\notag\\
 &\hspace{4cm}\to Z_{O}^{(0)}(r)O^a(\bm{r},\bm{R})+Z_{S}^{(2)}(r)r\bm{r}\cdot g\Eb^a(\bm{R})S(\bm{r},\bm{R})+\mathcal{O}(r^3)\,,
\end{align}
where the Wilson line $W$ acts as a gauge link from the position of the heavy quark to that of the heavy antiquark. Correlators of those interpolating fields in both theories give the same result, which determines the coefficients $Z$ as well as the matching coefficients.

\subsection{Coordinate transformations for quarkonium fields}\label{pNRQCDcoord}

The interpolating fields determine how the heavy quark-antiquark fields behave under spacetime symmetries. In fact, in the limit $g\to0$ one can neglect the Wilson lines and just determine the transformation of singlet and octet from different color projections of $Q=\psi\chi^\dagger$.\footnote{In this limit there is no longer any interaction between the heavy quark and antiquark, and they cannot form a bound state. Although we thus lose in this way the justification for the assumed hierarchy of scales, nevertheless this is not relevant as long as we are interested only in coordinate transformations.} The coordinate transformations do not depend on the color representation, so we use $Q$ for both singlet and octet.

First, we give here the transformations under the discrete symmetries~\cite{Brambilla:2004jw}:\footnote{Note that $C$ would not be a symmetry if we allowed different flavors for the heavy quark and antiquark.}
\begin{align}
 Q(t,\bm{r},\bm{R})&\stackrel{P}{\longrightarrow}-Q(t,-\bm{r},-\bm{R})\,,\\
 Q(t,\bm{r},\bm{R})&\stackrel{C}{\longrightarrow}\sigma_2Q^T(t,-\bm{r},\bm{R})\sigma_2\,,\\
 Q(t,\bm{r},\bm{R})&\stackrel{T}{\longrightarrow}\sigma_2Q(-t,\bm{r},\bm{R})\sigma_2\,,
\end{align}
where the transpose on the charge conjugated field refers both to color and spin space. Also note that charge conjugation exchanges the positions of the quark and the antiquark fields, so that $\bm{r}$ goes into $-\bm{r}$. For the behavior under spacetime translations and rotations, we refer to Appendix~\ref{SpacetimepNRQCD}.

Under boosts, the coordinate transformations are composed of the individual boosts of the heavy quark and antiquark fields [see Eqs.~\eqref{kpsi} and~\eqref{kchi}] located at $\bm{x}_1=\bm{R}+\bm{r}/2$ and $\bm{x}_2=\bm{R}-\bm{r}/2$ respectively:
\begin{align}\label{quarkoniumcoordinatetrans}
 \psi&(t,\bm{x}_1)\chi^\dagger(t,\bm{x}_2)\stackrel{K}{\longrightarrow}\psi(t,\bm{x}_1)\chi^\dagger(t,\bm{x}_2)-iM\bm{\eta}\cdot(\bm{x}_1+\bm{x}_2)\psi(t,\bm{x}_1)\chi^\dagger(t,\bm{x}_2)\notag\\
 & \hspace{1cm} +\bigl[\bm{\eta}\cdot(t\bm{\nabla}_1+\bm{x}_1\partial_0),\psi(t,\bm{x}_1)\bigr]\chi^\dagger(t,\bm{x}_2)+\psi(t,\bm{x}_1)\bigl[\bm{\eta}\cdot(t\bm{\nabla}_2+\bm{x}_2\partial_0),\chi^\dagger(t,\bm{x}_2)\bigr]+\dots\notag\\
 ={}&(1-2iM\bm{\eta}\cdot\bm{R})\psi(t,\bm{x}_1)\chi^\dagger(t,\bm{x}_2)+\bigl[\bm{\eta}\cdot(t\bm{\nabla}_R+\bm{R}\partial_0),\psi(t,\bm{x}_1)\chi^\dagger(t,\bm{x}_2)\bigr]\notag\\
 & \hspace{1cm} +\frac{1}{2}(\bm{\eta}\cdot\bm{r})\left(\bigl[\partial_0,\psi(t,\bm{x}_1)\bigr]\chi^\dagger(t,\bm{x}_2)-\psi(t,\bm{x}_1)\bigl[\partial_0,\chi^\dagger(t,\bm{x}_2)\bigr]\right)+\dots\,,
\end{align}
where the ellipsis stands for all terms of the boost transformation that are not related to the coordinate transformations [these are shown in Eq.~\eqref{quarkoniumnoncoordinatetrans}]. The first two terms on the right-hand side of the equality sign correspond to the usual coordinate transformations under boosts of a scalar field with mass $2M$, where only the center-of-mass coordinate participates in the boost and the relative distance remains unaffected. In the third term on the right-hand side, the time derivatives acting on the quark and antiquark fields cannot be written as one derivative acting on the whole quarkonium field because of the opposite signs. However, these time derivatives can be replaced by spatial derivatives through the equations of motion:
\begin{align}
 &\frac{1}{2}(\bm{\eta}\cdot\bm{r})\left(\bigl[\partial_0,\psi(t,\bm{x}_1)\bigr]\chi^\dagger(t,\bm{x}_2)-\psi(t,\bm{x}_1)\bigl[\partial_0,\chi^\dagger(t,\bm{x}_2)\bigr]\right)\notag\\
 &\hspace{5cm}=(\bm{\eta}\cdot\bm{r})\left[\frac{i}{4M}(\bm{\nabla}_1^2-\bm{\nabla}_2^2),\psi(t,\bm{x}_1)\chi^\dagger(t,\bm{x}_2)\right]+\mathcal{O}(M^{-3})\notag\\
 &\hspace{5cm}=(\bm{\eta}\cdot\bm{r})\left[\frac{i}{2M}\bm{\nabla}_R\cdot\bm{\nabla}_r,\psi(t,\bm{x}_1)\chi^\dagger(t,\bm{x}_2)\right]+\mathcal{O}(M^{-3})\,.
\end{align}
Thus, these terms give corrections of order $1/M$ and higher.

The other terms in the boost transformation of the quark and antiquark fields in the $g\to0$ limit can also be rewritten in terms of the center-of-mass and relative coordinates, $\Rb$ and $\rb$:
\begin{align}\label{quarkoniumnoncoordinatetrans}
 &\psi(t,\bm{x}_1)\chi^\dagger(t,\bm{x}_2)\stackrel{K}{\longrightarrow}\ldots+\frac{i}{2M}\bigl[\bm{\eta}\cdot(\bm{\nabla}_1+\bm{\nabla}_2),\psi(t,\bm{x}_1)\chi^\dagger(t,\bm{x}_2)\bigr]\notag\\
 &\hspace{12pt}-\frac{1}{4M}\bigl[(\bm{\eta}\times\bm{\nabla}_1)\cdot,\bm{\sigma}\psi(t,\bm{x}_1)\chi^\dagger(t,\bm{x}_2)\bigr]+\frac{1}{4M}\bigl[(\bm{\eta}\times\bm{\nabla}_2)\cdot,\psi(t,\bm{x}_1)\chi^\dagger(t,\bm{x}_2)\bm{\sigma}\bigr]+\mathcal{O}(M^{-2})\notag\\
 &=\ldots+\frac{i}{2M}\bigl[\bm{\eta}\cdot\bm{\nabla}_R,\psi(t,\bm{x}_1)\chi^\dagger(t,\bm{x}_2)\bigr]-\frac{1}{8M}\left(\bm{\sigma}^{(1)}+\bm{\sigma}^{(2)}\right)\cdot\bigl[(\bm{\eta}\times\bm{\nabla}_R),\psi(t,\bm{x}_1)\chi^\dagger(t,\bm{x}_2)\bigr]\notag\\
 &\hspace{12pt}-\frac{1}{4M}\left(\bm{\sigma}^{(1)}-\bm{\sigma}^{(2)}\right)\cdot\bigl[(\bm{\eta}\times\bm{\nabla}_r),\psi(t,\bm{x}_1)\chi^\dagger(t,\bm{x}_2)\bigr]+\mathcal{O}(M^{-3})\,.
\end{align}
Here the ellipsis denotes the terms shown in Eq.~\eqref{quarkoniumcoordinatetrans} due to the coordinate transformations. We have also introduced the convenient notation
\begin{equation}
 \bm{\sigma}^{(1)}Q=\bm{\sigma}Q\hspace{20pt}\mathrm{and}\hspace{20pt}\bm{\sigma}^{(2)}Q=-Q\bm{\sigma}\,.
 \label{sigma12}
\end{equation}
This is to say that $\bm{\sigma}^{(1)}$ acts on the spin of the heavy quark and $\bm{\sigma}^{(2)}$ acts on the spin of the heavy antiquark (they correspond to the respective generators of rotations, see Appendix~\ref{SpacetimepNRQCD}). Since $\sigma_2\bm{\sigma}\sigma_2=-\bm{\sigma}^T$ and $\bm{\sigma}^TQ^T=(Q\bm{\sigma})^T$, charge conjugation effectively exchanges $\bm{\sigma}^{(1)}\leftrightarrow\bm{\sigma}^{(2)}$.

From these expressions, we expect the boost generator in the $g\to0$ limit to behave like
\begin{align}
 \bm{k}_Q\stackrel{g\to0}{=}{}&it\bm{\nabla}_R+i\bm{R}\partial_0+2M\bm{R}-\frac{1}{4M}\bm{\nabla}_R-\frac{1}{4M}\bigl\{\bm{r},(\bm{\nabla}_R\cdot\bm{\nabla}_r)\bigr\}\notag\\
 &-\frac{i}{8M}\bm{\nabla}_R\times\left(\bm{\sigma}^{(1)}+\bm{\sigma}^{(2)}\right)-\frac{i}{4M}\bm{\nabla}_r\times\left(\bm{\sigma}^{(1)}-\bm{\sigma}^{(2)}\right)+\mathcal{O}(M^{-3})\,.
 \label{freeboost}
\end{align}
This limit is interesting for the ansatz we are going to make for the singlet and octet boost generators, since it determines which coefficients we expect to be of order $1+\mathcal{O}(\alpha_\mathrm{s})$. In the last two terms of the first line, we have used
\begin{equation}
 \bm{r}(\bm{\nabla}_R\cdot\bm{\nabla}_r)=\frac{1}{2}\bigl\{\bm{r},(\bm{\nabla}_R\cdot\bm{\nabla}_r)\bigr\}-\frac{1}{2}\bm{\nabla}_R\,,
\end{equation}
in order to obtain terms that are explicitly Hermitian or anti-Hermitian.

Finally, we list here how the boost generators are required to behave under the discrete symmetries parity, charge conjugation, and time reversal:
\begin{equation}
 P\bm{k}_Q =-\bm{k}_Q\,,\hspace{30pt}C\bm{k}_Q =\sigma_2\bm{k}_Q^T\sigma_2\,,\hspace{30pt}T\bm{k}_Q = \sigma_2\bm{k}_Q\sigma_2\,.
\end{equation}
Note that $P$ changes the sign of both $\bm{r}$ and $\bm{R}$, $C$ changes the sign of $\bm{r}$ only, $T$ changes the sign of $t$ and takes the complex conjugate. While for the singlet field the transpose operation required by the $C$ transformation is trivially realized in color space, for the octet field the boost transformation is consistent with charge conjugation if it is of the form (in matrix notation) 
\begin{equation}
 \mathrm{O}\stackrel{K}{\longrightarrow}\mathrm{O}'=\mathrm{O}-i\etab\cdot\left(\kb_{O}^{(A)}\mathrm{O}+\mathrm{O}\kb_{O}^{(B)}\right)\,,
\end{equation}
where the two parts $\kb_{O}^{(A)}$ and $\kb_{O}^{(B)}$ are exchanged under $C$ as $\kb_{O}^{(A)}\stackrel{C}{\longleftrightarrow}\sigma_2\left(\kb_{O}^{(B)}\right)^T\sigma_2$.

\subsection{Redundancies and field redefinitions}\label{pNRQCDredef}

In order to find the boost generators in pNRQCD, we will use the EFT approach and write down the most general form allowed by the symmetries of the theory. However, it turns out that several terms in this ansatz are redundant, in the sense that one can make a field redefinition that removes them from the boost generators without changing the form of the Lagrangian. Thus, there is no loss in generality if one chooses to work with boost generators where these redundant terms are absent. We will identify appropriate field redefinitions in this section. Since we calculate the transformation of the Lagrangian up to orders $M^0r^1$ and $M^{-1}r^0$ in the next section, it is necessary to include all terms of order $M^0r^2$ and $M^{-1}r^0$ in the boosts. We will use the notation $c^{(m,n)}$ for the Wilson coefficients of operators that are of order $M^{-m}r^n$.

\subsubsection{Singlet field}

Even though we work with a general ansatz, some terms may be omitted from the start, which is similar to the construction of the pNRQCD Lagrangian. A term like $\bm{r}\cdot\bm{\nabla}_r$, for example, is neutral with respect to any symmetry and also the power counting. In principle one could add an infinite number of these terms to any operator in the Lagrangian, which would mean that at each order in the power counting, one would have to match an infinite number of terms, making the construction of the EFT impossible. By comparison with NRQCD, however, one sees that each derivative appears with at least one power of $1/M$, so also in pNRQCD one can neglect any term where there are more derivatives than powers of $1/M$. The same argument applies to spin-dependent terms, where each Pauli matrix has to be suppressed by a power of $1/M$. The only exception to this are the kinetic energy terms that have one derivative more than powers of $1/M$.

By extension, these rules also apply to the construction of the boost generators in the following way. Operators leading to terms in the transformation of the Lagrangian that would have to be canceled by derivative or spin terms with an insufficient $1/M$ suppression are immediately ruled out. For instance, since the center-of-mass kinetic energy is of the form $\nablab_R^2/M$ [and not $(\bm{r}\cdot\nablab_R)^2/(M\,r^2)$] at order $1/M$, we can exclude from the start a term like  $\bm{r}(\bm{r}\cdot\nablab_R)/(M\,r^2)$ from the boost, which has no counterpart in the Lagrangian. These arguments apply to Hermitian and anti-Hermitian terms differently, as Hermitian terms lead to commutators in the boosted Lagrangian, which often reduce the number of derivatives, while anti-Hermitian terms lead to anticommutators.

Keeping this in mind and writing everything in terms of explicitly Hermitian or anti-Hermitian operators (where we stay close to the nomenclature in~\cite{Brambilla:2003nt}), a rather general ansatz for the boost generator of the singlet is given by:\footnote{The subscripts on the boost coefficients, $a'$, $a''$, $a'''$, $b$, $c$, etc., are labels used to distinguish between different operators with the same suppression in $1/M$ and $r$. Primes distinguish operators that differ only in the contraction of their vector indices.}
\begin{align}\label{general boost on singlet}
 \bm{k}_{S}={}&it\bm{\nabla}_R+i\bm{R}\partial_0+2M\Rb-\frac{k_{SD}^{(1,0)}}{4M}\nablab_R-\frac{1}{4M}\left\{k_{Sa'}^{(1,0)}\rb,(\nablab_R\cdot\nablab_r)\right\}\notag\\
 &-\frac{1}{4M}\left\{k_{Sa''}^{(1,0)}(\rb\cdot\nablab_R),\nablab_r\right\}-\frac{1}{4M}\left\{k_{Sa'''}^{(1,0)}\rb\cdot,\nablab_r\right\}\bm{\nabla}_R\notag\\
 &-\frac{1}{4M}\left\{\frac{k_{Sb}^{(1,0)}}{r^2}\rb(\rb\cdot\nablab_R)r_i,(\nabla_r)_i\right\}-\frac{ik_{Sc}^{(1,0)}}{8M}\nablab_R\times\left(\sigmab^{(1)}+\sigmab^{(2)}\right)\notag\\
 &-\frac{ik_{Sd''}^{(1,0)}}{8Mr^2}\left(\rb\cdot\nablab_R\right)\left(\rb\times\left(\sigmab^{(1)}+\sigmab^{(2)}\right)\right)-\frac{ik_{Sd'''}^{(1,0)}}{8Mr^2}\left((\bm{r}\times\bm{\nabla}_R)\cdot\left(\bm{\sigma}^{(1)}+\bm{\sigma}^{(2)}\right)\right)\bm{r}\notag\\
 &-\frac{i}{8M}\left\{k_{Sa}^{(1,-1)},\nablab_r\times\left(\sigmab^{(1)}-\sigmab^{(2)}\right)\right\}+\frac{i}{8M}\left[\frac{k_{Sb'}^{(1,-1)}}{r^2}\left(\rb\cdot\left(\sigmab^{(1)}-\sigmab^{(2)}\right)\right)\rb\times,\nablab_r\right]\notag\\
 &-\frac{i}{8M}\left\{\frac{k_{Sb''}^{(1,-1)}}{r^2}\rb\times\left(\sigmab^{(1)}-\sigmab^{(2)}\right)r_i,(\nabla_r)_i\right\}+\mathcal{O}\left(M^{-2}r^0,M^{-1}r^1,M^0r^3\right)\,.
\end{align}
Note that we cannot write any color singlet operator at $\mathcal{O}\left(M^{0}r^2\right)$; also time derivatives do not yet appear at this order, apart from the coordinate transformations. Since the Wilson coefficients here depend on $r$, they have to be included inside the anticommutators with the derivative $\nablab_r$. We have used the identity
\begin{equation}
 \delta_{ij}\epsilon_{klm}=\delta_{ik}\epsilon_{jlm}+\delta_{il}\epsilon_{kjm}+\delta_{im}\epsilon_{klj} \,,
 \label{D3ident}
\end{equation}
in order to eliminate several terms. A term like $ik_{Sd'}^{(1,0)}(\rb\times\nablab_R)(\rb\cdot(\sigmab^{(1)}+\sigmab^{(2)}))/(8Mr^2)$, for instance, can be expressed in terms of the operators of $k_{Sc}^{(1,0)}$, $k_{Sd''}^{(1,0)}$, and $k_{Sd'''}^{(1,0)}$ through this identity; this can be shown by multiplying Eq.~\eqref{D3ident} with $r_ir_j(\nabla_R)_k(\sigma^{(1)}+\sigma^{(2)})_l$. A similar relation can be found between the operators of the coefficients, $k_{Sa}^{(1,-1)}$, $k_{Sb'}^{(1,-1)}$, and $k_{Sb''}^{(1,-1)}$, and an omitted $\{(ik_{Sb'''}^{(1,-1)}/8Mr^2)\rb\left[\rb\times\left(\sigmab^{(1)}-\sigmab^{(2)}\right)\right]_i,(\nabla_r)_i\}$ term.

Not all the terms in the boost generator, Eq.~\eqref{general boost on singlet}, are necessary if one exploits the freedom to perform field redefinitions; in other words, one can always redefine the fields as long as the symmetry properties of the fields are not altered. In order to keep the form of the Lagrangian intact after the field redefinitions, we will only consider unitary transformations $\mathcal{U}_S=\exp[u_S]$ ($u_S$ is anti-Hermitian), for which the new singlet field $\widetilde{\mathrm{S}}$ is related to the old $\mathrm{S}$ via $\mathrm{S}=\mathcal{U}_S\widetilde{\mathrm{S}}$~\cite{Brambilla:2003nt}.

In order to find a suitable unitary transformation, we need to look for terms which are anti-Hermitian and $P$, $C$, and $T$ invariant. Such terms can be easily found by multiplying the Hermitian terms in $\bm{k}_S$ with $\bm{\nabla}_R/M$, which explains the nomenclature we use for $\mathcal{U}_S$:
\begin{align}
 \mathcal{U}_S=\exp&\left[-\frac{1}{4M^2}\left\{q_{Sa''}^{(1,0)}\rb\cdot\nablab_R,\nablab_r\cdot\nablab_R\right\}-\frac{1}{4M^2}\left\{q_{Sa'''}^{(1,0)}\rb\cdot,\nablab_r\right\}\nablab_R^2\right.\notag\\
 &-\frac{1}{4M^2}\left\{\frac{q_{Sb}^{(1,0)}}{r^2}(\rb\cdot\nablab_R)^2\rb\cdot,\nablab_r\right\}-\frac{iq_{Sd'''}^{(1,0)}}{8M^2r^2}\left(\rb\cdot\nablab_R\right)\left((\rb\times\nablab_R)\cdot\left(\sigmab^{(1)}+\sigmab^{(2)}\right)\right)\notag\\
 &+\frac{i}{8M^2}\left\{q_{Sa}^{(1,-1)},(\nablab_r\times\nablab_R)\cdot\left(\sigmab^{(1)}-\sigmab^{(2)}\right)\right\}\notag\\
 &-\frac{i}{8M^2}\left\{\frac{q_{Sb'}^{(1,-1)}}{r^2}\left(\rb\cdot\left(\sigmab^{(1)}-\sigmab^{(2)}\right)\right)\left(\rb\times\nablab_R\right)\cdot,\nablab_r\right\}\notag\\
 &+\left.\frac{i}{8M^2}\left\{\frac{q_{Sb''}^{(1,-1)}}{r^2}\left((\rb\times\nablab_R)\cdot\left(\sigmab^{(1)}-\sigmab^{(2)}\right)\right)\rb\cdot,\nablab_r\right\}+\dots\right]\,,
 \label{singletredef}
\end{align}
where the ellipsis stands for higher order terms, which do not affect the calculations of this paper. The coefficients $q^{(m,n)}_S$ are free parameters.

We can work out the transformation of the new singlet field $\widetilde{\mathrm{S}}$ under boosts in the following way:
\vspace{-12pt}
\begin{align}
 \widetilde{\mathrm{S}}'&=\mathcal{U'}^{\dagger}_S\mathrm{S}'=\mathcal{U'}^{\dagger}_S(1-i\etab\cdot\kb_S)\mathcal{U}_S\widetilde{\mathrm{S}}=\left[1-\mathcal{U}^\dagger_S(i\etab\cdot\kb_S)\mathcal{U}_S+\left(\delta\mathcal{U}_S^\dagger\right)\mathcal{U}_S\right]\widetilde{\mathrm{S}}\notag\\
 &\equiv\left(1-i\etab\cdot\widetilde{\bm{k}}_S\right)\widetilde{\mathrm{S}}\,,
\end{align}
\vspace{-30pt}
\begin{align}
 \delta\mathcal{U}_S^\dagger(\nablab_R,\Eb,\Bb)={}&\left[\etab\cdot(t\nablab_R+\bm{R}\partial_0),\mathcal{U}_S^\dagger(\nablab_R,\Eb,\Bb)\right]\notag\\
 &+\mathcal{U}_S^\dagger(\nablab_R+\etab\partial_0,\Eb+\etab\times\Bb,\Bb-\etab\times\Eb)-\mathcal{U}_S^\dagger(\nablab_R,\Eb,\Bb)\,,
\end{align}
with the second line expanded to linear order in $\etab$. The transformed boost generator $\widetilde{\bm{k}}_S$ has to be expanded to the same order as the original $\bm{k}_S$:

\begin{align}\label{singleredeftrans}
 \widetilde{\bm{k}}_S={}&\bm{k}_S+\left[\bm{\hat{k}}_S,u_S\right]\notag-u_S(\nablab_R+\etab\partial_0,\Eb+\etab\times\Bb,\Bb-\etab\times\Eb)+u_S(\nablab_R,\Eb,\Bb)\\
 &+\frac{1}{2}\left[\left[\bm{\hat{k}}_S,u_S\right]\notag-u_S(\nablab_R+\etab\partial_0,\Eb+\etab\times\Bb,\Bb-\etab\times\Eb)+u_S(\nablab_R,\Eb,\Bb),u_S\right]+\dots\notag\\
 ={}&\bm{k}_S+[2M\bm{R},u_S]+\mathcal{O}(M^{-2})\,.
\end{align}
Inserting the explicit field redefinition from Eq.~\eqref{singletredef}, we obtain
\begin{align}
 \widetilde{\bm{k}}_S=\bm{k}_S&+\frac{1}{2M}\left\{q_{Sa''}^{(1,0)}\rb,(\nablab_R\cdot\nablab_r)\right\}+\frac{1}{2M}\left\{q_{Sa''}^{(1,0)}(\rb\cdot\nablab_R),\nablab_r\right\}\notag\\
 &+\frac{1}{M}\left\{q_{Sa'''}^{(1,0)}\rb\cdot,\nablab_r\right\}\bm{\nabla}_R+\frac{1}{M}\left\{\frac{q_{Sb}^{(1,0)}}{r^2}\rb(\rb\cdot\nablab_R)r_i,(\nabla_r)_i\right\}\notag\\
 &-\frac{iq_{Sd'''}^{(1,0)}}{4Mr^2}\left(\rb\cdot\nablab_R\right)\left(\rb\times\left(\sigmab^{(1)}+\sigmab^{(2)}\right)\right)+\frac{iq_{Sd'''}^{(1,0)}}{4Mr^2}\left(\left(\bm{r}\times \bm{\nabla}_R\right)\cdot\left(\bm{\sigma}^{(1)}+\bm{\sigma}^{(2)}\right)\right)\bm{r}\notag\\
 &+\frac{i}{4M}\left\{q_{Sa}^{(1,-1)},\nablab_r\times\left(\sigmab^{(1)}-\sigmab^{(2)}\right)\right\}-\frac{i}{4M}\left[\frac{q_{Sb'}^{(1,-1)}}{r^2}\left(\rb\cdot\left(\sigmab^{(1)}-\sigmab^{(2)}\right)\right)\rb\times,\nablab_r\right]\notag\\
 &+\frac{i}{4M}\left\{\frac{q_{Sb''}^{(1,-1)}}{r^2}\rb\times\left(\sigmab^{(1)}-\sigmab^{(2)}\right)r_i,(\nabla_r)_i\right\}+\mathcal{O}\left(M^{-2}\right)\,.
\end{align}
These extra terms can be absorbed in the operators already present in Eq.~\eqref{general boost on singlet} by changing the coefficients in the following way:
\vspace{-5pt}
\begin{align}\label{singlet boost constraints}
 &\widetilde{k}_{Sa'}^{(1,0)}=k_{Sa'}^{(1,0)}-2q_{Sa''}^{(1,0)}\,, & &\widetilde{k}_{Sa''}^{(1,0)}=k_{Sa''}^{(1,0)}-2q_{Sa''}^{(1,0)}\,, & &\widetilde{k}_{Sa'''}^{(1,0)}=k_{Sa'''}^{(1,0)}-4q_{Sa'''}^{(1,0)}\,,\notag\\
 &\widetilde{k}_{Sb}^{(1,0)}=k_{Sb}^{(1,0)}-4q_{Sb}^{(1,0)}\,, & &\widetilde{k}_{Sd''}^{(1,0)}=k_{Sd''}^{(1,0)}+2q_{Sd'''}^{(1,0)}\,, & &\widetilde{k}_{Sd'''}^{(1,0)}=k_{Sd'''}^{(1,0)}-2q_{Sd'''}^{(1,0)}\,,\notag\\
 &\widetilde{k}_{Sa}^{(1,-1)}=k_{Sa}^{(1,-1)}-2q_{Sa}^{(1,-1)}\,, & &\widetilde{k}_{Sb'}^{(1,-1)}=k_{Sb'}^{(1,-1)}-2q_{Sb'}^{(1,-1)}\,, & &\widetilde{k}_{Sb''}^{(1,-1)}=k_{Sb''}^{(1,-1)}-2q_{Sb''}^{(1,-1)}\,.
\end{align}

\vspace{-5pt}
The seven free parameters $q^{(m,n)}_S$ in the unitary transformation can be chosen in any convenient way. Comparing this to the expected result in the $g\to0$ limit from Eq.~\eqref{freeboost}, we choose to set $\widetilde{k}_{Sa''}^{(1,0)}$, $\widetilde{k}_{Sa'''}^{(1,0)}$, $\widetilde{k}_{Sb}^{(1,0)}$, $\widetilde{k}_{Sd'''}^{(1,0)}$, $\widetilde{k}_{Sb'}^{(1,-1)}$ and $\widetilde{k}_{Sb''}^{(1,-1)}$ equal to zero, as well as to fix $\widetilde{k}_{Sa}^{(1,-1)}=1$. Then, after dropping the tilde notation for the new field, the boost transformation becomes 
\begin{align}
 \bm{k}_S={}&it\bm{\nabla}_R+i\bm{R}\partial_0+2M\Rb-\frac{k_{SD}^{(1,0)}}{4M}\nablab_R-\frac{1}{4M}\left\{k_{Sa'}^{(1,0)}\rb,(\nablab_R\cdot\nablab_r)\right\}\notag\\
 &-\frac{ik_{Sc}^{(1,0)}}{8M}\nablab_R\times\left(\sigmab^{(1)}+\sigmab^{(2)}\right)-\frac{ik_{Sd''}^{(1,0)}}{8Mr^2}\left(\rb\cdot\nablab_R\right)\left(\rb\times\left(\sigmab^{(1)}+\sigmab^{(2)}\right)\right)\notag\\
 &-\frac{i}{4M}\nablab_r\times\left(\sigmab^{(1)}-\sigmab^{(2)}\right)+\mathcal{O}\left(M^{-2}r^0,M^{-1}r^1,M^0r^3\right)\,,
\end{align}
in which only four coefficients, $k_{SD}^{(1,0)}$, $k_{Sa'}^{(1,0)}$, $k_{Sc}^{(1,0)}$, and $k_{Sd''}^{(1,0)}$, remain undetermined.

\subsubsection{Octet field}

\vspace{-10pt}
In a similar fashion, one can proceed to determine the most general form of the boost transformation for the octet field. The main difference from the singlet is that all center-of-mass derivatives (except for the coordinate transformations) have to be replaced by covariant derivatives in the adjoint representation, $\Db^{ab}=\delta^{ab}\nablab_R-f^{abc}g\bm{A}^c$, due to the color charge of the octet field. There are no operators at order $M^0r^2$ for the singlet field, but for the case of the octet, one can write two operators involving the chromoelectric field. There are no new terms at order $M^{-1}r^0$.

We write now the color components of the octet field explicitly instead of using the matrix notation, for which the boost transformation reads
\begin{equation}
 O^a\stackrel{K}{\longrightarrow}O^{a\,\prime}=\bigl(\delta^{ab}-i\etab\cdot\kb_O^{ab}\bigr)O^b\,.
\end{equation}
The parity transformation of the boost generator in component notation is the same as in matrix notation. For the charge conjugation and time reversal transformations, we introduce a sign factor $\zeta^a$ through $(T^a)^T=(T^a)^*\equiv\zeta^aT^a$ (the double appearance of the color index~$a$ in any instance of $\zeta^a$ does not imply its summation). With this the fields in the adjoint representation transform as
\begin{align}
 &O^a\stackrel{C}{\longrightarrow}\sigma_2\zeta^aO^a\sigma_2\,, & &E^a\stackrel{C}{\longrightarrow}-\zeta^aE^a\,, & &B^a\stackrel{C}{\longrightarrow}-\zeta^aB^a\,,\\
 &O^a\stackrel{T}{\longrightarrow}\sigma_2\zeta^aO^a\sigma_2\,, & &E^a\stackrel{T}{\longrightarrow}\zeta^aE^a\,, & &B^a\stackrel{T}{\longrightarrow}-\zeta^aB^a\,.
\end{align}
The boost generator in component notation has to transform like
\begin{equation}
 \kb_O^{ab}\stackrel{C}{\longrightarrow}\zeta^a\zeta^b\sigma_2\left(\kb_O^{ab}\right)^T\sigma_2\,,\hspace{30pt}\kb_O^{ab}\stackrel{T}{\longrightarrow}\zeta^a\zeta^b\sigma_2\kb_O^{ab}\sigma_2\,.
\end{equation}
For the sign factors, one can use the following identities:
\begin{equation}
 \zeta^a\zeta^b\delta^{ab}=\delta^{ab}\,,\hspace{20pt}\zeta^a\zeta^b\zeta^cf^{abc}=-f^{abc}\,,\hspace{20pt}\zeta^a\zeta^b\zeta^cd^{abc}=d^{abc}\,,
\end{equation}
which follow from the commutation relations of the color matrices.

A general ansatz for the boost generator of the octets is then:
\begin{align}\label{general octet boost}
 \kb_O^{ab}={}&\delta^{ab}(it\nablab_R+i\bm{R}\partial_0+2M\Rb)-\frac{k_{OD}^{(1,0)}}{4M}\Db_R^{ab}+\frac{i}{8}f^{abc}k_{Oa}^{(0,2)}(\rb\cdot g\Eb^c)\rb+\frac{i}{8}f^{abc}k_{Ob}^{(0,2)}r^2g\Eb^c\notag\\
 &-\frac{1}{4M}\left\{k_{Oa'}^{(1,0)}\rb,(\Db_R^{ab}\cdot\nablab_r)\right\}-\frac{1}{4M}\left\{k_{Oa''}^{(1,0)}(\rb\cdot\Db_R^{ab}),\nablab_r\right\}-\frac{1}{4M}\left\{k_{Oa'''}^{(1,0)}\rb\cdot,\nablab_r\right\}\Db_R^{ab}\notag\\
 &-\frac{1}{4M}\left\{\frac{k_{Ob}^{(1,0)}}{r^2}\rb(\rb\cdot\Db_R^{ab})r_i,(\nabla_r)_i\right\}-\frac{ik_{Oc}^{(1,0)}}{8M}\Db_R^{ab}\times\left(\sigmab^{(1)}+\sigmab^{(2)}\right)\notag\\
 &-\frac{ik_{Od''}^{(1,0)}}{8Mr^2}\left(\rb\cdot\Db_R^{ab}\right)\left(\rb\times\left(\sigmab^{(1)}+\sigmab^{(2)}\right)\right)-\frac{ik_{Od'''}^{(1,0)}}{8Mr^2}\left((\bm{r}\times\Db_R^{ab})\cdot\left(\bm{\sigma}^{(1)}+\bm{\sigma}^{(2)}\right)\right)\bm{r}\notag\\
 &-\frac{i\delta^{ab}}{8M}\left\{k_{Oa}^{(1,-1)},\nablab_r\times\left(\sigmab^{(1)}-\sigmab^{(2)}\right)\right\}+\frac{i\delta^{ab}}{8M}\left[\frac{k_{Ob'}^{(1,-1)}}{r^2}\left(\rb\cdot\left(\sigmab^{(1)}-\sigmab^{(2)}\right)\right)\rb\times,\nablab_r\right]\notag\\
 &-\frac{i\delta^{ab}}{8M}\left\{\frac{k_{Ob''}^{(1,-1)}}{r^2}\rb\times\left(\sigmab^{(1)}-\sigmab^{(2)}\right)r_i,(\nabla_r)_i\right\}+\mathcal{O}\left(M^{-2}r^0,M^{-1}r^1,M^0r^3\right)\,.
\end{align}
We can again perform a redefinition of the octet field through a unitary transformation $\widetilde{O}^a=\mathcal{U}_O^{ab}O^b$ with $\mathcal{U}_O=\exp[u_O]$, in order to reduce the number of coefficients in $\kb_O$. For this transformation matrix, the same arguments apply as in the singlet case, so that we write the anti-Hermitian operator $u_O$ as:
\begin{align}\label{unitaryoctet}
 u_O^{ab}={}&-\frac{q_{Oa}^{(0,2)}}{16M}\bigl\{(\rb\cdot\Db_R),(\rb\cdot g\Eb)\bigr\}^{ab}+\frac{q_{Ob}^{(0,2)}}{16M}r^2\bigl\{\Db_R\cdot, g\Eb\bigr\}^{ab}\notag\\
 &-\frac{1}{4M^2}\left\{q_{Oa''}^{(1,0)}(\rb\cdot\Db_R),(\nablab_r\cdot\Db_R)\right\}^{ab}-\frac{1}{4M^2}\left\{q_{Oa'''}^{(1,0)}\rb\cdot, \nablab_r\right\}(\Db_R^2)^{ab}\notag\\
 &-\frac{1}{4M^2}\left\{\frac{q_{Ob}^{(1,0)}}{r^2}\left((\rb\cdot\Db_R)^2\right)^{ab}\rb\cdot,\nablab_r\right\}\notag\\
 &-\frac{iq_{Od'''}^{(1,0)}}{16M^2r^2}\left\{(\rb\cdot\Db_R),\left((\rb\times\Db_R)\cdot\left(\sigmab^{(1)}+\sigmab^{(2)}\right)\right)\right\}^{ab}\notag\\
 &+\frac{i}{8M^2}\left\{q_{Oa}^{(1,-1)},(\nablab_r\times\Db_R^{ab})\cdot\left(\sigmab^{(1)}-\sigmab^{(2)}\right)\right\}\notag\\
 &-\frac{i}{8M^2}\left\{\frac{q_{Ob'}^{(1,-1)}}{r^2}\left(\rb\cdot\left(\sigmab^{(1)}-\sigmab^{(2)}\right)\right)(\rb\times\Db_R^{ab})\cdot,\nablab_r\right\}\notag\\
 &+\frac{i}{8M^2}\left\{\frac{q_{Ob''}^{(1,-1)}}{r^2}\left((\rb\times\Db_R^{ab})\cdot\left(\sigmab^{(1)}-\sigmab^{(2)}\right)\right)\rb\cdot,\nablab_r\right\}+\dots\,,
\end{align}
where $\{A,B\}^{ab}=A^{ab'}B^{b'b}+B^{ab'}A^{b'b}$, and it is understood that $\Eb^{ab}=-if^{abc}\Eb^c$.

Just like in the singlet case, the new boost generator (for the new octet field) after this transformation is given by:
\begin{align}
 \widetilde{\kb}_O^{ab}={}&\kb_O^{ab}+\bigl[2M\Rb,u_O^{ab}\bigr]+\mathcal{O}\left(M^{-2}\right)\notag\\
 ={}&\kb_O^{ab}-\frac{i}{8}f^{abc}q_{Oa}^{(0,2)}(\rb\cdot g\Eb^c)\rb-\frac{i}{8}f^{abc}q_{Ob}^{(0,2)}r^2g\Eb^c\notag\\
 &+\frac{1}{2M}\left\{q_{Oa''}^{(1,0)}\rb,(\Db_R^{ab}\cdot\nablab_r)\right\}+\frac{1}{2M}\left\{q_{Oa''}^{(1,0)}(\rb\cdot\Db_R^{ab}),\nablab_r\right\}\notag\\
 &+\frac{1}{M}\left\{q_{Oa'''}^{(1,0)}\rb\cdot,\nablab_r\right\}\Db_R^{ab}+\frac{1}{M}\left\{\frac{q_{Ob}^{(1,0)}}{r^2}\rb(\rb\cdot\Db_R^{ab})r_i,(\nabla_r)_i\right\}\notag\\
 &-\frac{iq_{Od'''}^{(1,0)}}{4Mr^2}\left(\rb\times\left(\sigmab^{(1)}+\sigmab^{(2)}\right)\right)(\rb\cdot\Db_R^{ab})+\frac{iq_{Od'''}^{(1,0)}}{4Mr^2}\left((\rb\times\Db_R^{ab})\cdot\left(\sigmab^{(1)}+\sigmab^{(2)}\right)\right)\rb\notag\\
 &+\frac{i\delta^{ab}}{4M}\left\{q_{Oa}^{(1,-1)},\nablab_r\times\left(\sigmab^{(1)}-\sigmab^{(2)}\right)\right\}-\frac{i\delta^{ab}}{4M}\left[\frac{q_{Ob'}^{(1,-1)}}{r^2}\left(\rb\cdot\left(\sigmab^{(1)}-\sigmab^{(2)}\right)\right)\rb\times,\nablab_r\right]\notag\\
 &+\frac{i\delta^{ab}}{4M}\left\{\frac{q_{Ob''}^{(1,-1)}}{r^2}\left(\rb\times\left(\sigmab^{(1)}-\sigmab^{(2)}\right)\right)r_i,(\nabla_r)_i\right\}+\mathcal{O}(M^{-2})\,.
\end{align}
This gives formally the same relations for the transformed boost coefficients as in the singlet case, with the addition of two new relations for the coefficients of the terms with the chromoelectric field:
\begin{align}
 &\widetilde{k}_{Oa}^{(0,2)}=k_{Oa}^{(0,2)}-2q_{Oa}^{(0,2)}\,, & &\widetilde{k}_{Ob}^{(0,2)}=k_{Ob}^{(0,2)}-2q_{Ob}^{(0,2)}\,, & & \notag\\
 &\widetilde{k}_{Oa'}^{(1,0)}=k_{Oa'}^{(1,0)}-2q_{Oa''}^{(1,0)}\,, & &\widetilde{k}_{Oa''}^{(1,0)}=k_{Oa''}^{(1,0)}-2q_{Oa''}^{(1,0)}\,, & &\widetilde{k}_{Oa'''}^{(1,0)}=k_{Oa'''}^{(1,0)}-4q_{Oa'''}^{(1,0)}\,,\notag\\
 &\widetilde{k}_{Ob}^{(1,0)}=k_{Ob}^{(1,0)}-4q_{Ob}^{(1,0)}\,, & &\widetilde{k}_{Od''}^{(1,0)}=k_{Od''}^{(1,0)}+2q_{Od'''}^{(1,0)}\,, & &\widetilde{k}_{Od'''}^{(1,0)}=k_{Od'''}^{(1,0)}-2q_{Od'''}^{(1,0)}\,,\notag\\
 &\widetilde{k}_{Oa}^{(1,-1)}=k_{Oa}^{(1,-1)}-2q_{Oa}^{(1,-1)}\,, & &\widetilde{k}_{Ob'}^{(1,-1)}=k_{Ob'}^{(1,-1)}-2q_{Ob'}^{(1,-1)}\,, & &\widetilde{k}_{Ob''}^{(1,-1)}=k_{Ob''}^{(1,-1)}-2q_{Ob''}^{(1,-1)}\,.
\label{reparameterizationofkO}
\end{align}
We choose the parameters  $q^{(m,n)}_O$ to set $\widetilde{k}_{Oa''}^{(1,0)}$, $\widetilde{k}_{Oa'''}^{(1,0)}$, $\widetilde{k}_{Ob}^{(1,0)}$, $\widetilde{k}_{Od'''}^{(1,0)}$, $\widetilde{k}_{Ob'}^{(1,-1)}$, $\widetilde{k}_{Ob''}^{(1,-1)}$ and $\widetilde{k}_{Ob}^{(0,2)}$ equal to zero, as well as to fix $\widetilde{k}_{Oa}^{(0,2)}=1$ and $\widetilde{k}_{Oa}^{(1,-1)}=1$.\footnote{The choice $\widetilde{k}_{Oa}^{(0,2)}=1$ is dictated by the tree level matching result of~\cite{Brambilla:2003nt}.} Then, after dropping the tilde notation, the boost transformation simplifies to
\begin{align}\label{boostoctet}
 \bm{k}_O^{ab}={}&\delta^{ab}(it\bm{\nabla}_R+i\bm{R}\partial_0+2M\Rb)-\frac{k_{OD}^{(1,0)}}{4M}\Db_R^{ab}+\frac{i}{8}f^{abc}(\rb\cdot g\Eb^c)\rb-\frac{1}{4M}\left\{k_{Oa'}^{(1,0)}\rb,(\nablab_r\cdot\Db_R^{ab})\right\}\notag\\
 &-\frac{ik_{Oc}^{(1,0)}}{8M}\Db_R^{ab}\times\left(\sigmab^{(1)}+\sigmab^{(2)}\right)-\frac{ik_{Od''}^{(1,0)}}{8Mr^2}\left(\rb\cdot\Db_R^{ab}\right)\left(\rb\times\left(\sigmab^{(1)}+\sigmab^{(2)}\right)\right)\notag\\
 &-\frac{i\delta^{ab}}{4M}\nablab_r\times\left(\sigmab^{(1)}-\sigmab^{(2)}\right)+\mathcal{O}\left(M^{-2}r^0,M^{-1}r^1,M^0r^3\right)\,,
\end{align}
in which only four undetermined coefficients $k_{OD}^{(1,0)}$, $k_{Oa'}^{(1,0)}$, $k_{Oc}^{(1,0)}$, and $k_{Od''}^{(1,0)}$ remain, just like in the case of the singlet. These coefficients, as well as the ones from the singlet, will be constrained in the next section.

Finally, we observe that with respect to~\cite{Brambilla:2003nt} the unitary transformation $u_O^{ab}$ contains two more terms: the first two terms in the right-hand side of Eq.~\eqref{unitaryoctet}, which are proportional to the chromoelectric field. These two new terms allow us to choose the octet field in such a way that its Lorentz boost contains the chromoelectric field exactly in the form $if^{abc}(\rb\cdot g\Eb^c)\rb/8$, see Eq.~\eqref{boostoctet}. As we will see in the next section, this form of the boost is convenient, for it leads to stricter constraints on the Wilson coefficients of the Lagrangian in the octet sector than in~\cite{Brambilla:2003nt}.

\subsection{Invariance of the Lagrangian}\label{pNRQCDinv}
\subsubsection{Singlet sector}

The boost generators have to satisfy the commutation relation Eq.~\eqref{commutation relation}, which for the singlet at leading order in $1/M$ corresponds to
\begin{align}
 (\xib&\times\etab)\cdot(\bm{R}\times\nablab_R)-\bigl[\xib\cdot\bm{\hat{k}}_S,2M\etab\cdot\bm{R}\bigr]+\bigl[\etab\cdot\bm{\hat{k}}_S,2M\xib\cdot\bm{R}\bigr]+\mathcal{O}(M^{-1})\notag\\
 ={}&(\xib\times\etab)\cdot(\bm{R}\times\nablab_R)+(\xib\times\etab)\cdot(k_{Sa'}^{(1,0)}\rb\times\nablab_r)+\frac{ik_{Sc}^{(1,0)}}{2}(\xib\times\etab)\cdot(\sigmab^{(1)}+\sigmab^{(2)})\notag\\
 &-\frac{ik_{Sd''}^{(1,0)}}{2r^2}(\xib\times\etab)\cdot\left(\rb\times\left(\rb\times\left(\sigmab^{(1)}+\sigmab^{(2)}\right)\right)\right)+\mathcal{O}(M^{-1})\notag\\
 \stackrel{!}{=}{}&(\xib\times\etab)\cdot\left(\bm{R}\times\bm{\nabla}_R+\bm{r}\times\bm{\nabla}_r+\frac{i}{2}\left(\sigmab^{(1)}+\sigmab^{(2)}\right)\right).
\end{align}
This fixes three further coefficients: $k_{Sa'}^{(1,0)}=k_{Sc}^{(1,0)}=1$, and $k_{Sd''}^{(1,0)}=0$. Note that the term $\rb\times\nablab_r$, which generates rotations of the relative distance coordinate, is obtained from terms that we have included in $\bm{\hat{k}}_S$, as opposed to $\Rb\times\nablab_R$, which comes from the coordinate transformations.

The last remaining coefficient $k_{SD}^{(1,0)}$ is fixed when we apply the boost transformation to the singlet sector of the Lagrangian up to $\mathcal{O}(M^{-2})$ (we follow the notation from Ref.~\cite{Brambilla:2003nt})
\begin{align}\label{ss lagrangian}
 \mathcal{L}_\mathrm{pNRQCD}^{(S)}={}&\int d^3r\,\mathrm{Tr}\biggl[\mathrm{S}^{\dagger}\biggl(i\partial_0+\frac{1}{2M}\left\{c_S^{(1,-2)},\nablab_r^2\right\}+\frac{c_S^{(1,0)}}{4M}\nablab_R^2-V_S^{(0)}-\frac{V_S^{(1)}}{M}+\frac{V_{rS}}{M^2}\notag\\
 &+\frac{V_{p^2Sa}}{4M^2}\nablab_R^2+\frac{1}{2M^2}\left\{V_{p^2Sb},\nablab_r^2\right\}+\frac{V_{L^2 Sa}}{4M^2r^2}\left(\rb\times\nablab_R\right)^2+\frac{V_{L^2 Sb}}{M^2r^2}\left(\rb\times\nablab_r\right)^2\notag\\
 &-\frac{V_{S_{12}S}}{M^2r^2}\left(3(\rb\cdot\sigmab^{(1)})(\rb\cdot\sigmab^{(2)})-r^2(\sigmab^{(1)}\cdot\sigmab^{(2)})\right)-\frac{V_{S^2S}}{4M^2}\sigmab^{(1)}\cdot\sigmab^{(2)}\notag\\
 &+\frac{iV_{LSSa}}{4M^2}(\rb\times\nablab_R)\cdot(\sigmab^{(1)}-\sigmab^{(2)})+\frac{iV_{LSSb}}{2M^2}(\rb\times\nablab_r)\cdot(\sigmab^{(1)}+\sigmab^{(2)})\biggr)\mathrm{S}\biggr]\,,
\end{align}
where the subscripts $a$ and $b$ (later also $c$, $d$, and $e$) on the potentials are labels used to distinguish different operators of the same type. The difference between the transformed Lagrangian and the original $\mathcal{L}_\mathrm{pNRQCD}^{(S)}$ needs to be a derivative. We obtain:
\begin{align}
 \partial_\mu\widehat{\Delta}^\mu\mathcal{L}^{(S)}={}&\int d^3r\,\mathrm{Tr}\left[\etab\cdot\mathrm{S}^{\dagger}\left(i\left(1-c_S^{(1,0)}\right)\nablab_R-\frac{1}{2M}\left(k_{SD}^{(1,0)}-c_S^{(1,0)}\right)\nablab_R\partial_0\right.\right.\notag\\
 &-\frac{i}{M}\left(V_{p^2Sa}+V_{L^2Sa}+\frac{1}{2}V_S^{(0)}\right)\nablab_R+\frac{i}{Mr^2}\left(V_{L^2Sa}+\frac{r}{2}V_S^{(0)\,\prime}\right)\rb(\rb\cdot\nablab_R)\notag\\
 &+\left.\left.\frac{1}{2M}\left(V_{LSSa}+\frac{1}{2r}V_S^{(0)\,\prime}\right)\left(\sigmab^{(1)}-\sigmab^{(2)}\right)\times\rb\right)\mathrm{S}\right]\,,
\end{align}
where we have neglected the terms from the coordinate transformations. The prime on a potential $V$ denotes derivative with respect to $r$.

None of these terms has the form of an overall derivative, so all coefficients have to vanish, which gives the following constraints:
\begin{align}
 k_{SD}^{(1,0)}&=c_S^{(1,0)}=1\,, & V_{p^2Sa}&=\frac{r}{2}V_S^{(0)\prime}-\frac{1}{2}V_S^{(0)}\,,\notag\\
 V_{L^2Sa}&=-\frac{r}{2}V_S^{(0)\prime}\,, & V_{LSSa}&=-\frac{1}{2r}V_S^{(0)\prime}\,.
 \label{singletconstraints}
 \end{align}
These coincide with the results obtained in~\cite{Brambilla:2003nt}. The constraints for the singlet spin dependent potential $V_{LSSa}$ and for the singlet spin independent potentials $V_{p^2Sa}$ and $V_{L^2Sa}$ were first obtained in~\cite{Gromes:1984ma} and~\cite{Barchielli:1988zp} respectively by boosting the potentials expressed in terms of Wilson loops; a more recent derivation can be found in~\cite{Brambilla:2001xk}. Note that with the last remaining boost coefficient $k_{SD}^{(1,0)}$ fixed to unity, the boost generator for the singlet field up to this order is exactly the same as in the $g\to0$ limit; in other words, there are no loop corrections to any of the coefficients. It is important to remember, however, that this form of the boost generator has been a particular choice obtained through certain field redefinitions. Other choices are equally valid and may change the constraints derived above. Our choice corresponds to the one made in~\cite{Brambilla:2003nt}.

\subsubsection{Octet sector}\label{sectionoctettransformation}

The calculation of the commutator of two boosts for the octet fields is analogous to that for the singlet fields, so we have $k_{Oa'}^{(1,0)}=k_{Oc}^{(1,0)}=1$ and $k_{Od''}^{(1,0)}=0$ for the octet. The only remaining boost coefficient is then $k_{OD}^{(1,0)}$.

The octet sector of the pNRQCD Lagrangian up to $\mathcal{O}(M^{-2})$ can be written as~\cite{Brambilla:2003nt}, 
\begin{align}\label{oo lagrangian}
 \mathcal{L}_\mathrm{pNRQCD}^{(O)}={}&\int d^3r\,\mathrm{Tr}\left\{\mathrm{O}^{\dagger}\left(iD_0+\frac{1}{2M}\left\{c_O^{(1,-2)},\nablab_r^2\right\}+\frac{c_O^{(1,0)}}{4M}\Db_R^2-V_O^{(0)}-\frac{V_O^{(1)}}{M}-\frac{V_{rO}}{M^2}\right.\right.\notag\\
 &+\frac{V_{p^2Oa}}{4M^2}\Db_R^2+\frac{1}{2M^2}\bigl\{V_{p^2Ob},\nablab_r^2\bigr\}+\frac{V_{L^2Oa}}{4M^2r^2}(\rb\times\Db_R)^2+\frac{V_{L^2Ob}}{M^2r^2}(\rb\times\nablab_r)^2\notag\\
 &-\frac{V_{S_{12}O}}{M^2r^2}\left(3\left(\rb\cdot\sigmab^{(1)}\right)\left(\rb\cdot\sigmab^{(2)}\right)-r^2\left(\sigmab^{(1)}\cdot\sigmab^{(2)}\right)\right)-\frac{V_{S^2O}}{4M^2}\sigmab^{(1)}\cdot\sigmab^{(2)}\notag\\
 &+\left.\frac{iV_{LSOa}}{4M^2}(\rb\times\Db_R)\cdot\left(\sigmab^{(1)}-\sigmab^{(2)}\right)+\frac{iV_{LSOb}}{2M^2}(\rb\times\nablab_r)\cdot\left(\sigmab^{(1)}+\sigmab^{(2)}\right)\right)\mathrm{O}\notag\\
 &+\left[\mathrm{O}^\dagger\left(\frac{V_{OO}^{(0,1)}}{2}\rb\cdot g\Eb+\frac{V_{OOa}^{(0,2)}}{8}\bigl[(\rb\cdot\Db_R),(\rb\cdot g\Eb)\bigr]+\frac{V_{OOb}^{(0,2)}}{8}r^2\bigl[\Db_R\cdot,g\Eb\bigr]\right.\right.\notag\\
 &+\frac{iV^{(1,0)}_{OOa}}{8M}\bigl\{\bm{\nabla}_r\cdot,\bm{r}\times g\Bb\bigr\}+\frac{c_FV_{OOb}^{(1,0)}}{2M}g\Bb\cdot\sigmab^{(1)}-\frac{V_{O\otimes Ob}^{(1,0)}}{2M}g\Bb\cdot\sigmab^{(2)}\notag\\
 &+\frac{V_{OOc}^{(1,0)}}{2Mr^2}(\rb\cdot g\Bb)\left(\rb\cdot\sigmab^{(1)}\right)-\frac{V_{O\otimes Oc}^{(1,0)}}{2Mr^2}(\rb\cdot g\Bb)\left(\rb\cdot\sigmab^{(2)}\right)+\frac{V_{OOd}^{(1,0)}}{2Mr}\rb\cdot g\Eb\notag\\
 &-\frac{iV_{OO}^{(1,1)}}{8M}\{(\rb\times\Db_R)\cdot,g\Bb\}\notag\\
 &+\frac{ic_SV_{OOa}^{(2,0)}}{16M^2}\bigl[\Db_R\times,g\Eb\bigr]\cdot\sigmab^{(1)}-\frac{iV_{O\otimes Oa}^{(2,0)}}{16M^2}\bigl[\Db_R\times,g\Eb\bigr]\cdot\sigmab^{(2)}\notag\\
 &+\frac{iV_{OOb'}^{(2,0)}}{16M^2r^2}\bigl\{(\rb\times\Db_R)\cdot,g\Eb\bigr\}(\rb\cdot\sigmab^{(1)})-\frac{iV_{OOb''}^{(2,0)}}{16M^2r^2}\left\{\left((\rb\times\Db_R)\cdot\sigmab^{(1)}\right),(\rb\cdot g\Eb)\right\}\notag\\
 &-\frac{iV_{O\otimes Ob'}^{(2,0)}}{16M^2r^2}\bigl\{(\rb\times\Db_R)\cdot,g\Eb\bigr\}(\rb\cdot\sigmab^{(2)})+\frac{iV_{O\otimes Ob''}^{(2,0)}}{16M^2r^2}\left\{\left((\rb\times\Db_R)\cdot\sigmab^{(2)}\right),(\rb\cdot g\Eb)\right\}\notag\\
 &+\frac{1}{16M^2}\left\{V^{(2,0)}_{OOc'}(\bm{r}\cdot g\Eb),\left(\bm{\nabla}_r\cdot\bm{D}_R\right)\right\}+\frac{1}{16M^2}\left\{V^{(2,0)}_{OOc''}r_igE_j,(\nabla_r)_j(D_R)_i\right\}\notag\\
 &+\frac{1}{16M^2}\left\{V^{(2,0)}_{OOc'''}r_igE_j,(\nabla_r)_i(D_R)_j\right\}+\frac{1}{16M^2}\left\{\frac{V^{(2,0)}_{OOd}}{r^2}r_ir_j(\bm{r}\cdot g\Eb),(\nabla_r)_i(D_R)_j\right\}\notag\\
 &-\left.\left.\left.\frac{iV_{OOe}^{(2,0)}}{8M^2r}\bigl\{(\rb\times\Db_R)\cdot,g\Bb\bigr\}\right)\mathrm{O}+c.c.\right]\right\}\,,
\end{align}
where $c.c.$ refers to the charge conjugate of every term inside the square brackets. In the terms of order $M^{-1}r^1$ and $M^{-2}r^0$, we include only those that contain a covariant derivative acting on the octet field, because otherwise they do not contribute to the boost transformation at the order we are interested in.\footnote{Note that in Ref.~\cite{Brambilla:2003nt} the identity~\eqref{D3ident} was not used, so the operators considered there were not all linearly independent. In particular, there are two more potentials $V_{OOb'''}^{(2,0)}$ and $V_{O\otimes Ob'''}^{(2,0)}$ there, which we do not have here as the corresponding operators are linear combinations of those in~\eqref{oo lagrangian}. Ultimately, these potentials were found to be zero in~\cite{Brambilla:2003nt}, showing that the results are unaffected by the choice.}

Applying the boost transformation to this Lagrangian, we obtain the following difference with respect to the original Lagrangian:
\begin{align}
 \partial_\mu\widehat{\Delta}^\mu\mathcal{L}^{(O)}={}&\int d^3r\,\mathrm{Tr}\left\{\mathrm{O}^{\dagger}\left(i\left(1-c_O^{(1,0)}\right)(\etab\cdot\Db_R)-\frac{1}{4M}\left(k_{OD}^{(1,0)}-c_O^{(1,0)}\right)\etab\cdot\bigl\{D_0,\Db_R\bigr\}\right.\right.\notag\\
 &-\frac{i}{M}\left(V_{p^2Oa}+V_{L^2Oa}+\frac{1}{2}k_{OD}^{(1,0)}V_O^{(0)}\right)(\etab\cdot\Db_R)\notag\\
 &+\frac{i}{Mr^2}\left(V_{L^2Oa}+\frac{r}{2}V_O^{(0)\prime}\right)(\etab\cdot\rb)(\rb\cdot\Db_R)\notag\\
 &-\left.\frac{1}{2M}\left(V_{LSOa}+\frac{1}{2r}V_O^{(0)\prime}\right)(\etab\times\rb)\cdot\left(\sigmab^{(1)}-\sigmab^{(2)}\right)\right)\mathrm{O}\notag\\
 &+\left[\mathrm{O}^\dagger\left(\frac{1}{2}\left(V_{OO}^{(1,1)}-V_{OO}^{(0,1)}\right)(\etab\times\rb)\cdot g\Bb\right.\right.\notag\\
 &+\frac{1}{4M}\left(c_SV_{OOa}^{(2,0)}-2c_FV_{OOb}^{(1,0)}+\frac{1}{2}V_{OO}^{(0,1)}+\frac{1}{2}\right)(\etab\times g\Eb)\cdot\sigmab^{(1)}\notag\\
 &-\frac{1}{4M}\left(V_{O\otimes Oa}^{(2,0)}-2V_{O\otimes Ob}^{(1,0)}+\frac{1}{2}V_{OO}^{(0,1)}-\frac{1}{2}\right)(\etab\times g\Eb)\cdot\sigmab^{(2)}\notag\\
 &-\frac{1}{4Mr^2}\left(V_{OOb'}^{(2,0)}-2V_{OOc}^{(1,0)}\right)((\etab\times\rb)\cdot g\Eb)\left(\rb\cdot\sigmab^{(1)}\right)\notag\\
 &+\frac{1}{4Mr^2}\left(V_{O\otimes Ob'}^{(2,0)}-2V_{O\otimes Oc}^{(1,0)}\right)((\etab\times\rb)\cdot g\Eb)\left(\rb\cdot\sigmab^{(2)}\right)\notag\\
 &+\frac{1}{4Mr^2}\left(V_{OOb''}^{(2,0)}+\frac{r}{2}V_{OO}^{(0,1)\prime}\right)\left((\etab\times\rb)\cdot\sigmab^{(1)}\right)(\rb\cdot g\Eb)\notag\\
 &-\frac{1}{4Mr^2}\left(V_{O\otimes Ob''}^{(2,0)}+\frac{r}{2}V_{OO}^{(0,1)\prime}\right)\left((\etab\times\rb)\cdot\sigmab^{(2)}\right)(\rb\cdot g\Eb)\notag\\
 &-\frac{i}{8M}\left\{\left(V_{OOc'}^{(2,0)}+V_{OOa}^{(1,0)}-1\right)(\rb\cdot g\Eb),(\etab\cdot\nablab_r)\right\}\notag\\
 &-\frac{i}{8M}\left\{\left(V_{OOc''}^{(2,0)}-V_{OOa}^{(1,0)}+1\right)(\etab\cdot\rb),(\nablab_r\cdot g\Eb)\right\}\notag\\
 &-\frac{i}{8M}\left\{V_{OOc'''}^{(2,0)}(\etab\cdot g\Eb)\rb\cdot,\nablab_r\right\}-\frac{i}{8M}\left\{\frac{V_{OOd}^{(2,0)}}{r^2}(\etab\cdot\rb)(\rb\cdot g\Eb)\rb\cdot,\nablab_r\right\}\notag\\
 &+\left.\left.\left.\frac{1}{2Mr}\left(V_{OOe}^{(2,0)}-V_{OOd}^{(1,0)}\right)(\etab\times\rb)\cdot g\Bb\right)\mathrm{O}+c.c.\right]\right\}\,,
\end{align}
where we have neglected the terms from the coordinate transformations. Because this difference is not a derivative, all coefficients need to vanish, from which the following constraints are derived:
\begin{align}
 k_{OD}^{(1,0)}&=c_O^{(1,0)}=1\,, & V_{p^2Oa}&=\frac{r}{2}V_O^{(0)\prime}-\frac{1}{2}V_O^{(0)}\,,\notag\\
 V_{L^2Oa}&=-\frac{r}{2}V_O^{(0)\prime}\,, & V_{LSOa}&=-\frac{1}{2r}V_O^{(0)\prime}\,,\notag\\
 c_SV_{OOa}^{(2,0)}&=2c_FV_{OOb}^{(1,0)}-\frac{1}{2}V_{OO}^{(0,1)}-\frac{1}{2}\,, & V_{O\otimes Oa}^{(2,0)}&=2V_{O\otimes Ob}^{(1,0)}-\frac{1}{2}V_{OO}^{(0,1)}+\frac{1}{2}\,,\notag\\
 V_{OOb'}^{(2,0)}&=2V_{OOc}^{(1,0)}\,, & V_{O\otimes Ob'}^{(2,0)}&=2V_{O\otimes Oc}^{(1,0)}\,,\notag\\
 V_{OOb''}^{(2,0)}&=-\frac{r}{2}V_{OO}^{(0,1)\prime}\,, & V_{O\otimes Ob''}^{(2,0)}&=-\frac{r}{2}V_{OO}^{(0,1)\prime}\,,\notag\\
 V_{OOe}^{(2,0)}&=V_{OOd}^{(1,0)}\,, & V_{OO}^{(1,1)}&=V_{OO}^{(0,1)}\,,\notag\\
 V_{OOc'}^{(2,0)}&=1-V_{OOa}^{(1,0)}\,, & V_{OOc''}^{(2,0)}&=V_{OOa}^{(1,0)}-1\,,\notag\\
 V_{OOc'''}^{(2,0)}&=0\,, & V_{OOd}^{(2,0)}&=0\,.
\label{octetconstraints}
\end{align} 
They are in agreement with~\cite{Brambilla:2003nt}. Moreover, because of the particular choice of octet fields made in this work, whose boosts are of the form~\eqref{boostoctet}, four of the constraints of~\cite{Brambilla:2003nt} have simplified significantly. These are the last four of~\eqref{octetconstraints}.\footnote{\label{footnoteoctet}
If we had tuned $q_{Oa}^{(0,2)}$ such that $\widetilde{k}_{Oa}^{(0,2)}=0$, the third and fourth last constraints of~\eqref{octetconstraints} would change into $V_{OOc'}^{(2,0)}=-V_{OOa}^{(1,0)}$ and $V_{OOc''}^{(2,0)}=V_{OOa}^{(1,0)}-2$.}

\subsubsection{Singlet-octet sector}

Finally, moving on to the singlet-octet sector, several terms that appear in the octet sector are absent because of charge conjugation invariance. In accordance with~\cite{Brambilla:2003nt}, the pNRQCD Lagrangian in the singlet-octet sector up to $\mathcal{O}(M^{-2})$ is given by
\begin{align}\label{soh lagrangian}
 \mathcal{L}_\mathrm{pNRQCD}^{(SO,\,h)}={}&\int d^3r\,\mathrm{Tr}\left[\mathrm{S}^{\dagger}\left(V_{SO}^{(0,1)}\rb\cdot g\Eb+\frac{c_FV_{SOb}^{(1,0)}}{2M}g\Bb\cdot\left(\sigmab^{(1)}-\sigmab^{(2)}\right)\right.\right.\notag\\
 &+\frac{V_{SOc}^{(1,0)}}{2Mr^2}(\rb\cdot g\Bb)\left(\rb\cdot\left(\sigmab^{(1)}-\sigmab^{(2)}\right)\right)+\frac{V_{SOd}^{(1,0)}}{Mr}\rb\cdot g\Eb\notag\\
 &-\frac{i}{4M}V_{SO}^{(1,1)}\bigl\{(\rb\times\Db_R)\cdot,g\Bb\bigr\}+\frac{ic_S}{16M^2}V_{SOa}^{(2,0)}\bigl[\Db_R\times,g\Eb\bigr]\cdot\left(\sigmab^{(1)}-\sigmab^{(2)}\right)\notag\\
 &+\frac{iV_{SOb'}^{(2,0)}}{16M^2r^2}\bigl\{(\rb\times\Db_R)\cdot,g\Eb\bigr\}\left(\rb\cdot\left(\sigmab^{(1)}-\sigmab^{(2)}\right)\right)\notag\\
 &-\frac{iV_{SOb''}^{(2,0)}}{16M^2r^2}\left\{\left((\rb\times\Db_R)\cdot\left(\sigmab^{(1)}-\sigmab^{(2)}\right)\right),(\rb\cdot g\Eb)\right\}\notag\\
 &-\left.\left.\frac{iV_{SOe}^{(2,0)}}{4M^2r}\bigl\{(\rb\times\Db_R)\cdot,g\Bb\bigr\}\right)\mathrm{O}+H.c.\right]\,,
\end{align}
where again, at orders $M^{-1}r^1$ and $M^{-2}r^0$, we only include terms with covariant derivatives acting on the quarkonium fields.\footnote{We also do not write the operator with potential $V_{SOb'''}^{(2,0)}$ of~\cite{Brambilla:2003nt}, which is a linear combination of the ones in~\eqref{soh lagrangian}.} As all operators between the large round brackets are Hermitian, we have labeled the Lagrangian with $h$. Such operators are the only ones that are allowed in the pure singlet or octet sectors.

In the singlet-octet sector, on the other hand, one may in principle also add anti-Hermitian operators. Instead of canceling, they give terms of the form $\mathrm{S}^\dagger a\mathrm{O}-\mathrm{O}^\dagger a\mathrm{S}$, where $a$ indicates the anti-Hermitian operator. We are not aware of any argument that would exclude such terms a priori, so we give here also the singlet-octet Lagrangian for anti-Hermitian operators in the large round brackets:
\begin{align}\label{soa lagrangian}
 \mathcal{L}_\mathrm{pNRQCD}^{(SO,\,a)}={}&\int d^3r\,\mathrm{Tr}\left[\mathrm{S}^{\dagger}\left(\frac{1}{2M}\left\{rV_{SOe}^{(1,0)},\nablab_r\cdot g\Eb\right\}+\frac{iV_{SOf}^{(1,0)}}{2Mr}(\rb\times g\Eb)\cdot\left(\sigmab^{(1)}+\sigmab^{(2)}\right)\right.\right.\notag\\
 &-\frac{i}{4M^2}\left\{rV_{SOf}^{(2,0)}g\Bb\cdot,(\nablab_r\times\Db_R)\right\}+\frac{V_{SOg'}^{(2,0)}}{16M^2r}\left\{(\rb\cdot g\Bb),\left(\Db_R\cdot\left(\sigmab^{(1)}+\sigmab^{(2)}\right)\right)\right\}\notag\\
 &+\frac{V_{SOg''}^{(2,0)}}{16M^2r}\left\{(\rb\cdot\Db_R),\left(g\Bb\cdot\left(\sigmab^{(1)}+\sigmab^{(2)}\right)\right)\right\}\notag\\
 &+\left.\left.\frac{V_{SOg'''}^{(2,0)}}{16M^2r}\left(\rb\cdot\left(\sigmab^{(1)}+\sigmab^{(2)}\right)\right)\left\{\Db_R\cdot,g\Bb\right\}\right)\mathrm{O}+H.c.\right]\,, 
\end{align}
Such terms were not included in the analysis of~\cite{Brambilla:2003nt}.

A boost transformation generates the following new terms in the singlet-octet Lagrangian (we neglect again the terms from the coordinate transformations):
\begin{align}
 \partial_\mu\widehat{\Delta}^\mu\mathcal{L}^{(SO,\,h)}={}&\int d^3r\,\mathrm{Tr}\left[\mathrm{S}^{\dagger}\left(\left(V_{SO}^{(1,1)}-V_{SO}^{(0,1)}\right)(\etab\times\rb)\cdot g\Bb\right.\right.\notag\\
 &+\frac{1}{4M}\left(c_SV_{SOa}^{(2,0)}-2c_FV_{SOb}^{(1,0)}+V_{SO}^{(0,1)}\right)(\etab\times g\Eb)\left(\sigmab^{(1)}-\sigmab^{(2)}\right)\notag\\
 &-\frac{1}{4Mr^2}\left(V_{SOb'}^{(2,0)}-2V_{SOc}^{(1,0)}\right)((\etab\times\rb)\cdot g\Eb)\left(\rb\cdot\left(\sigmab^{(1)}-\sigmab^{(2)}\right)\right)\notag\\
 &+\frac{1}{4Mr^2}\left(V_{SOb''}^{(2,0)}+rV_{SO}^{(0,1)\prime}\right)\left((\etab\times\rb)\cdot\left(\sigmab^{(1)}-\sigmab^{(2)}\right)\right)(\rb\cdot g\Eb)\notag\\
 &+\left.\left.\frac{1}{Mr}\left(V_{SOe}^{(2,0)}-V_{SOd}^{(1,0)}\right)(\etab\times\rb)\cdot g\Bb\right)\mathrm{O}+H.c.\right]\,,
\end{align}
\begin{align}
 \partial_\mu\widehat{\Delta}^\mu\mathcal{L}^{(SO,\,a)}={}&\int d^3r\,\mathrm{Tr}\left[\mathrm{S}^{\dagger}\left(\frac{1}{2M}\left\{r\left(V_{SOf}^{(2,0)}-V_{SOe}^{(1,0)}\right),(\etab\times\nablab_r)\cdot g\Bb\right\}\right.\right.\notag\\
 &-\frac{i}{4Mr}\left(V_{SOg'}^{(2,0)}-2V_{SOf}^{(1,0)}\right)\left(\etab\cdot\left(\sigmab^{(1)}+\sigmab^{(2)}\right)\right)(\rb\cdot g\Bb)\notag\\
 &-\frac{i}{4Mr}\left(V_{SOg''}^{(2,0)}+2V_{SOf}^{(1,0)}\right)(\etab\cdot\rb)\left(g\Bb\cdot\left(\sigmab^{(1)}+\sigmab^{(2)}\right)\right)\notag\\
 &-\left.\left.\frac{iV_{SOg'''}^{(2,0)}}{4Mr}(\etab\cdot g\Bb)\left(\rb\cdot\left(\sigmab^{(1)}+\sigmab^{(2)}\right)\right)\right)\mathrm{O}+H.c.\right]\,,
\end{align}
which leads to the constraints
\begin{align}
 V_{SO}^{(0,1)}&=V_{SO}^{(1,1)}\,, & c_SV_{SOa}^{(2,0)}&=2c_FV_{SOb}^{(1,0)}-V_{SO}^{(0,1)}\,, & &V_{SOb'}^{(2,0)}=2V_{SOc}^{(1,0)}\,,\notag\\
 V_{SOb''}^{(2,0)}&=-rV_{SO}^{(0,1)\prime}\,, & V_{SOe}^{(2,0)}&=V_{SOd}^{(1,0)}\,, & &V_{SOf}^{(2,0)}=V_{SOe}^{(1,0)}\,,\notag\\
 V_{SOg'}^{(2,0)}&=2V_{SOf}^{(1,0)}\,, & V_{SOg''}^{(2,0)}&=-2V_{SOf}^{(1,0)}\,, & &V_{SOg'''}^{(2,0)}=0\,.
 \label{singletoctetconstraints}
\end{align}
Again, these are in agreement with~\cite{Brambilla:2003nt}, except for the relations between the potentials $V_{SOe}^{(1,0)}$, $V_{SOf}^{(1,0)}$, $V_{SOf}^{(2,0)}$, $V_{SOg'}^{(2,0)}$, $V_{SOg''}^{(2,0)}$, and $V_{SOg'''}^{(2,0)}$, which are the Wilson coefficients from $\mathcal{L}_\mathrm{pNRQCD}^{(SO,\,a)}$ that are new. As already noticed in~\cite{Brambilla:2003nt}, the above constraints require the chromoelectric field to enter the Lagrangian in the combination $\rb \cdot \left( g \Eb - i \{\Db_R\times,g\Bb\}/(4M)\right)$, i.e., as in the Lorentz force.

\section{Conclusions}\label{discussion}

\subsection{Summary and implications}

In the paper, we have investigated boost transformations of nonrelativistic fields in low energy EFTs for heavy (anti)quarks, by starting from the general form allowed by charge conjugation, parity, and time reversal, while exploiting the freedom to remove redundant terms through field redefinitions. Relations among the Wilson coefficients have been derived by applying those transformations to the corresponding Lagrangian up to a certain order in the expansion, and requiring that they leave the action invariant as well as that they satisfy the Poincar\'e algebra. The results confirm known relations from the literature~\cite{Manohar:1997qy,Brambilla:2003nt,Brambilla:2008zg,Heinonen:2012km}, in both NRQCD and pNRQCD (in the equal-flavor case), and add new ones. They can be found in Eqs.~\eqref{consNRQCDafterboost1}, \eqref{consNRQCDafterboost2}, \eqref{4fermionconstraints1}, \eqref{4fermionconstraints2}, \eqref{4fermionconstraints3}, \eqref{s88relation} for NRQCD and~\eqref{singletconstraints}, \eqref{octetconstraints}, \eqref{singletoctetconstraints} for pNRQCD. Note that restricting to the singlet sector of pNRQCD provides also the relations for the strongly-coupled case. Finally, the obtained relations may be translated into relations among Wilson coefficients of NRQED~\cite{Caswell:1985ui,Kinoshita:1995mt} and potential NRQED~(pNRQED)~\cite{Pineda:1997bj,Pineda:1998kn}.

The present approach is complementary to previous methods and provides new insights into them. The derivation of the boost transformation via the induced representation in~\cite{Heinonen:2012km} gives some intuitive understanding of the form of several but not all terms appearing in the boost. In particular, it leaves open the question of how to systematically generate terms that are not present in the free case, include quantum corrections in the form of Wilson coefficients for these terms,\footnote{Examples of both issues are the terms $-c_Dg\Eb/(8M^2)-ic_Sg\Eb\times\sigmab/(8M^2)$ and the $1/M^3$ terms proportional to the chromomagnetic field in the boost transformations of the heavy (anti)quark fields in~NRQCD.} or reduce nonminimal sets of operators. The method presented here solves these issues by adopting for the boost the same approach as used in constructing the effective field theory. It consists in allowing in the boost all terms consistent with the symmetries, and in factorizing for each term possible high energy contributions into suitable Wilson coefficients. We have shown that the symmetries of the action and the Poincar\'e algebra are sufficient to fix the form of the boost generators in pNRQCD and NRQCD (at the order we have worked) and to constrain the Wilson coefficients of the EFT Lagrangians. In pNRQCD we have seen an example where several equivalent boost generators are available, and we have resolved the ambiguity by removing redundant terms via a redefinition of the heavy (anti)quark fields. This amounts to having chosen to work with fields that transform in a specific way under boosts. Different choices lead to different constraints on the Wilson coefficients in the Lagrangian, hence the constraints depend on how the fields transform under Lorentz transformations (see footnote~\ref{footnoteoctet} for a concrete example). Similar ambiguities are also expected to become relevant in NRQCD at higher orders.

The method of~\cite{Brambilla:2003nt,Brambilla:2008zg}, although similar to the one presented here, is based on the computation of the Noether charges. Constraints on the Wilson coefficients follow from requiring that the Noether charges fulfill the Poincar\'e algebra assuming canonical (anti)commutation relations for the fields. With respect to that method, the one presented in this paper provides an alternative approach where one computes boosts for each field individually and requires the Poincar\'e algebra for these boost generators (rather than the Noether charges) as well as the invariance of the action. Computing the boosts of the nonrelativistic fields has a value in itself as it may prove useful in different applications. In this way, we could extend the set of constraints derived in~\cite{Brambilla:2003nt,Brambilla:2008zg} both in NRQCD and in pNRQCD. The present method is well suited for automatization in programs capable of symbolic manipulation. Most of the results presented in this work have been checked in this way. 

In summary, the method presented here provides a straightforward way to obtain the constraints on the Wilson coefficients of a nonrelativistic EFT (where Poincar\'e invariance is not manifestly realized) induced by the underlying relativistic theory (where Poincar\'e invariance is manifestly realized). Moreover, the examples worked out in the paper lead us to conjecture that all the Wilson coefficients of the boost can be fixed, through the Poincar\'e algebra and the invariance of the action, in terms of either constants or linear combinations of Wilson coefficients appearing in the Lagrangian up to field redefinitions.

\subsection{Outlook}

The suggested method is general enough to allow for several possible extensions.
First, it may be used to constrain the nonrelativistic EFTs to orders even higher than the ones considered here. As an example, the present work has all the necessary ingredients to constrain the NRQCD Lagrangian at $\mathcal{O}(M^{-4})$ in the two-fermion sector, which has been derived in~\cite{Gunawardana:2017zix}. This would also allow comparisons to the results for NRQED presented in~\cite{Hill:2012rh}.

Second, our results can easily be extended to the case of pNRQCD with heavy quark and antiquark of different flavors. Although charge conjugation can no longer be used to constrain the form of the Lagrangian or the boost transformation and so one has to add $C$-odd terms to both, as long as one restricts the analysis purely to QCD contributions, which are insensitive to flavor, one may still use $C$ symmetry combined with a flavor exchange to eliminate terms. In addition, one has to pay attention to the fact that there are now two mass scales, $M_1$ and $M_2$, which will typically appear in the combinations of the total mass, $M_1+M_2$, or the reduced mass, $M_1M_2/(M_1+M_2)$, and the $1/M$ terms of this paper need to be adapted accordingly. Constraints for the potentials up to order $1/M_{1,2}^2$ have been known for a long time~\cite{Gromes:1984ma,Barchielli:1988zp,Brambilla:2001xk}; constraints beyond that have not been investigated so far.

Recently, reparametrization invariance has been used to organize the resummation of certain classes of higher-order operators in the HQET~\cite{Mannel:2018mqv}. It is conceivable that such a resummation may be extended to include operators related by Poincar\'e invariance in NRQCD and pNRQCD.

Furthermore, we expect that our method can be applied to theories of weakly interacting massive particles (or WIMPs). There have been various suggestions about the properties of dark matter (DM) using WIMPs during the last few decades, such as supersymmetric dark matter (SUSY-DM), axions, sterile neutrinos, etc.~\cite{Jungman:1995df,Duffy:2009ig,Drewes:2013gca}. As the mass of the DM candidates is assumed to be greater than currently accessible energy scales, the study of DM production and annihilation is largely based on nonrelativistic EFTs. The direct detection via nucleon-DM scattering processes was investigated in~\cite{Fitzpatrick:2012ix}, where the operators were constructed based on Galilean invariance and the EFT formalism.\footnote{This nonrelativistic approach has also been used to study the $X(3872)$ unconventional hadron in~\cite{Braaten:2015tga}.} Instead of Galilean invariance, one can construct the Lagrangian from Poincar\'e invariance, such as it has been done in~\cite{Hill:2014yxa, Chen:2018uqz}, and ask whether Poincar\'e invariance can give different and/or additional constraints on the operators with respect to Galilean invariance. This is justified by the fact that the underlying theory (whose explicit formulation is yet to be found) is supposed to be Poincar\'e invariant. Similarly, the method may be used to constrain operators showing up in nonrelativistic EFTs for heavy Majorana neutrinos that may be relevant in the framework of leptogenesis~\cite{Biondini:2013xua}.

The constraints derived from exploiting the spacetime symmetries of nonrelativistic EFTs can be utilized for another (conjectured) EFT of QCD, which is valid in the nonperturbative regime, the effective string theory (EST)~\cite{Nambu:1978bd,Luscher:1980fr,Kogut:1981gm,Luscher:2002qv,PerezNadal:2008vm,Brambilla:2014eaa}. The EST provides an analytic description of the gluodynamics of a static quark-antiquark system at long distances $r\Lambda_\text{QCD}\gg 1$ with the transverse vibrations of the string between the heavy quark and antiquark as the degrees of freedom. Assuming that the expectation value of a rectangular Wilson loop in the large time limit can be expressed in terms of the string partition function, one can establish a one-to-one mapping between the heavy quark potentials and  correlators of the string vibrational modes. These last ones depend on some parameters describing the unknown short distance behaviour of the EST~\cite{Kogut:1981gm,PerezNadal:2008vm,Brambilla:2014eaa}. The potentials in the long distance regime are then translated into EST correlators, which are computed analytically. Relations among the potentials due to the Poincar{\'e} invariance of QCD can be used to constrain the short distance parameters of the EST. Analyses along these lines have been carried out in~\cite{PerezNadal:2008vm,Brambilla:2014eaa,Hwang:2017kck,ESTHwangPhD}.

Finally, we would like to argue that it should be possible to extend the method presented in this work also to different theories and to different (nonmanifestly realized) symmetries. Principally, this would require adapting the symmetry argument used to restrict the allowed operators in the nonlinear field transformations to the new symmetries.

\acknowledgments

N.B.\ acknowledges the support of the DFG grant BR 4058/2-1. A.V.\ acknowledges the support of the Sino-German CRC 110 ``Symmetries and the Emergence of Structure in QCD''. This work was supported by the DFG cluster of excellence ``Origin and structure of the universe'' (www.universe-cluster.de). S.H.\ thanks Aneesh Manohar, and M.B.\ thanks Richard Hill and Mikhail Solon for discussions. S.H.\ was supported by the International Max Planck Research School of the Max-Planck-Institut f\"ur Physik (Werner-Heisenberg-Institut) M\"unchen, and M.B.\ acknowledges support by the Japanese Society for the Promotion of Science (JSPS).

\appendix

\section{Spacetime translations and rotations}\label{PoincareAlgebra}

\subsection{Transformations in NRQCD}\label{PoincareAlgebraNRQCD}

Spacetime translations act only on the coordinates, shifting the origin by a constant vector $a^\mu$. The transformed field in the new coordinate system corresponds to the original field at the coordinates before the transformation. The form of the translation generator $P_\mu$ for a generic field $\phi(x)$ can then be obtained from a Taylor expansion to first order:
\begin{equation}
 \phi(x)\stackrel{P_\mu}{\longrightarrow}\phi'(x)=\phi(x+a)=\left[1+a^\mu\partial_\mu+\mathcal{O}\left(a^2\right)\right]\phi(x)\equiv\left[1-ia^\mu P_\mu+\mathcal{O}\left(a^2\right)\right]\phi(x)\,.
\end{equation}
From this we take $P_\mu=i\partial_\mu$, or in nonrelativistic notation $P_0=i\partial_0$ and $\bm{P}=-i\nablab$. This is already the final form of the translation generator for the light quark and gluon fields, but for the heavy (anti)quark fields we need to include the effect of the field redefinitions performed to remove the mass term in the Lagrangian. This modifies the generator to
\begin{equation}
 P_\mu=e^{\pm iMt}(i\partial_\mu)e^{\mp iMt}=i\partial_\mu\pm\delta_{\mu0}M\,,
\end{equation}
so $P_0\psi=(i\partial_0+M)\psi$ and $P_0\chi=(i\partial_0-M)\chi$.

Rotations act both on the coordinates and on the field components. The coordinates are transformed under infinitesimal rotations such that $\bm{r}$ in the new coordinate system corresponds to $\bm{r}+\bm{\alpha}\times\bm{r}$ in the old, where the direction of $\bm{\alpha}$ gives the rotation axis and its absolute value gives the infinitesimal rotation angle. The components of the Pauli spinor fields are rotated with the Pauli matrix $\bm{\sigma}/2$, while the gauge fields transform as vectors, whose behavior follows directly from the coordinate transformations and Eq.~\eqref{gentrafo}:
\begin{align}
 A_0(x)&\stackrel{J}{\longrightarrow}A'_0(x) =A_0(x)+\bigl[\bm{\alpha}\cdot(\bm{r}\times\bm{\nabla}),A_0(x)\bigr] \equiv\left(1+i\bm{\alpha}\cdot\bm{j}_{A_0}\right)A_0(x)\,,\\
 \psi(x)&\stackrel{J}{\longrightarrow}\psi'(x)=\left(1+\frac{i}{2}\bm{\alpha}\cdot\bm{\sigma}\right)\psi(x)+\bigl[\bm{\alpha}\cdot(\bm{r}\times\bm{\nabla}),\psi(x)\bigr] \equiv\left(1+i\bm{\alpha}\cdot\bm{j}_{\psi}\right)\psi(x)\,,\\
 \chi(x)&\stackrel{J}{\longrightarrow}\chi'(x)=\left(1+\frac{i}{2}\bm{\alpha}\cdot\bm{\sigma}\right)\chi(x)+\bigl[\bm{\alpha}\cdot(\bm{r}\times\bm{\nabla}),\chi(x)\bigr] \equiv\left(1+i\bm{\alpha}\cdot\bm{j}_{\chi}\right)\chi(x)\,,\\
 \bm{A}(x)&\stackrel{J}{\longrightarrow}\bm{A}'(x)=\bm{A}(x)-\bm{\alpha}\times\bm{A}(x)+\bigl[\bm{\alpha}\cdot(\bm{r}\times\bm{\nabla}),\bm{A}(x)\bigr] \equiv\left(1+i\bm{\alpha}\cdot\bm{j}_{\bm{A}}\right)\bm{A}(x)\,,
\end{align}
where again we have written the universal term $\bm{r}\times(-i\nablab)$ for the coordinate transformations in the form of a commutator. We use a capital $\bm{J}$ to denote the generators of rotations in general, and a lowercase $\bm{j}$ for the particular representation.

As we have done for the boost, we can convert the transformation of the gauge field $A_\mu$ under rotations into a transformation of the covariant derivatives:
\begin{align}
 D'_0&=\partial_0+igA'_0=D_0+\bigl[\bm{\alpha}\cdot(\bm{r}\times\bm{\nabla}),D_0\bigr]\,,\\
 \bm{D}'&=\bm{\nabla}-ig\bm{A}'=\bm{D}-\bm{\alpha}\times\bm{D}+\bigl[\bm{\alpha}\cdot(\bm{r}\times\bm{\nabla}),\bm{D}\bigr]\,.
\end{align}
From these, it also follows that the chromoelectric and chromomagnetic fields $\Eb$ and $\Bb$ transform as vectors under rotations, i.e., $\bm{j_E}=\bm{j_B}=\bm{j_A}$. Expressions for the Noether charges of spacetime translations and rotations in NRQCD can be found in~\cite{Brambilla:2003nt,Vairo:2003gx}.

\subsection{Transformations in pNRQCD}\label{SpacetimepNRQCD}

It is useful to look also here at the transformation properties of $Q=\psi\chi^\dagger$, which includes both singlet and octet fields as different color projections. Time translations are straightforward in pNRQCD; $\psi$ and $\chi^\dagger$ are evaluated at the same time, so the time argument of the quarkonium fields is shifted in the same way. The additional mass terms introduced through the field redefinitions of $\psi$ and $\chi$ add up, which gives the following transformation
\begin{equation}
 Q(t,\bm{r},\bm{R})\stackrel{P_0}{\longrightarrow}Q'(t,\bm{r},\bm{R})=(1-2iMa_0)Q(t,\bm{r},\bm{R})+\bigl[a_0\partial_0,Q(t,\bm{r},\bm{R})\bigr]\,.
\end{equation}
We have assumed that the quark and the antiquark fields have the same mass $M$, so that the generator of time translations is $P_0=i\partial_0+2M$.

Space translations act only on the center-of-mass coordinate $\bm{R}$: both the heavy quark and antiquark are shifted by the same amount, so the relative coordinate remains unaffected. This means
\begin{equation}
Q(t,\bm{r},\bm{R})\stackrel{P_i}{\longrightarrow}Q'(t,\bm{r},\bm{R})=Q(t,\bm{r},\bm{R})+\bigl[\bm{a}\cdot\bm{\nabla}_R,Q(t,\bm{r},\bm{R})\bigr]\,,
\end{equation}
with the generator for space translations $\bm{P}=-i\bm{\nabla}_R$.

Under rotations, both the center-of-mass and the relative coordinates transform in the same way. The component transformations of $\psi$ and $\chi$ lead to a commutator with the quark-antiquark field and the Pauli matrices:
\begin{equation}
Q(t,\bm{r},\bm{R})\stackrel{J}{\longrightarrow}Q'(t,\bm{r},\bm{R})=Q(t,\bm{r},\bm{R})+\left[\bm{\alpha}\cdot\left(\bm{R}\times\bm{\nabla}_R+\bm{r}\times\bm{\nabla}_r+\frac{i}{2}\,\bm{\sigma}\right),Q(t,\bm{r},\bm{R})\right].
\end{equation}
With the convention for the Pauli matrices of Eq.~\eqref{sigma12}, this gives the generator of rotations as $\bm{j}_Q=\bm{R}\times(-i\bm{\nabla}_R)+\bm{r}\times(-i\bm{\nabla}_r)+\left(\bm{\sigma}^{(1)}+\bm{\sigma}^{(2)}\right)/2$. From this, it is straightforward to see that
\begin{equation}
 Q_1=\frac{1}{\sqrt{2}}\mathrm{Tr}[Q]\qquad\mathrm{and}\qquad \bm{Q}_3=\frac{1}{\sqrt{2}}\mathrm{Tr}[\bm{\sigma}Q]
\end{equation}
transform as a singlet (scalar) and triplet (vector) respectively under rotations (the trace is understood only in spin space). We can then decompose the matrix valued quark-antiquark field into
\begin{equation}
 Q=\frac{1}{\sqrt{2}}Q_1\mathbbm{1}+\frac{1}{\sqrt{2}}\bm{Q}_3\cdot\bm{\sigma}\,.
\end{equation}
The bilinears in the Lagrangian then give
\begin{equation}
 \mathrm{Tr}\left[Q^\dagger Q\right]=Q_1^\dagger Q_1+\bm{Q}_3^\dagger\cdot\bm{Q}_3\,,
\end{equation}
\begin{equation}
 \mathrm{Tr}\left[Q^\dagger \frac{i\left(\bm{\sigma}^{(1)}+\bm{\sigma}^{(2)}\right)}{2}Q\right]=\bm{Q}_3^\dagger\times\bm{Q}_3\qquad\mathrm{and}\qquad\mathrm{Tr}\left[Q^\dagger \frac{\bm{\sigma}^{(1)}-\bm{\sigma}^{(2)}}{2}Q\right]=\bm{Q}_3^\dagger Q_1+Q_1^\dagger\bm{Q}_3\,.
\end{equation}
Expressions for the Noether charges of spacetime translations and rotations in pNRQCD can be found in~\cite{Brambilla:2003nt,Vairo:2003gx}.

\section{Constraints in the four-fermion sector of NRQCD}\label{appendix}

At $\mathcal{O}\left(M^{-4}\right)$, one has to include also heavy (anti)quark fields in $\kb_\psi$ and $\kb_\chi$. The terms affecting the four-fermion Lagrangian given in Sec.~\ref{NRQCD4f} can be parametrized as follows:
\begin{align}
 \bm{\hat{k}}_\psi\Bigr|_{2f}={}&\frac{a_{11}}{M^4}\Dlr\chi\chi^\dagger+\frac{a_{12}}{M^4}\chi\nablab\chi^\dagger+\frac{a_{13}}{M^4}\chi\chi^\dagger\Dlr\notag\\
 &+\frac{a_{81}}{M^4}\Dlr T^a\chi\chi^\dagger T^a+\frac{a_{82}}{M^4}T^a\chi\Db^{ab}\chi^\dagger T^b+\frac{a_{83}}{M^4}T^a\chi\chi^\dagger\Dlr T^a\notag\\
 &+\frac{ib_{11}}{M^4}\Dlr\times\sigmab\chi\chi^\dagger-\frac{ib_{12}}{M^4}\sigmab\chi\times\nablab\chi^\dagger-\frac{ib_{13}}{M^4}\sigmab\chi\times\chi^\dagger\Dlr\notag\\*
 &+\frac{ib_{14}}{M^4}\Dlr\chi\times\chi^\dagger\sigmab+\frac{ib_{15}}{M^4}\chi\nablab\times\chi^\dagger\sigmab+\frac{ib_{16}}{M^4}\chi\chi^\dagger\Dlr\times\sigmab\notag\\
 &+\frac{ib_{81}}{M^4}\Dlr\times\sigmab T^a\chi\chi^\dagger T^a-\frac{ib_{82}}{M^4}\sigmab T^a\chi\times\Db^{ab}\chi^\dagger T^b-\frac{ib_{83}}{M^4}\sigmab T^a\chi\times\chi^\dagger\Dlr T^a\notag\\
 &+\frac{ib_{84}}{M^4}\Dlr T^a\chi\times\chi^\dagger\sigmab T^a+\frac{ib_{85}}{M^4}T^a\chi\Db^{ab}\times\chi^\dagger\sigmab T^b+\frac{ib_{86}}{M^4}T^a\chi\chi^\dagger\Dlr\times\sigmab T^a\notag\\
 &+\frac{c_{11}}{M^4}(\Dlr\cdot\sigmab)\chi\chi^\dagger\sigmab+\frac{c_{12}}{M^4}\sigma_i\chi\nabla_i\chi^\dagger\sigmab+\frac{c_{13}}{M^4}\sigma_i\chi\chi^\dagger\overleftrightarrow{D_i}\sigmab\notag\\
 &+\frac{c_{14}}{M^4}\overleftrightarrow{D_i}\sigmab\chi\chi^\dagger\sigma_i+\frac{c_{15}}{M^4}\sigmab\chi\nabla_i\chi^\dagger\sigma_i+\frac{c_{16}}{M^4}\sigmab\chi\chi^\dagger(\Dlr\cdot\sigmab)\notag\\
 &+\frac{c_{17}}{M^4}\Dlr\sigma_i\chi\chi^\dagger\sigma_i+\frac{c_{18}}{M^4}\sigma_i\chi\nablab\chi^\dagger\sigma_i+\frac{c_{19}}{M^4}\sigma_i\chi\chi^\dagger\Dlr\sigma_i\notag\\
 &+\frac{c_{81}}{M^4}(\Dlr\cdot\sigmab)T^a\chi\chi^\dagger\sigmab T^a+\frac{c_{82}}{M^4}\sigma_i T^a\chi D^{ab}_i\chi^\dagger\sigmab T^b+\frac{c_{83}}{M^4}\sigma_iT^a\chi\chi^\dagger\overleftrightarrow{D_i}\sigmab T^a\notag\\
 &+\frac{c_{84}}{M^4}\overleftrightarrow{D_i}\sigmab T^a\chi\chi^\dagger\sigma_i T^a+\frac{c_{85}}{M^4}\sigmab T^a\chi D^{ab}_i\chi^\dagger\sigma_i T^b+\frac{c_{86}}{M^4}\sigmab T^a\chi\chi^\dagger(\Dlr\cdot\sigmab)T^a\notag\\
 &+\frac{c_{87}}{M^4}\Dlr\sigma_iT^a\chi\chi^\dagger\sigma_i T^a+\frac{c_{88}}{M^4}\sigma_i T^a\chi\Db^{ab}\chi^\dagger\sigma_i T^b+\frac{c_{89}}{M^4}\sigma_iT^a\chi\chi^\dagger\Dlr\sigma_iT^a\,,\label{appB1}\\
 \bm{\hat{k}}_\chi \Bigr|_{2f}  = {}&\bm{\hat{k}}_\psi \Bigr|_{2f} (\psi\leftrightarrow\chi)\label{appB2}\,.
\end{align}
Here we understand the left-right derivatives on the left hand side of $\chi\chi^\dagger$ as
\begin{equation}
 \Dlr T\chi\chi^\dagger T\psi=T(\Db\chi)\chi^\dagger T\psi+\Db(T\chi\chi^\dagger T\psi)\,,
\end{equation}
and implicitly perform an integration by parts on the second term. The overall spatial derivatives introduced by this integration are irrelevant for everything that will be discussed in this paper, so we will ignore them. This definition then also implies that the left-derivative part of $\Dlr$ acts also on the terms outside the bilinear in which it appears. The left-right derivatives on the right hand side of $\chi\chi^\dagger$ are defined as above. As an example we give the boost transformation proportional to $a_{11}$ and $a_{13}$ due to the $\chi$ field in $\psi^\dagger\chi\chi^\dagger\psi$:
\begin{equation}
 \psi^\dagger(\bm{\hat{k}}_\chi\chi)\chi^\dagger\psi \Bigr|_{a_{11},a_{13}} = \frac{a_{11}}{M^4}(\psi^\dagger\Dlr\psi)\psi^\dagger\chi\chi^\dagger\psi-\frac{a_{11}}{M^4}\psi^\dagger\psi\psi^\dagger\chi(\nablab\chi^\dagger\psi)+\frac{a_{13}}{M^4}\psi^\dagger\psi(\psi^\dagger\Dlr\chi)\chi^\dagger\psi+\dots\,.
\end{equation}

When we calculate the commutator of two boosts at $\mathcal{O}(M^{-3})$ and consider only the two-fermion part, we get some constraints on the boost coefficients $a_{nm}$, $b_{nm}$ and $c_{nm}$. At this order, only the terms with a center-of-mass derivative do not cancel automatically, and none of the $a_{nm}$ coefficients can appear, because they do not give terms antisymmetric in $\xib$ and~$\etab$.

There are again two contributions to this commutator; at $\mathcal{O}(M^{-3})$, the first of them is
\begin{align}
 -{}&\left[\xib\cdot\bm{\hat{k}}_\psi \Bigr|_{2f} ,M\etab\cdot\rb\right]+\left[\etab\cdot\bm{\hat{k}}_\psi \Bigr|_{2f},M\xib\cdot\rb\right]\notag\\
 ={}&-\frac{2i}{M^3}(b_{11}+b_{12}+b_{13})(\xib\times\etab)\cdot\sigmab\chi\chi^\dagger-\frac{2i}{M^3}(b_{14}+b_{15}+b_{16})\chi\chi^\dagger(\xib\times\etab)\cdot\sigmab\notag\\
 &-\frac{2i}{M^3}(b_{81}+b_{82}+b_{83})(\xib\times\etab)\cdot\sigmab T^a\chi\chi^\dagger T^a-\frac{2i}{M^3}(b_{84}+b_{85}+b_{86})T^a\chi\chi^\dagger(\xib\times\etab)\cdot\sigmab T^a\notag\\
 &+\frac{1}{M^3}(c_{11}+c_{12}+c_{13}-c_{14}-c_{15}-c_{16})(\xib\times\etab)\cdot(\sigmab\chi\times\chi^\dagger\sigmab)\notag\\
 &+\frac{1}{M^3}(c_{81}+c_{82}+c_{83}-c_{84}-c_{85}-c_{86})(\xib\times\etab)\cdot(\sigmab T^a\chi\times\chi^\dagger\sigmab T^a)\,.
\end{align}
The second contribution comes from the transformation of the $\chi$ fields inside $\bm{\hat{k}}_\psi|_{2f} $
\begin{align}
 -\,&\etab\cdot\frac{\partial}{\partial\bm{\tilde{\eta}}}\left[i\xib\cdot\bm{\hat{k}}_\psi\Bigr|_{2f}(D_0,\Db,\Eb,\Bb,(1+iM\bm{\tilde{\eta}}\cdot\rb)\chi,\psi)\right]_{\tilde{\eta}=0}\notag\\
 +\,&\xib\cdot\frac{\partial}{\partial\bm{\tilde{\xi}}}\left[i\etab\cdot\bm{\hat{k}}_\psi\Bigr|_{2f}(D_0,\Db,\Eb,\Bb,(1+iM\bm{\tilde{\xi}}\cdot\rb)\chi,\psi)\right]_{\tilde{\xi}=0}\notag\\
 ={}&\frac{2i}{M^3}(b_{11}-b_{12}+b_{13})(\xib\times\etab)\cdot\sigmab\chi\chi^\dagger+\frac{2i}{M^3}(b_{14}-b_{15}+b_{16})\chi\chi^\dagger(\xib\times\etab)\cdot\sigmab\notag\\
 &+\frac{2i}{M^3}(b_{81}-b_{82}+b_{83})(\xib\times\etab)\cdot\sigmab T^a\chi\chi^\dagger T^a+\frac{2i}{M^3}(b_{84}-b_{85}+b_{86})T^a\chi\chi^\dagger(\xib\times\etab)\cdot\sigmab T^a\notag\\
 &-\frac{1}{M^3}(c_{11}-c_{12}+c_{13}-c_{14}+c_{15}-c_{16})(\xib\times\etab)\cdot(\sigmab\chi\times\chi^\dagger\sigmab)\notag\\
 &-\frac{1}{M^3}(c_{81}-c_{82}+c_{83}-c_{84}+c_{85}-c_{86})(\xib\times\etab)\cdot(\sigmab T^a\chi\times\chi^\dagger\sigmab T^a)\,,
\end{align}
where we have kept only terms linear in $\xib$ and $\etab$.

The sum of these contributions has to vanish, thus we have
\begin{align}\label{2fboostcomm}
 0=&-\frac{4i}{M^3}b_{12}(\xib\times\etab)\cdot\sigmab\chi\chi^\dagger-\frac{4i}{M^3}b_{82}(\xib\times\etab)\cdot\sigmab T^a\chi\chi^\dagger T^a\notag\\
 &-\frac{4i}{M^3}b_{15}\chi\chi^\dagger
 (\xib\times\etab)\cdot\sigmab-\frac{4i}{M^3}b_{85}T^a\chi\chi^\dagger(\xib\times\etab)\cdot\sigmab T^a\notag\\
 &+\frac{2}{M^3}\left(c_{12}-c_{15}\right)(\xib\times\etab)\cdot(\sigmab\chi\times\chi^\dagger\sigmab)\notag\\
 &+\frac{2}{M^3}\left(c_{82}-c_{85}\right)(\xib\times\etab)\cdot(\sigmab T^a\chi\times\chi^\dagger\sigmab T^a)\,,
\end{align}
which fixes the two-fermion boost parameters to be
\begin{equation}\label{commrelM4}
 b_{12}=b_{15}=b_{82}=b_{85}=0\,,\hspace{15pt}c_{12}=c_{15}\,,\hspace{15pt}c_{82}=c_{85}\,.
\end{equation}

\newpage
At $\mathcal{O}(M^{-4})$ there is no relevant new information from the boost commutator. As the boost generator for the heavy (anti)quark field does not contain an $\mathcal{O}(M^0)$ term, the only contribution to the commutator at this order comes from the boost of the derivatives in Eq.~\eqref{appB1}. As before, all terms with coefficients $a_{mn}$ and the last row of $c_{mn}$ coefficients in both singlet and octet sectors are proportional to $\xib\cdot\etab$ and cancel in the commutator. For each of the remaining terms there is a corresponding contribution from the $\mathcal{O}(M^{-5})$ boost generator, which can be obtained by inserting $i\overleftrightarrow{D_0}$ or $i\partial_0$ at the respective position (left of $\chi$, right of $\chi^\dagger$, or between them) into a term from Eq.~\eqref{appB1} with a $b_{m2}$, $b_{m5}$, $c_{m2}$, or $c_{m5}$ coefficient. The commutator with $\pm iM\etab\cdot\rb$ will then cancel the spatial derivative and give a term of the same form as the $\mathcal{O}(M^{-4})$ boost of $\bm{\hat{k}}_\psi|_{2f}$. There are other boost terms at $\mathcal{O}(M^{-5})$ with a left-right spatial derivative instead of a center-of-mass derivative, or with a chromoelectric field instead of the two derivatives, but in all these cases the contributions from the $\pm iM\etab\cdot\rb$ terms to the commutator cancel each other like they did in Eq.~\eqref{2fboostcomm}. So the only constraints we get from the boost commutator at this order are ones that fix $\mathcal{O}(M^{-5})$ coefficients in terms of lower order boost coefficients.

In order to get constraints from the boost transformation of $\mathcal{L}$ at $\mathcal{O}(M^{-4})$, we need all four-fermion operators of $\mathcal{O}(M^{-4})$, 
most of which can be found in~\cite{Brambilla:2008zg} (the last two operators multiplying $s_{8-8}(^1S_0,{^3S_1})$ and $s_{8-8}(^3S_1,{^3S_1})$ are new),
\begin{align}
 \mathcal{L}^{(4)}\Bigr|_{4f} = {}&-\frac{g_1(^1S_0)}{8M^4}\left(\psi^\dagger\Dlr^2\chi\chi^\dagger\psi+\psi^\dagger\chi\chi^\dagger\Dlr^2\psi\right)\notag\\
 &-\frac{g_1(^3S_1)}{8M^4}\left(\psi^\dagger(\Dlr^2)\sigmab\chi\cdot\chi^\dagger\sigmab\psi+\psi^\dagger\sigmab\chi\chi^\dagger(\Dlr^2)\sigmab\psi\right)\notag\\
 &-\frac{g_1(^3S_1,{^3D_1})}{8M^4}\left(\frac{1}{2}\psi^\dagger\left\{(\Dlr\cdot\sigmab),\Dlr\right\}\chi\cdot\chi^\dagger\sigmab\psi -\frac{1}{3}\psi^\dagger(\Dlr^2)\sigmab\chi\cdot\chi^\dagger\sigmab\psi +H.c.\right) \notag\\ 
 &-\frac{g_8(^1S_0)}{8M^4}\left(\psi^\dagger\Dlr^2T^a\chi\chi^\dagger T^a\psi+\psi^\dagger\chi\chi^\dagger\Dlr^2T^a\psi\right)\notag\\
 &-\frac{g_8(^3S_1)}{8M^4}\left(\psi^\dagger(\Dlr^2)\sigmab T^a\chi\cdot\chi^\dagger\sigmab T^a\psi+\psi^\dagger\sigmab T^a\chi\chi^\dagger(\Dlr^2)T^a\sigmab\psi\right)\notag\\
 &-\frac{g_8(^3S_1,{^3D_1})}{8M^4}\left(\frac{1}{2}\psi^\dagger\left\{(\Dlr\cdot\sigmab),\Dlr\right\}T^a\chi\cdot\chi^\dagger\sigmab T^a\psi \right. \notag\\
 &\phantom{-\frac{g_8(^3S_1,{^3D_1})}{8M^4}}\left. -\frac{1}{3}\psi^\dagger(\Dlr^2)\sigmab T^a\chi\cdot\chi^\dagger\sigmab T^a\psi + H.c.\right)\notag\\ 
 &-\frac{f_1(^1P_1)}{4M^4}\psi^\dagger\Dlr\chi\cdot\chi^\dagger\Dlr\psi\notag\\
 &-\frac{f_1(^3P_0)}{12M^4}\psi^\dagger(\Dlr\cdot\sigmab)\chi\chi^\dagger(\Dlr\cdot\sigmab)\psi\notag\\
 &-\frac{f_1(^3P_1)}{8M^4}\left(\psi^\dagger\overleftrightarrow{D_i}\sigma_j\chi\chi^\dagger\overleftrightarrow{D_i}\sigma_j\psi-\psi^\dagger\overleftrightarrow{D_i}\sigma_j\chi\chi^\dagger\overleftrightarrow{D_j}\sigma_i\psi\right)\notag\\
 &-\frac{f_1(^3P_2)}{4M^4}\left(\frac{1}{2}\psi^\dagger\overleftrightarrow{D_i}\sigma_j\chi\chi^\dagger
\overleftrightarrow{D_i}\sigma_j\psi+\frac{1}{2}\psi^\dagger\overleftrightarrow{D_i}\sigma_j\chi\chi^\dagger\overleftrightarrow{D_j}\sigma_i\psi\right.\notag\\
 &\phantom{-\frac{f_1(^3P_2)}{4M^4}}\left.-\frac{1}{3}\psi^\dagger(\Dlr\cdot\sigmab)\chi\chi^\dagger(\Dlr\cdot\sigmab)\psi\right)\notag\\
 &-\frac{f_8(^1P_1)}{4M^4}\psi^\dagger\Dlr T^a\chi\cdot\chi^\dagger\Dlr T^a\psi\notag\\
 &-\frac{f_8(^3P_0)}{12M^4}\psi^\dagger(\Dlr\cdot\sigmab)T^a\chi\chi^\dagger(\Dlr\cdot\sigmab)T^a\psi\notag\\
 &-\frac{f_8(^3P_1)}{8M^4}\left(\psi^\dagger\overleftrightarrow{D_i}\sigma_jT^a\chi\chi^\dagger\overleftrightarrow{D_i}\sigma_jT^a\psi-\psi^\dagger\overleftrightarrow{D_i}\sigma_jT^a\chi\chi^\dagger\overleftrightarrow{D_j}\sigma_iT^a\psi\right)\notag\\
 &-\frac{f_8(^3P_2)}{4M^4}\left(\frac{1}{2}\psi^\dagger\overleftrightarrow{D_i}\sigma_jT^a\chi\chi^\dagger\overleftrightarrow{D_i}\sigma_jT^a\psi
+\frac{1}{2}\psi^\dagger\overleftrightarrow{D_i}\sigma_jT^a\chi\chi^\dagger\overleftrightarrow{D_j}\sigma_iT^a\psi\right.\notag\\
 &\phantom{-\frac{f_8(^3P_2)}{4M^4}}\left.-\frac{1}{3}\psi^\dagger(\Dlr\cdot\sigmab)T^a\chi\chi^\dagger(\Dlr\cdot\sigmab)T^a\psi\right)\notag\\
 &-\frac{if_{1\,{\rm cm}}}{2M^4}\bigl(\psi^\dagger(\Dlr\times\sigmab)\chi\cdot\nablab\chi^\dagger\psi+(\nablab\psi^\dagger\chi)\cdot\chi^\dagger(\Dlr\times\sigmab)\psi\bigr)\notag\\
 &+\frac{if'_{1\,{\rm cm}}}{2M^4}\bigl(\psi^\dagger\Dlr\chi\cdot(\nablab\times\chi^\dagger\sigmab\psi)+(\nablab\times\psi^\dagger\sigmab\chi)\cdot\chi^\dagger\Dlr\psi\bigr)\notag\\
 &-\frac{if_{8\,{\rm cm}}}{2M^4}\bigl(\psi^\dagger(\Dlr\times\sigmab)T^a\chi\cdot\Db^{ab}\chi^\dagger T^b\psi+(\Db^{ab}\psi^\dagger T^b\chi)\cdot\chi^\dagger(\Dlr\times\sigmab)T^a\psi\bigr)\notag\\
 &+\frac{if'_{8\,{\rm cm}}}{2M^4}\bigl(\psi^\dagger\Dlr T^a\chi\cdot(\Db^{ab}\times\chi^\dagger\sigmab T^b\psi)+(\Db^{ab}\times\psi^\dagger\sigmab T^b\chi)\cdot\chi^\dagger\Dlr T^a\psi\bigr)\notag\\
 &+\frac{g_{1a\,{\rm cm}}}{M^4}\left(\nabla_i\psi^\dagger\sigma_j\chi\right)\left(\nabla_i\chi^\dagger\sigma_j\psi\right)+\frac{g_{8a\,{\rm cm}}}{M^4}\left(D_i^{ab}\psi^\dagger\sigma_jT^b\chi\right)\left(D_i^{ac}\chi^\dagger\sigma_jT^c\psi\right)\notag\\
 &+\frac{g_{1b\,{\rm cm}}}{M^4}\left(\nablab\cdot\psi^\dagger\sigmab\chi\right)\left(\nablab\cdot\chi^\dagger\sigmab\psi\right)
+\frac{g_{8b\,{\rm cm}}}{M^4}\left(\Db^{ab}\cdot\psi^\dagger\sigmab T^b\chi\right)\left(\Db^{ac}\cdot\chi^\dagger\sigmab T^c\psi\right)\notag\\
 &+\frac{g_{1c\,{\rm cm}}}{M^4}\left(\nablab\psi^\dagger\chi\right)\cdot\left(\nablab\chi^\dagger\psi\right)+\frac{g_{8c\,{\rm cm}}}{M^4}\left(\Db^{ab}\psi^\dagger T^b\chi\right)\cdot\left(\Db^{ac}\chi^\dagger T^c\psi\right)\notag\\
 &+\frac{s_{1-8}(^1S_0,{^3S_1})}{2M^4}\left(\psi^\dagger g\Bb\cdot\sigmab\chi\chi^\dagger\psi+\psi^\dagger\chi\chi^\dagger g\Bb\cdot\sigmab\psi\right)\notag\\
 &+\frac{s_{1-8}(^3S_1,{^1S_0})}{2M^4}\left(\psi^\dagger g\Bb\chi\cdot\chi^\dagger\sigmab\psi+\psi^\dagger\sigmab\chi\cdot\chi^\dagger g\Bb\psi\right)\notag\\
 &+\frac{s_{8-8}(^1S_0,{^3S_1})}{2M^4}d^{abc}g\Bb^a\cdot\left(\psi^\dagger\sigmab T^b\chi\chi^\dagger T^c\psi+\psi^\dagger T^b\chi\chi^\dagger\sigmab T^c\psi\right)\notag\\
 &+\frac{s_{8-8}(^3S_1,{^3S_1})}{2M^4}f^{abc}g\Bb^a\cdot\left(\psi^\dagger\sigmab T^b\chi\times\chi^\dagger\sigmab T^c\psi\right),
\end{align}
and all four-fermion operators with a center-of-mass derivative of $\mathcal{O}(M^{-5})$, which are new,
\begin{align}
\mathcal{L}^{(5)}\Bigr|_{4f,\,{\rm cm}} = {}&\frac{is_{1-8\,{\rm cm}}}{2M^5}\left(\psi^\dagger g\Eb\times\sigmab\chi\cdot\nablab\chi^\dagger\psi-(\nablab\psi^\dagger\chi)\cdot\chi^\dagger g\Eb\times\sigmab\psi\right)\notag\\
 &-\frac{is'_{1-8\,{\rm cm}}}{2M^5}\left(\psi^\dagger g\Eb\chi\cdot(\nablab\times\chi^\dagger\sigmab\psi)-(\nablab\times\psi^\dagger\sigmab\chi)\cdot\chi^\dagger g\Eb\psi\right)\notag\\
 &+\frac{is_{8-8\,{\rm cm}}}{2M^5}d^{abc}g\Eb^a\cdot\left(\psi^\dagger\sigmab T^b\chi\times\Db^{cd}\chi^\dagger T^d\psi+(\Db^{bd}\psi^\dagger T^d\chi)\times\chi^\dagger\sigmab T^c\psi\right)\notag\\
 &+\frac{is'_{8-8\,{\rm cm}}}{2M^5}f^{abc}gE_i^a\left(\psi^\dagger\sigma_iT^b\chi(\Db^{cd}\cdot\chi^\dagger\sigmab T^d\psi)+(\Db^{bd}\cdot\psi^\dagger\sigmab T^d\chi)\chi^\dagger\sigma_iT^c\psi\right)\,.
\end{align}

For dimensional reasons the $\mathcal{O}(M^{-5})$ four-fermion Lagrangian can either contain three derivatives or one derivative and one gluon field. 
Parity allows only the combination of a chromoelectric field and a derivative. 
As stated above, only operators with center-of-mass derivatives are relevant for this order of the boost transformation.
In principle, one can write more operators with a center-of-mass derivative, but, once integrated by parts and neglecting overall derivatives, those operators reduce to ones with a derivative acting on the chromoelectric field, e.g., 
\begin{equation}
 if^{abc}g\Eb^a\cdot\left(\psi^\dagger T^b\chi\Db^{cd}\chi^\dagger T^d\psi+(\Db^{bd}\psi^\dagger T^d\chi)\chi^\dagger T^c\psi\right)=-(\Db^{ad}\cdot g\Eb^d)\,if^{abc}\psi^\dagger T^b\chi\chi^\dagger T^c\psi\,.
\end{equation}
Such terms do not contribute to the boost transformation of the Lagrangian at~$\mathcal{O}(M^{-4})$. 

After a somewhat lengthy calculation of the boost transformation of the Lagrangian at $\mathcal{O}(M^{-4})$, which we have checked with the help of a code for symbolic calculations, we obtain the following constraints:
\begin{equation*}
 a_{11}=\frac{1}{4}g_1(^1S_0)\,,\hspace{10pt}a_{12}=-\frac{1}{4}f_1(^1S_0)\,,\hspace{10pt}a_{13}=\frac{1}{4}f_1(^1P_1)\,,
\end{equation*}
\begin{equation*}
 a_{81}=\frac{1}{4}g_8(^1S_0)\,,\hspace{10pt}a_{82}=-\frac{1}{4}f_8(^1S_0)\,,\hspace{10pt}a_{83}=\frac{1}{4}f_8(^1P_1)\,,
\end{equation*}
\begin{equation*}
 b_{12}=b_{15}=0\,,\hspace{10pt}b_{13}=-\frac{1}{4}f_1(^3S_1)+b_{14}\,,\hspace{10pt}b_{16}=-\frac{1}{4}f_1(^1S_0)+b_{11}\,,
\end{equation*}
\begin{equation*}
 b_{82}=b_{85}=0\,,\hspace{10pt}b_{83}=-\frac{1}{4}f_8(^3S_1)+b_{84}\,,\hspace{10pt}b_{86}=-\frac{1}{4}f_8(^1S_0)+b_{81}\,,
\end{equation*}
\begin{equation*}
 c_{11}=\frac{1}{8}\left(g_1(^3S_1,{^3D_1})-f_1(^3S_1)\right)\,,\hspace{30pt}c_{13}=\frac{1}{8}\left(f_1(^3P_2)-f_1(^3P_1)\right)\,,
\end{equation*}
\begin{equation*}
 c_{14}=\frac{1}{8}\left(g_1(^3S_1,{^3D_1})+f_1(^3S_1)\right)\,,\hspace{30pt}c_{16}=\frac{1}{12}\left(f_1(^3P_0)-f_1(^3P_2)\right)\,,
\end{equation*}
\begin{equation*}
 c_{17}=\frac{1}{12}\left(3g_1(^3S_1)-g_1(^3S_1,{^3D_1})\right)\,,\hspace{30pt}c_{19}=\frac{1}{8}\left(f_1(^3P_1)+f_1(^3P_2)\right)\,,
\end{equation*}
\begin{equation*}
 c_{81}=\frac{1}{8}\left(g_8(^3S_1,{^3D_1})-f_8(^3S_1)\right)\,,\hspace{30pt}c_{83}=\frac{1}{8}\left(f_8(^3P_2)-f_8(^3P_1)\right)\,,
\end{equation*}
\begin{equation*}
 c_{84}=\frac{1}{8}\left(g_8(^3S_1,{^3D_1})+f_8(^3S_1)\right)\,,\hspace{30pt}c_{86}=\frac{1}{12}\left(f_8(^3P_0)-f_8(^3P_2)\right)\,, 
\end{equation*}
\begin{equation*}
 c_{87}=\frac{1}{12}\left(3g_8(^3S_1)-g_8(^3S_1,{^3D_1})\right)\,,\hspace{30pt}c_{89}=\frac{1}{8}\left(f_8(^3P_1)+f_8(^3P_2)\right)\,,
\end{equation*}
\begin{equation*}
 c_{15}=-c_{12}\,,\hspace{10pt}c_{18}=-\frac{1}{4}f_1(^3S_1)\,,\hspace{10pt}c_{85}=-c_{82}\,,\hspace{10pt}c_{88}=-\frac{1}{4}f_8(^3S_1)\,,
\end{equation*}
\begin{equation*}
 s_{1-8\,{\rm cm}}-\frac{1}{2}s_{1-8}(^1S_0,{^3S_1})-\frac{c_F-1}{2}f_1(^1S_0)-\frac{c_F-1}{12}f_8(^3S_1)-2b_{11}-\frac{1}{3}b_{84}=0\,,
\end{equation*}
\begin{equation*}
 s'_{1-8\,{\rm cm}}-\frac{1}{2}s_{1-8}(^3S_1,{^1S_0})-\frac{c_F-1}{2}f_1(^3S_1)-\frac{c_F-1}{12}f_8(^1S_0)-2b_{14}-\frac{1}{3}b_{81}=0\,,
\end{equation*}
\begin{equation*}
 s_{8-8\,{\rm cm}}-\frac{1}{2}s_{8-8}(^1S_0,{^3S_1})-\frac{c_F-1}{4}f_8(^1S_0)-\frac{c_F-1}{4}f_8(^3S_1)-b_{81}-b_{84}=0\,,
\end{equation*}
\begin{equation}
 s'_{8-8\,{\rm cm}}+\frac{1}{2}s_{8-8}(^3S_1,{^3S_1})-\frac{4c_F-3}{8}f_8(^3S_1)+c_{82}=0\,,
\end{equation}
where we have also used the already obtained constraints~\eqref{consNRQCDafterboost2} and~\eqref{4fermionconstraints1}-\eqref{4fermionconstraints3} in order to express everything in terms of Wilson coefficients of the lowest order.

So far none of these constraints involves only Wilson coefficients of the Lagrangian, they rather define the boost parameters of $\bm{\hat{k}}_\psi|_{2f}$ and $\bm{\hat{k}}_\chi|_{2f}$. There remain two unconstrained boost parameters, $c_{12}$ and one of either $b_{11}$, $b_{14}$, $b_{81}$, or $b_{84}$. But if we combine them with the relations obtained from the commutator of two boosts, we get
\begin{equation}\label{s88relation}
 c_{12}=c_{15}=c_{82}=c_{85}=0\,,\hspace{15pt}s'_{8-8\,{\rm cm}}+\frac{1}{2}s_{8-8}(^3S_1,{^3S_1})-\frac{4c_F-3}{8}f_8(^3S_1)=0\,.
\end{equation}
The last equation now gives a new constraint on the Wilson coefficients without any parameters from the boost. 
The other relations that we derived for $b_{12}$, $b_{15}$, $b_{82}$ and $b_{85}$ in Eq.~\eqref{commrelM4} from the commutator of two boosts 
are consistent with the ones obtained from the transformation of the Lagrangian at $\mathcal{O}(M^{-4})$ and $\mathcal{O}(M^{-3})$.

\bibliography{library}
 
\end{document}